\documentclass[twocolumn,preprintnumbers,amsmath,amssymb,pra]{revtex4}
\usepackage{epstopdf}
\usepackage[colorlinks,linkcolor=blue,anchorcolor=blue,citecolor=blue]{hyperref}
\usepackage{txfonts}
\usepackage{subfigure}
\usepackage{graphicx,hyperref}
\usepackage{dcolumn}
\usepackage{bm}
\usepackage{extarrows}
\usepackage{setspace}
\usepackage{float}
 \allowdisplaybreaks
\usepackage{multirow} 
\usepackage{booktabs} 
\def\arctanh{\operatorname{arctanh}}
\def\sech{\operatorname{sech}}
\def\aa{\operatorname{a}}
\def\exp{\operatorname{e}}
 \allowdisplaybreaks

\begin{document}
\bibliographystyle{plainnat}
\title{Scattering of one-dimensional quantum droplets by a reflectionless potential well}
\author{Xiaoxiao Hu$^{1}$}
\author{Zhiqiang Li$^{1}$}
\author{Yu Guo$^{2}$}
\author{Yajiang Chen$^{1}$}
\author{Xiaobing Luo$^{1,3}$}
\altaffiliation{Corresponding author: xiaobingluo2013@aliyun.com}
\affiliation{$^{1}$Department of Physics, Zhejiang Sci-Tech University, Hangzhou, 310018, China}
\affiliation{$^{2}$Hunan Provincial Key Laboratory of Flexible Electronic Materials Genome Engineering, School of Physics and Electronic Science, Changsha University of Science and Technology, Changsha 410114, China}
\affiliation{$^{3}$School of Mathematics and Physics, Jinggangshan University, Ji'an 343009, China}

\date{\today}
\begin{abstract}
We investigate, both analytically and numerically, the scattering of one-dimensional quantum droplets by a P\"{o}schl-Teller reflectionless potential well, confirming that there is a sharp transition between full reflection and full transmission at a certain critical incident speed for both small droplets and large flat-top droplets. We observe sharp differences between small quantum droplet scattering and large quantum droplet scattering. The scattering of small quantum droplets is similar to that of solitons, where a spatially symmetric trapped mode is formed at the critical speed, whereas for large quantum droplets a spatially asymmetric trapped mode is formed. Additionally, a nonmonotonous dependence of the critical speed on the atom number is identified$:$ on the small-droplet side, the critical speed increases with the atom number, while in the flat-top regime, the critical speed decreases with increasing the atom number. Strikingly, the scattering excites internal modes below the particle-emission threshold, preventing the quantum droplets from emitting radiation upon interaction with the potential. Analysis of the small-amplitude excitation spectrum shows that as the number of particles increases, it becomes increasingly difficult to emit particles outside the droplet during scattering, while radiation from solitons cannot be completely avoided. Finally, we study the collision of two quantum droplets at the reflectionless potential, revealing the role of the $\pi$-phase difference ``generator'' played by the reflectionless potential.
\end{abstract}
\maketitle
\section{Introduction}
Van der Waals theory shows that at high densities the liquid state is created by a balance of interatomic attraction and short-range repulsion. However, quantum droplet generation in ultracold and very dilute atomic gases does not follow this classical liquid concept \cite{Cabrera2018,Cheiney2018,Semeghini2018,Errico2019,Ferrier-Barbut2016,B_ottcher2019,Chomaz2019}, and the generation of quantum droplets is purely a manifestation of quantum nature, where the competition between mean-field (MF) interactions and beyond-mean-field (BMF) corrections leads to stabilization of the system in self-bound droplet states \cite{Chomaz2016,Cappellaro2018,Bruun2018}. The BMF contribution accounts for the leading correction to the ground-state energy for a weakly interacting Bose gas \cite{Lee1957}, pioneered by Lee, Huang, and Yang (LHY), which stabilizes the gas to prevent collapse due to MF effects \cite{Petrov2015,ttcher2020,Luo2020,Malomed2021}, hence termed the LHY correction. The new type of self-bound quantum liquid states has been realized experimentally in short-range interacting bosonic mixtures of homonuclear \cite{Cabrera2018,Cheiney2018,Semeghini2018} and heteronuclear \cite{Errico2019}, and in dipolar gases \cite{Ferrier-Barbut2016,B_ottcher2019,Chomaz2019} and their mixtures \cite{ttcher2020,Luo2020}, manifesting the crucial role played by the beyond-mean-field quantum fluctuation \cite{ttcher2020,Luo2020,Malomed2021}. With a focus on short-range interacting droplets, experiments have investigated their dynamical formation \cite{Semeghini2018,Ferioli2020}, the crossover from droplet to gaseous BEC \cite{Cheiney2018,Cappellaro2018,Cui2021}, and collision dynamics \cite{Ferioli2019}. The dimensional crossover for the beyond-mean-field correction in Bose gases has also been analyzed in detail. The BMF contribution comes from the zero point energy of all Bogliubov modes, which is heavily dependent on the density of states and thus on the dimensionality of the system. Correspondingly, the sign and the structure of the beyond-mean-field terms differ in the different dimensional geometries \cite{Zin2018,Ilg2018,Lavoine2021}. One-dimensional (1D) droplets, governed by attractive LHY corrections, differ fundamentally from three-dimensional (3D) and two-dimensional (2D) droplets (with repulsive LHY corrections) \cite{Ferioli2020}. In particular, the stability of 1D quantum droplets has significant advantages, as the three-body loss in a 1D droplet is greatly reduced compared to its 3D counterpart \cite{Lavoine2021,Astrakharchik2006}. To date, within the framework of the modified Gross-Pitaevskii equation\cite{Ferioli2019,Petrov2016,Astrakharchik2018,Otajonov2019,Tylutki2020,Mithun2020,Fort2021}, the behaviors of 1D quantum droplets, such as collective excitations and dynamics \cite{Lavoine2021,Astrakharchik2018,Otajonov2019,Tylutki2020,Mithun2020}, confinement \cite{Pathak2022,Debnath2020,Englezos2023}, the effects of Rabi \cite{Chiquillo2019} or spin-orbit couplings \cite{Tononi2019,Gangwar2022}, and nonequilibrium properties\cite{Guebli2021,Mithun2021}, have been extensively studied. Interestingly, however, scattering of one-dimensional quantum droplets remains largely unexplored \cite{Debnath2023}.

One of the most fascinating phenomena of bright solitons that occurs in nonautonomous nonlinear systems is quantum reflection, which portrays the wave nature of solitons when scattered by surfaces or steps \cite{Baronio2004,Friedrich2004,Cote1997,Lizunova2020}, potential barriers \cite{Sakaguchi2005,Weiss2009,Streltsov2009,Hansen2021,Marchukov2019,Dunjko2020,Cuevas2013}, potential wells \cite{Miroshnichenko2003,Stoychev2004,Lee2006,Ernst2010,Al-Marzoug2011,Khawaja2021}, and impurities \cite{Forinash1994,Cao1995,Frantzeskakis2002,Goodman2004,Brazhnyi2011,Alotaibi2019}. The wave-particle duality of the soliton makes it possible for the soliton to undergo a quantum reflection from an attractive potential well and yet still retain its particle-like integrity to a large extent \cite{Lee2006}. In such a phenomenon, even without classical turning points, quantum reflection may occur as the soliton approaches the potential, which can be understood by the formation of a trapped mode at the center of potential \cite{Ernst2010,Goodman2004}. A well-known example is the bright soliton of the nonlinear Schr\"{o}dinger equation (NLSE) scattering from the P\"{o}schl-Teller reflectionless potential well\cite{Lee2006,Al-Marzoug2011,Khawaja2021}. In the case of solitons scattering from the reflectionless potential, quantum reflection occurs only below a critical initial speed, with a sharp transition between quantum reflection and transmission. An accurate calculation of the critical speed has been proposed by determining the profile and energy of the trapped mode using a variational method with ansatz of a soliton whose density profile is spatially symmetric with respect to the potential center. Such studies enabled the understanding of the mechanism of soliton energy exchange during scattering and will help in the implementation of future all-optical technologies such as soliton diodes and logic gates \cite{Asad-uz-zaman2013,Khawaja2013,Khawaja2015}. One-dimensional quantum droplets, as macroscopic manifestations of quantum fluctuations, are predicted to exhibit a number of appealing properties, such as collisional features, collective excitations and shapes (e.g. a liquid-like incompressible phase with a top-flat density profile, highlighted by a uniform bulk density, appears at large particle numbers), which are fundamentally different from those of one-dimensional bright solitons. Given the striking differences in the properties of solitons and droplets, it is natural to ask the following questions of general interest regarding the scattering of quantum droplets from the reflectionless potential:  Does the sharp transition from quantum reflection to full transmission also occur in the scattering of quantum droplets, as it does in the scattering of bright solitons of the NLSE? If yes, what are the unconventional features in quantum droplet scattering?

 In the present work, we comprehensively study the scattering of quantum droplets by the P\"{o}schl-Teller reflectionless potential well. We confirm that quantum reflection occurs, with a sharp transition between full reflection and full transmission at a critical initial speed, both for small droplets with a $\sech^2$ shape and for large droplets with a broad flat-top plateau. Our study shows that quantum droplets can be either completely reflected or completely transmitted off the reflectionless potential without any splitting or particle emission. The $100\%$ reflectivity or $100\%$ transmittance without any radiation when interacting with the reflectionless potential is a peculiar feature of quantum droplets, which is not really accessible to the scattering of solitons (especially large solitons). Our numerical simulations show that the trapped mode (i.e. full trapping by the potential at the critical speed) of the quantum droplets turns out to be spatially asymmetric at large particle numbers, which differs from the counterpart of small quantum droplets and solitons, and can be captured by a variational method with a position-dependent trial wavefunction. We also precisely determine the critical speeds for the quantum reflection of the droplet, showing a different dependence of the critical speed on the atom number for small and large droplets, and explain why the quantum droplet experiences no radiation or particle loss during scattering by analyzing the collective excitations of the quantum droplet.

The rest of the paper is organized as follows. In Sec.~\ref{section:II}, we give the exact solution of quantum droplets in free space, and discuss the properties of quantum droplets in the two limits of large and small atom numbers. In Sec.~\ref{section:III}, we define the reflection and transmission coefficients, and study the scattering of quantum droplets numerically. In Sec.~\ref{section:IV}, we use three methods to calculate the critical speed below which quantum reflection occurs, and analyze the formation of the unbalanced trapped modes (the nonlinear single-node stationary states) of the large quantum droplets with flat-top density profile. In Sec.~\ref{section:V}, we discuss in detail the energy exchange mechanism of the quantum droplets during the scattering process. In Sec.~\ref{section:VI}, we analyze the collective excitations during the scattering of quantum droplets. In Sec.~\ref{section:VII}, we study the collisions between two quantum droplets at the P\"{o}schl-Teller reflectionless potential well. Finally, we summarize and discuss our main findings in Sec.~\ref{section:VIII}.

\section{The Solutions of Quantum Droplets}
\label{section:II}
We consider a two-component Bose-Bose mixture with equal mass symmetry, with mutual repulsion between atoms of the same component $g_1 = g_2 = g_0 > 0$
, with mutual attraction between atoms of different components $g_{12} < 0$ and equal number of particles of both components $N_1 = N_2 = \tilde{N}$. Near the MF collapse point ($\delta g = g_{12}+g_{0}\ll g_0$), the energy of the homogeneous mixture can be found and the Gross-Pitaevskii equation (GPE) is derived as \cite{Petrov2016}:
\begin{align}\label{con:1}
	i\hbar \frac{\partial \Psi(x,t)}{\partial t} =&-\frac{\hbar^{2}}{2 m} \frac{\partial^{2} \Psi(x, t)}{\partial x^{2}}+\delta g|\Psi(x, t)|^{2} \Psi(x, t) \\\notag
&-\frac{\sqrt{2 m}}{\pi \hbar} g_{0}^{\frac{3}{2}}|\Psi(x, t)| \Psi(x, t) .
\end{align}
where the  last  term on the right-hand side of Eq.\;\eqref{con:1} corresponds to the LHY term, which is the first-order BMF correction term accounting for the quantum many-body effect in the weakly interacting regime ($g_0/n\ll 1$). In experiments, it is possible tune $\delta {g}$ to positive and negative values. Here we define the relative change of the mean field intensity $g=\delta {g}/\delta{g}_0$ (where $\delta{g}_{0}>0$ is a constant) to discuss a more general solution of Eq.\;\eqref{con:1}. By defining the length $\xi_{0}=\frac{\pi \hbar^{2} \delta g_{0}^{1 / 2}}{\sqrt{2} m g_{0}^{3 / 2}}$, the time $t_{0}=\frac{\pi^{2} \hbar^{3} \delta g_{0}}{2 m g_{0}^{3}}$ and the energy $E_{0}=\frac{2 m g_{0}^{3}}{\pi^{2} \hbar^{2} \delta g_{0}}$ as the characteristic units, Eq.\;\eqref{con:1} can be written in a dimensionless form
\begin{equation}\label{con:2}
i \psi_{{t}}=-\frac{1}{2} \psi_{{x} {x}}+g|\psi|^{2} \psi-|\psi| \psi ,
\end{equation}
with the normalization condition $\int_{-\infty}^{\infty}|\psi({x})|^{2} d {x}=N=\frac{{\pi}\tilde{N}}{\sqrt{2}}(\delta g_{0}/g_{0})^{3/2}$. Regardless of the sign of $g$, the attractive BMF term allows the system to admit a self-bound ground state in the form of $\psi(x,t)=\psi_0(x)e^{-i \mu t}$ with
\begin{equation}\label{con:3}
\psi_0(x)=\frac{-3 \mu}{1+\sqrt{1+\frac{9 \mu g}{2}} \cosh \left(\sqrt{-2 \mu x^{2}}\right)}.
\end{equation}

The relation between chemical potential $\mu$ and norm $N$ of the droplet depends on the sign of $g$ and is given by
\begin{align}
\label{con:4} N_{g=0} & =3\sqrt{2}(-\mu)^{\frac{3}{2}},\\
\label{con:5} N_{g<0} & =n_{0} \sqrt{\frac{2}{\mu_{0}}}\left(\sqrt{-\frac{\mu}{\mu_{0}}}-\arctan \sqrt{-\frac{\mu}{\mu_{0}}}\right),\\
\label{con:6} N_{g>0} & =n_{0} \sqrt{-\frac{2}{\mu_{0}}}\left[\ln \frac{1+\sqrt{\mu / \mu_{0}}}{\sqrt{1-\mu / \mu_{0}}}-\sqrt{\frac{\mu}{\mu_{0}}}\right].
\end{align}
In Eqs.\;\eqref{con:5} and \eqref{con:6}, $\mu_{0} = -2/(9g)$  and $n_{0} = 4/(9g^2)$ are the chemical potential and the saturation density of a spatially uniform liquid at $g > 0$, respectively. When g = 0, Eq. \eqref{con:3} represents the known Korteweg-de Vries (KdV) type droplet solution,
\begin{equation}\label{con:7}
\psi_0(x)=\frac{-3 \mu}{2} \operatorname{sech}^{2}\left(\sqrt{\frac{-\mu}{2}} x\right).
\end{equation}

It is worth noting that for $g > 0$, $0\leq\sqrt{1+\frac{9 \mu g}{2}}<1$ in Eq.\;\eqref{con:3}, the characteristics of quantum droplets can be analyzed using the formula, $\tanh (a+X)+\tanh (a-X)=\frac{2 \tanh (2 a)}{1+\operatorname{sech}(2 a) \cosh (2 X)}$, by which the ground state solution of the droplet can be written as
\begin{equation}\label{con:8}
\psi_0(x)=\sqrt{A} [\tanh (a+\sqrt{Ag} x)+\tanh (a-\sqrt{Ag} x)],
\end{equation}
where
\begin{align}
\label{con:9} A&=\frac{-\mu }{2g},\\
\label{con:10} a&=\frac{1}{2}\arctanh[(-\frac{9}{2}\mu g )^{\frac{1}{2} }].
\end{align}

From Eqs.\;\eqref{con:6} and \eqref{con:10}, we have $\mu (N\rightarrow \infty )=-\frac{2}{9g}$, $\mu (N\rightarrow 0)=0$, $-\frac{2}{9}< \mu g<0$, and $a$ $\in$ $(0,\infty )$,  which increases with $N$. For a large quantum droplet, which corresponds to a large value of $a$, Eq.\;\eqref{con:8} gives rise to the wave function of large-size droplets characterized by the kink structures at the edges as well as a uniform flat-top structure in the middle,
\begin{align}
\label{con:11} &\psi_{\text {middle}} =2 \sqrt{A},\\
\label{con:12} &\psi_{\operatorname{kink}} =\sqrt{A}\left[\tanh \left(a \mp \sqrt{\frac{-\mu}{2}} x\right)+1\right].
\end{align}
For quantum droplets with small value of $a$ ($a\ll 1$), the solution \eqref{con:8} can be approximated by Taylor expansion as
\begin{equation}\label{con:13}
\psi_0(x)\approx 2a\sqrt{A}  \operatorname{sech}^{2}(\sqrt{\frac{-\mu}{2}} x),
\end{equation}
which is found to feature a similar density profile to the KdV-type droplets.
\par{We show analytically that, upon variation of $a$ (which vanishes when $g = 0$), two physically distinct regimes, small droplets of $\sech^2$ shape and large droplets with a flat-top plateau, can be identified. From the analytical expressions \eqref{con:11}-\eqref{con:13}, we observe sharp differences between the $\sech^2$-shaped and flat-top droplet. It can be seen that for small droplet with $a\ll1$, as $\sqrt{-\mu/2}$ grows with $N$, the width of the $\sech^2$-shaped droplets becomes narrower and they behave like bright solitons. In contrast, for large quantum droplets with $a\gg1$, the MF term repulsion effect becomes significant, leading to the separation of kink and anti-kink pairs and thus to the formation of kink structures at the edges and flat-top structures in the center. In this case, the width of flat-top droplets will instead increase with the norm $N$, which has a common similarity with the classical liquid, where a plateau in the density profile expands with the growth of the mass of the droplet.}
\section{Scattering of quantum droplets}
\label{section:III}
In the previous section, we explored analytically the ground states of a one-dimensional free-space quantum droplet system, and in this section we consider the scattering of strictly one-dimensional quantum droplets by a reflectionless P\"{o}schl-Teller attractive potential, $V(x)=-U_{0} \operatorname{sech}^{2} (\alpha x)$ ($\alpha=\sqrt{U_{0}}$  is necessary for reflectionless scattering of linear matter waves), whose dynamics is governed by the GPE containing the external potential,
\begin{equation}\label{con:14}
i \psi_{{t}}=-\frac{1}{2} \psi_{{x} {x}}+g|\psi|^{2} \psi-|\psi| \psi -U_{0} \operatorname{sech}^{2}(\alpha x)\psi,
\end{equation}
where $U_0$ is expressed in units of $E_0$. Note that $\delta g_{0}$ is an arbitrary constant used as a measure. In this paper we choose $g = 0$ (corresponding to $\delta g = 0$), and $g= \pm1$ by assuming $\delta g_{0} = \delta g$, so that the droplet properties can only be controlled by the rescaled norm $N$ together with the sign of $g$. Due to the invariance of the system under Galilean transformations, the exact movable quantum droplets solution of the dimensionless GPE \eqref{con:2} can be obtained  from the stationary solution as follows,
\begin{equation}\label{con:15}
\psi(x,t)=\frac{-3 \mu \exp^{i[(x-\tilde{x}_0)v+\frac{v^2}{2}t-\mu t)]}}{1+\sqrt{1+\frac{9 \mu g}{2}} \cosh \left(\sqrt{-2 \mu (x-\tilde{x}_0)^{2}}\right)},
\end{equation}
where $v$ and $\tilde{x}_0$ are the initial speed and initial position of the quantum droplet. We study the scattering of quantum droplets by numerically solving Eq.\;\eqref{con:14} with $\psi(x,0)$ in Eq.\;\eqref{con:15} as the initial profile, based on the split-step Fourier (SSF) method. To get started, we define the corresponding reflection $(R)$ and transmission $(T)$ coefficients as follows,
\begin{align}
\label{con:16} R & =(1 / N) \int_{-\infty}^{-l} |\psi(x, t_{f})|^{2} d x, \\
\label{con:17} T & =(1 / N) \int_{l}^{\infty} |\psi(x, t_{f})|^{2}d x,
\end{align}
where $l$ is the length greater than the width of the potential well, and  $t_{f}$ is an evolution time  needed to make the scattered quantum droplets sufficiently distant from the potential well.
\begin{figure}[htbp]
\includegraphics[width=4.2cm]{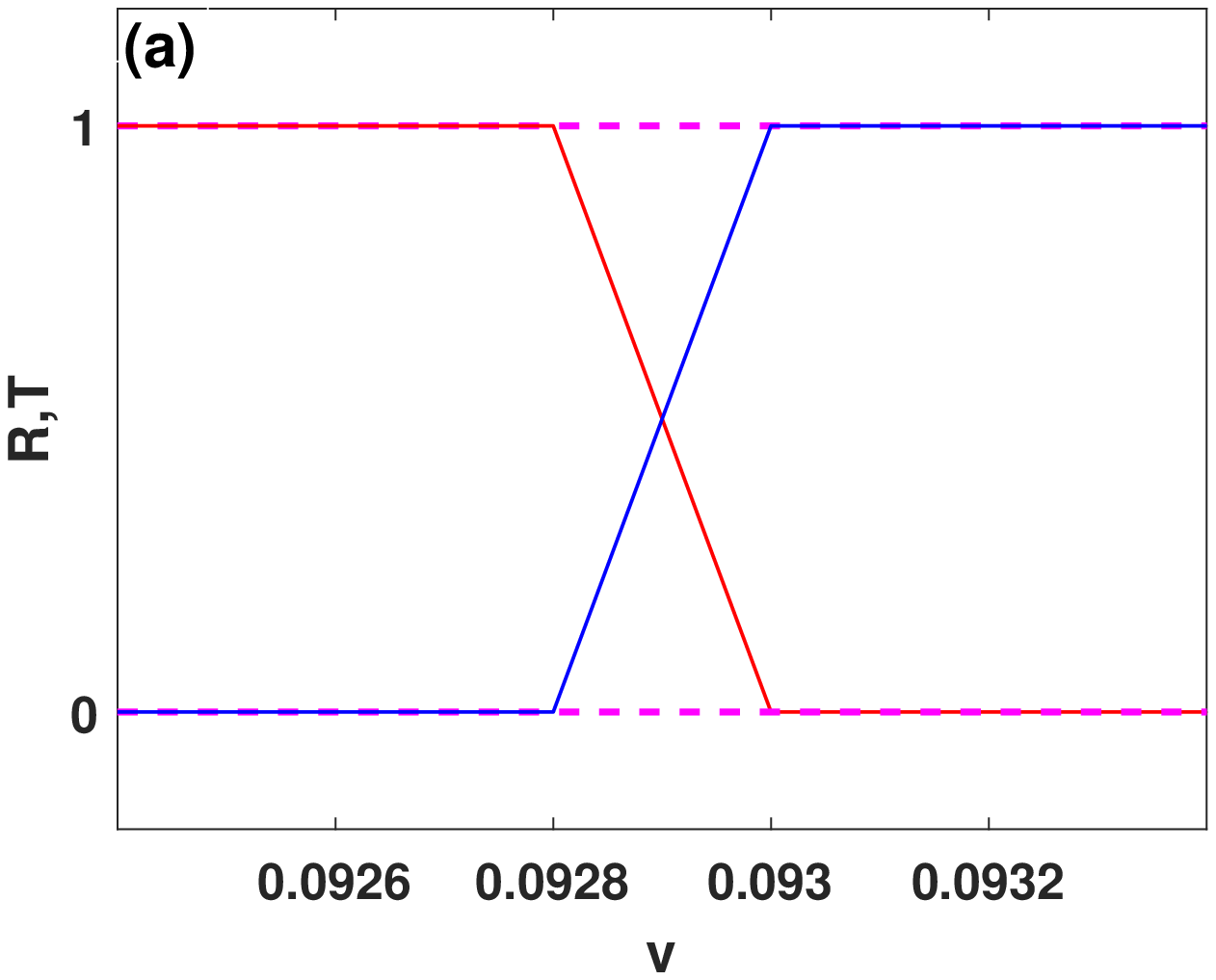}
\includegraphics[width=4.2cm]{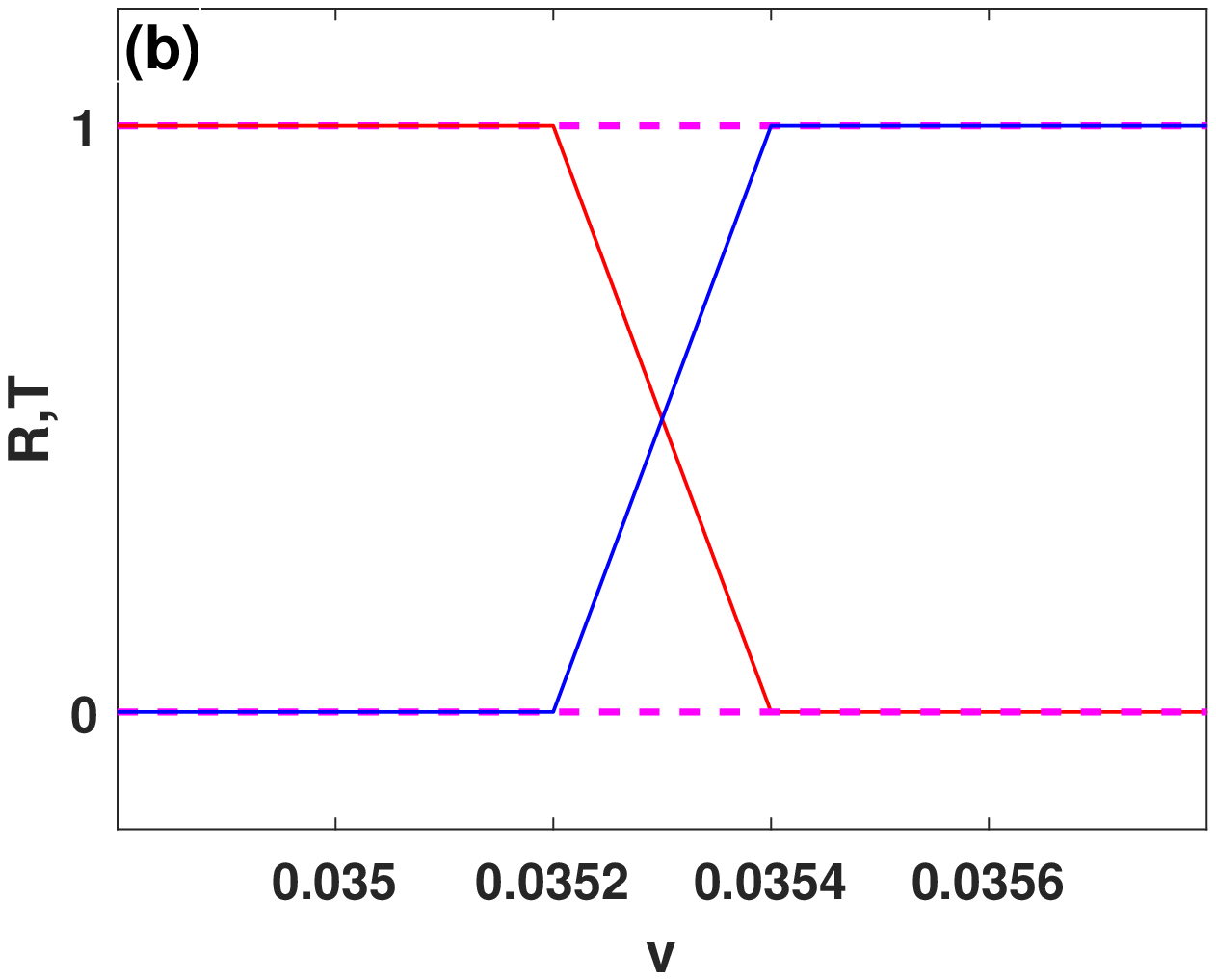}
\caption{Reflectance (red) and transmittance (blue) versus the initial speed of a quantum droplet scattering by a reflectionless potential. (a) is for a small quantum droplet with $N=1$ and (b) is for a large quantum droplet with $N=10$. Other parameters used are $U_0=4, \alpha=\sqrt{U_{0}}, g=1, \tilde{x}_0=-30.$}
\label{fig1}
\end{figure}
\begin{figure*}[htbp]
\includegraphics[width=5.3cm]{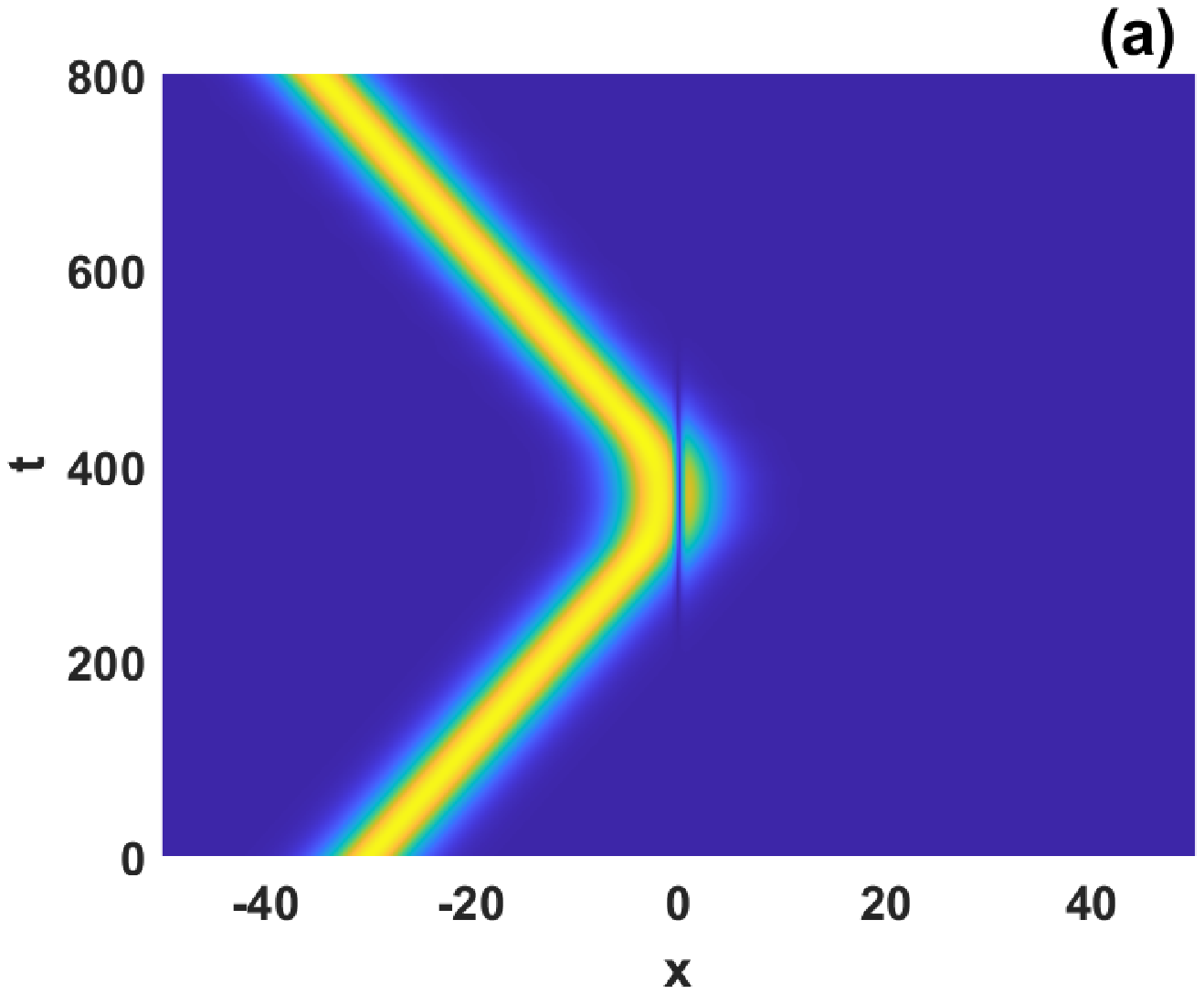}
\includegraphics[width=5.3cm]{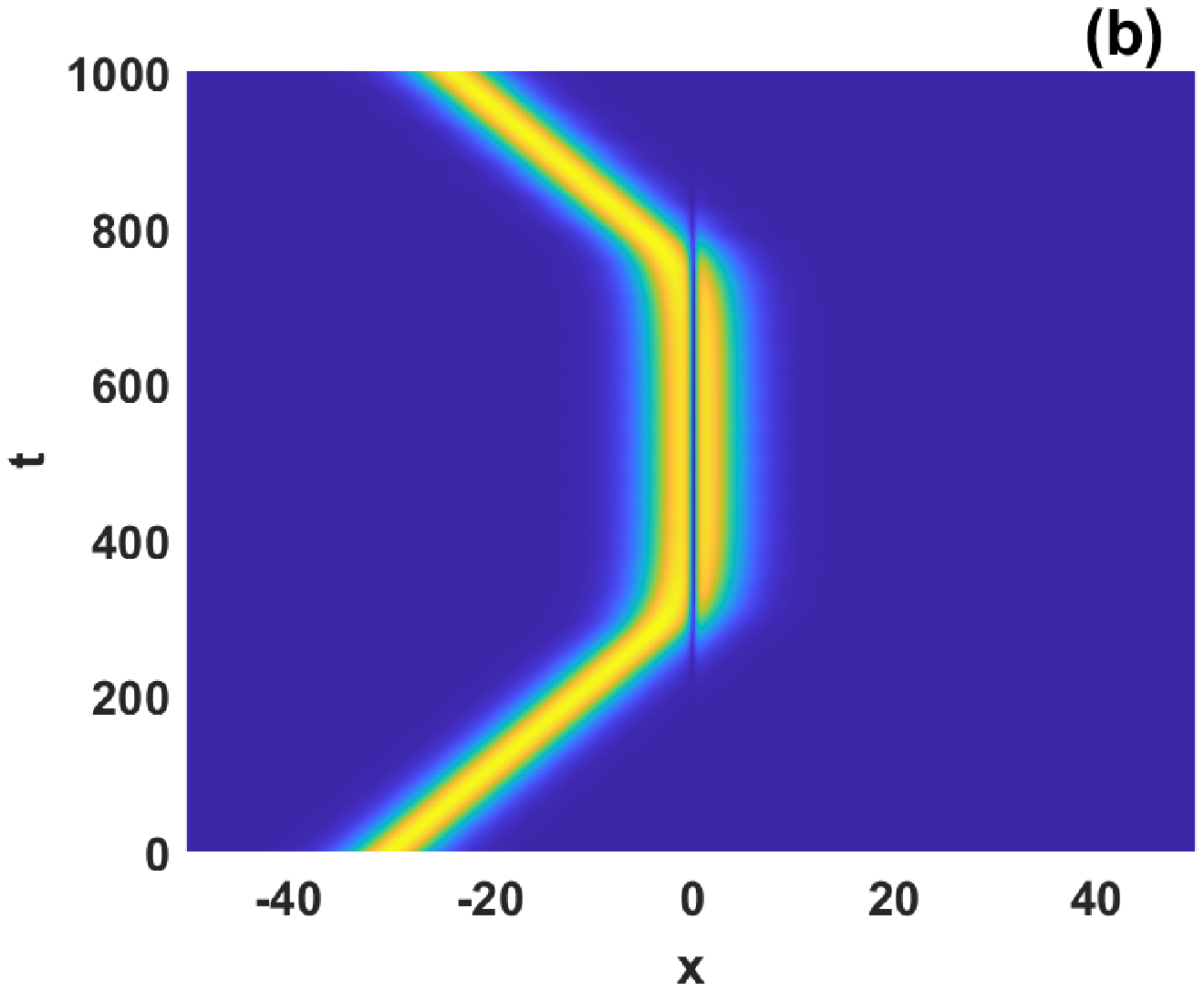}
\includegraphics[width=5.3cm]{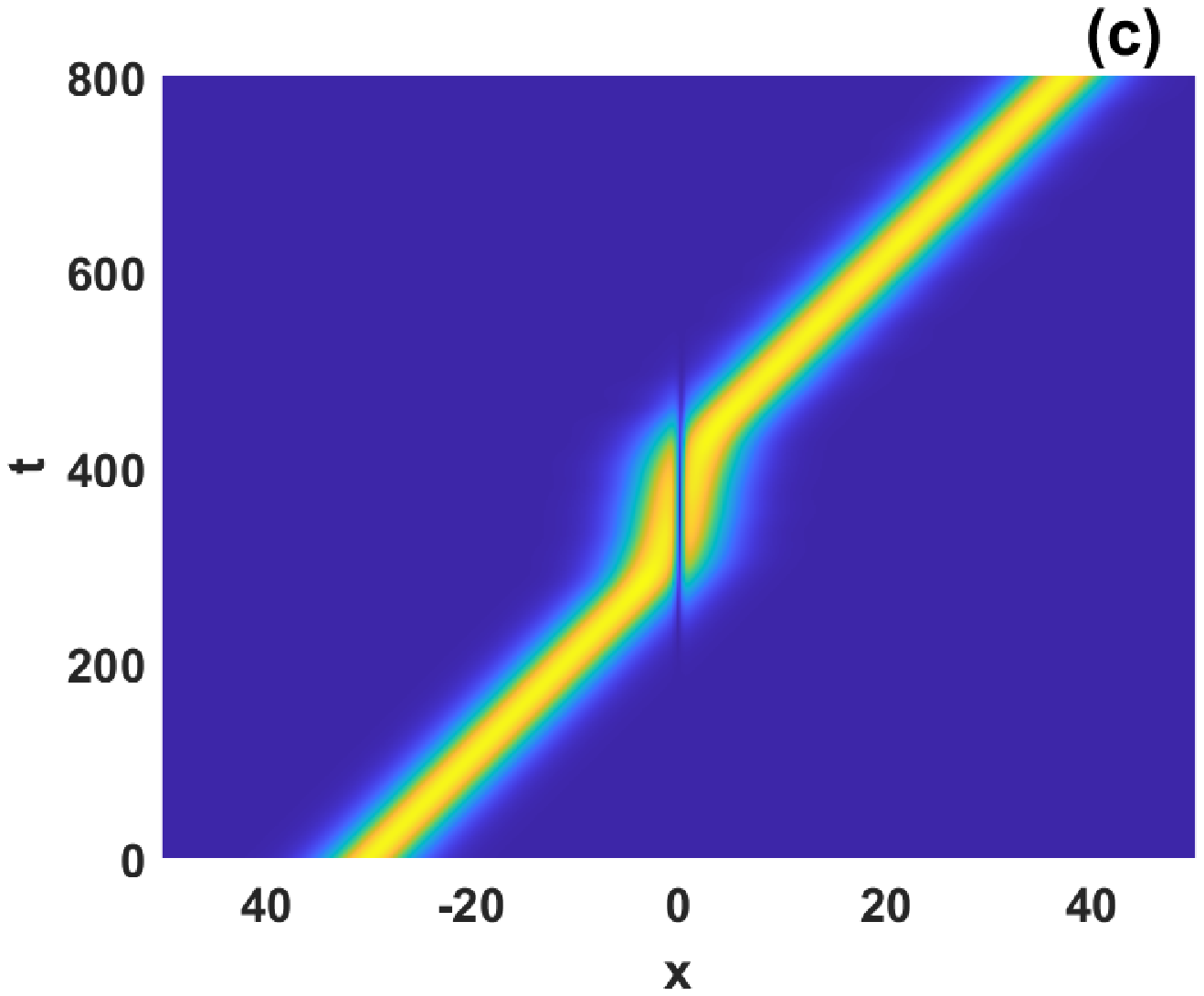}
\includegraphics[width=5.3cm]{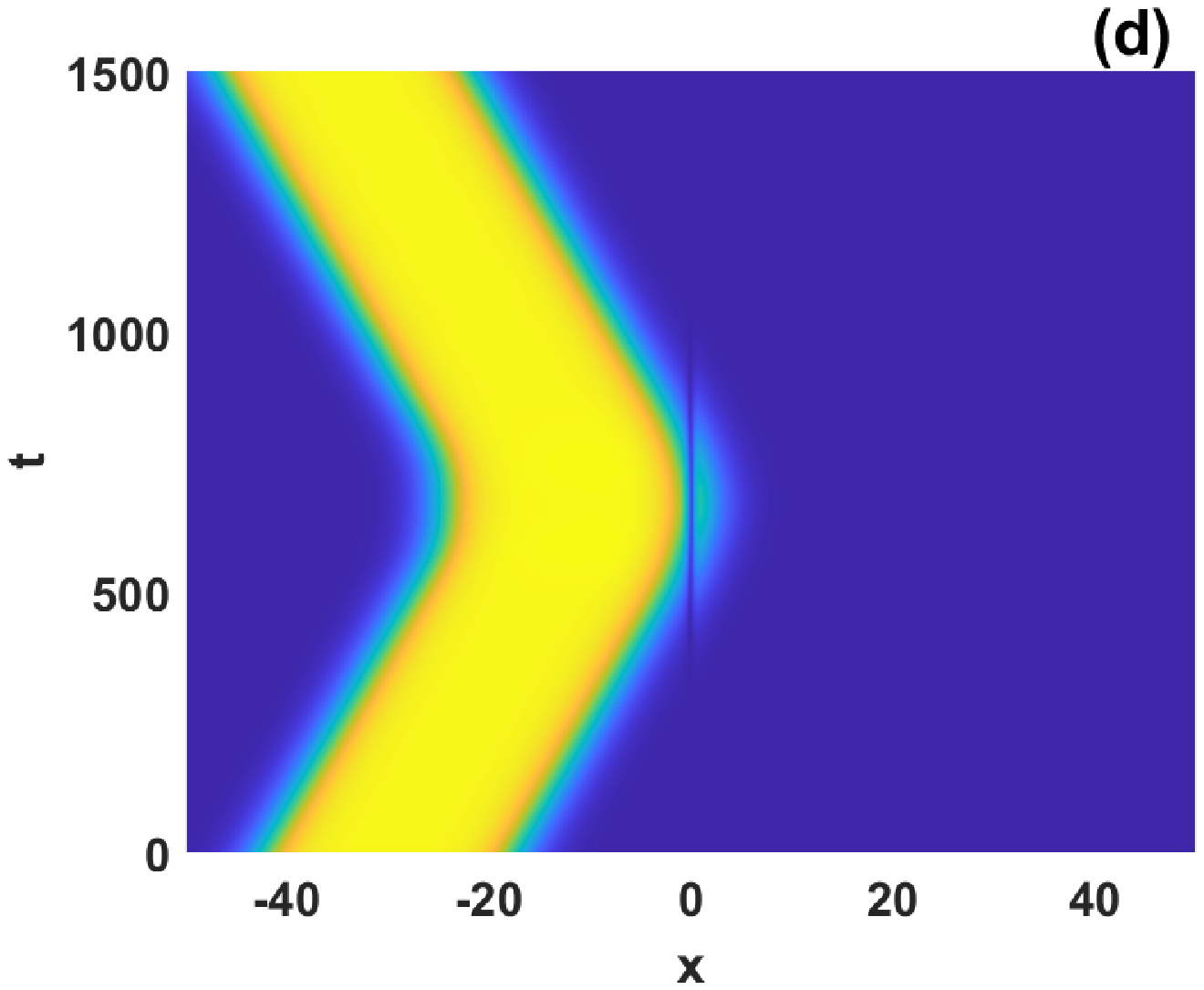}
\includegraphics[width=5.3cm]{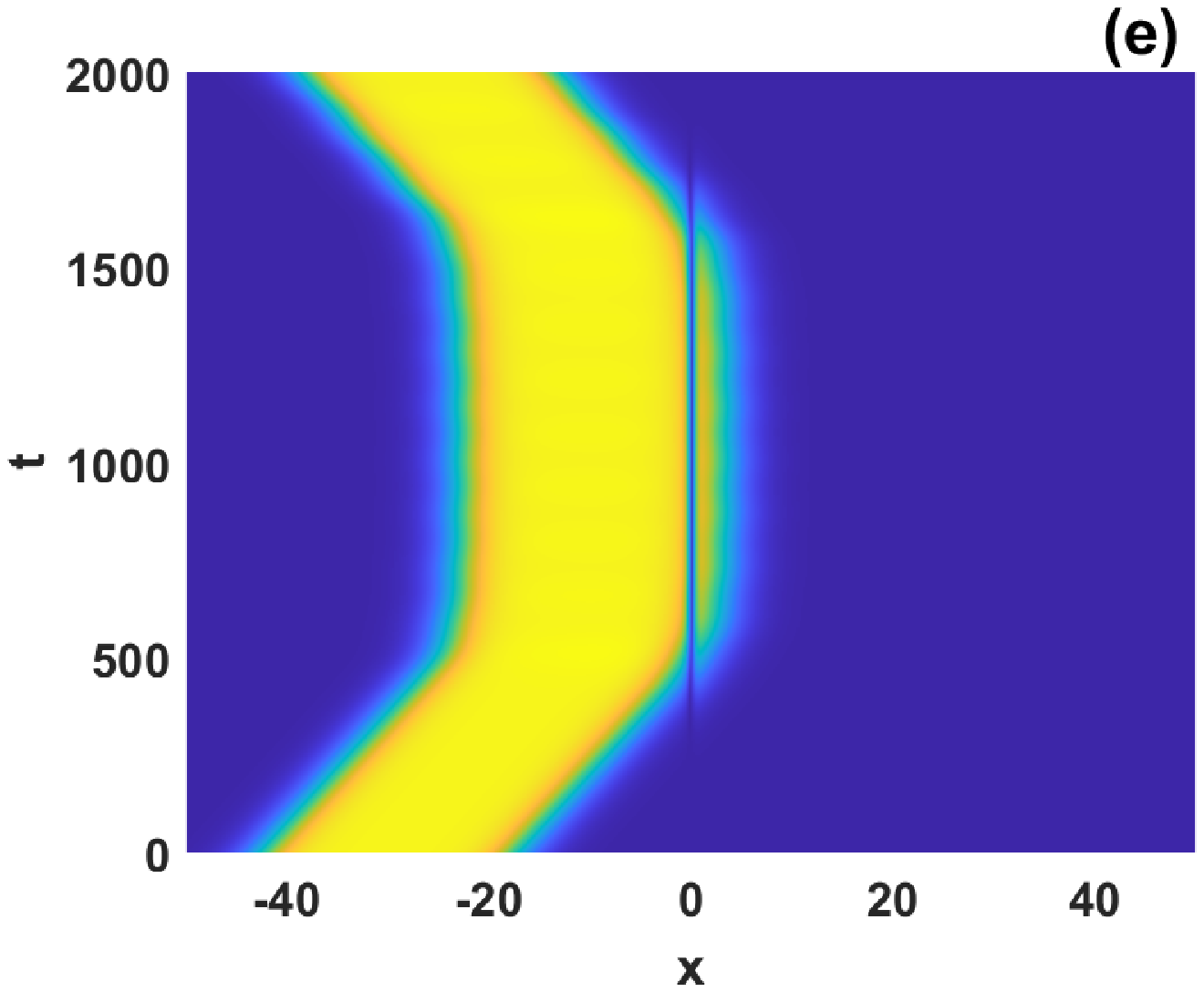}
\includegraphics[width=5.3cm]{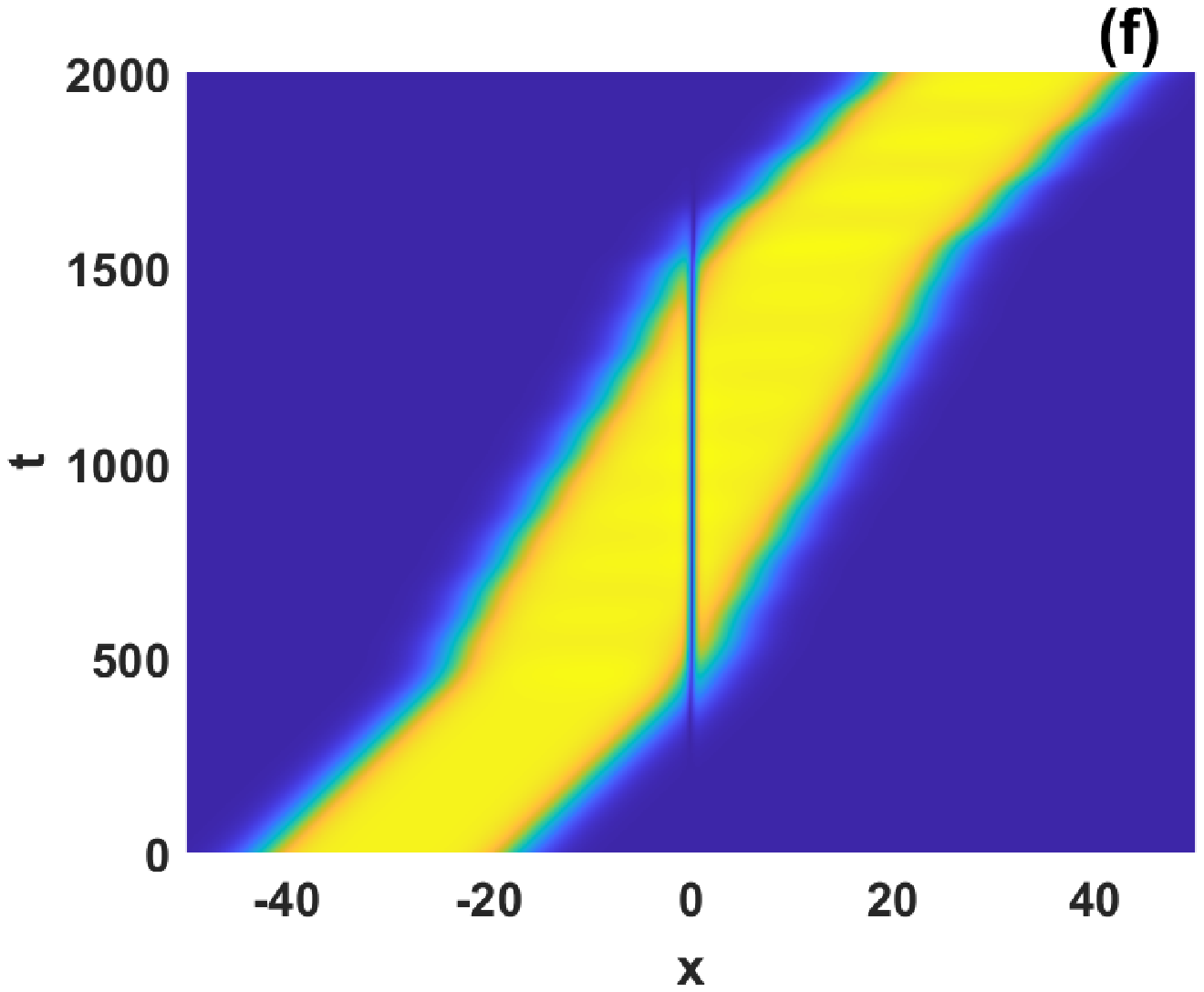}
    \caption{Scattering of quantum droplets at different speeds by a reflectionless potential centered at $x=0$ for small quantum droplets ($N=1$, upper row) and for large quantum droplets ($N=10$, lower row). Upper row: the initial speeds of small quantum droplets are (a) $v = 0.08$, (b) $v = v_c \approx 0.0928757$, and (c) $v = 0.095$. Lower row: (d) $v = 0.03$,  (e) $v = v_c \approx 0.03532$, and (f) $v = 0.04$. Other parameters used are $U_0=4, \alpha=\sqrt{U_{0}}, g=1, \tilde{x}_0=-30$.}
        \label{fig2}
\end{figure*}

FIG.\;\ref{fig1} shows the numerical results of reflection (red lines) and transmission (blue lines) coefficients versus the initial speed of a quantum droplet scattered by the reflectionless attractive potential. It can be seen that there is a sharp transition between full transmission and full reflection with a well-defined critical speed for both large and small quantum droplets. This is directly demonstrated by the spatiotemporal density plots of quantum droplets for two different norms $N=1$ [Fig.\;\ref{fig2}(a)-Fig.\;\ref{fig2}(c)] and $N=10$ [Fig.\;\ref{fig2}(e)-Fig.\;\ref{fig2}(d)], where two main outcomes of the scattering are distinguished, depending on the incident speeds. When the incident speed is below the critical speed, the quantum droplet experiences non-classical full reflection. The critical speed is then found numerically by increasing the incident speed until the duration of the quantum droplet trapped by the potential reaches the maximum length. When the incident speed exceeds the critical speed, the quantum droplet gets fully transmitted. Importantly, there is a striking difference in the trapped modes between the small and large droplets. The trapped modes of large droplets are found to be spatially asymmetric with respect to the center of the potential, which contrasts with the counterparts of small droplets. Another exotic property is that the large flat-top droplet, although it can be very wide, scatters from the potential well without any splitting or radiation, so that reflectance and transmittance can reach $100\%$.
\section{Critical modes in quantum droplets scattering process}
\label{section:IV}
The occurrence of quantum reflections indicates the existence of a zero-speed state during scattering, which represents the instantaneous state of the turning point of quantum reflection. Here we define the state of $v = 0$ (the velocity turning point) during the quantum reflection as the zero-speed state $\phi(x-x_0)$ (where $x_0$ is the position of droplet peak at the velocity turning point). The energy $E_z$ of the zero-speed state can be calculated from the energy functional,
\begin{equation}\label{con:18}
	E_{z}(x_0)=\int_{-\infty}^{+\infty}[\frac{1}{2}|\phi_{x}|^{2}+\frac{g}{2}|\phi|^{4}-\frac{2}{3}|\phi|^{3}+V(x)|\phi|^{2}] dx.
\end{equation}

As the quantum droplet is initially far away from the potential well, the potential energy is very small and negligible. Thus, the initial energy of the quantum droplet, calculated from Eq.\;\eqref{con:14} with the initial profile $\psi(x,0)$, can be given by
\begin{equation}\label{con:19}
\begin{aligned}
E_{d}&=\int_{-\infty}^{+\infty}\left(\frac{1}{2}\left|\psi_{x}\right|^{2}+\frac{g}{2}|\psi|^{4}-\frac{2}{3}|\psi|^{3}\right) dx \\
&=\frac{1}{2}Nv^{2}+\int_{-\infty}^{+\infty}\left[\frac{1}{2}\left|(\psi_0)_{x}\right|^{2}+\frac{g}{2}|\psi_0|^{4}-\frac{2}{3}|\psi_0|^{3}\right] dx \\
&=\frac{1}{2}Nv^{2}+E_{sd},
\end{aligned}
\end{equation}
where $E_{sd}$ is the energy of the stationary droplet in the form of Eq.\;\eqref{con:3}. According to the conservation of energy, the initial energy of the quantum droplet is equal to the energy at zero velocity (turning point of quantum reflection), so that the initial speed at which quantum reflection occurs can be calculated as
\begin{equation}\label{con:20}
v=\sqrt{\frac{2(E_z-E_{sd})}{N}}.
\end{equation}
The energy of zero-speed state is increased with the initial speed of the droplet. It reaches a maximum value at the critical speed, above which the droplet gets transmitted. Thus, the critical speed can be computed numerically by increasing the initial droplet speed such that the zero-speed state of maximum energy can be reached. The critical speed is the initial droplet speed at which the zero-speed state of maximum energy is attained,

\begin{equation}\label{con:21}
v_c=\sqrt{\frac{2[(E_z)_{max}-E_{sd}]}{N}}.
\end{equation}

Quantum droplets with different initial speeds produce different zero-speed states at different locations during quantum reflection. The energy of the zero-speed state is related to the location of the zero-speed state. To describe the zero-speed state at different positions, we propose the following position-dependent trial function,
\begin{equation}\label{con:22}
\phi(x)=A\psi_0[\gamma (x-x_0)]\tanh(\beta x),
\end{equation}
where $\gamma$, $\beta$ are the variational parameters with $\gamma$ accounting for the central slope and $\beta$ accounting for the overall width of the mode, $x_0$ denotes the position of the zero-speed state. We normalize the trial function to $N$, which yields
\begin{equation}\label{con:23}
A(\gamma,\beta,x_0)=\left(\frac{N}{\int_{-\infty}^{+\infty}|\psi_0[\gamma (x-x_0)]\tanh(\beta x)|^2dx}\right)^\frac{1}{2}.
\end{equation}

The energy functional for the zero-speed state can be calculated by substituting the normalized trial function \eqref{con:22} into Eq.\;\eqref{con:18}. However, the integration can not be obtained in analytical form. Thus, we make the integral in Eq.\;\eqref{con:23} and the similar integral in the energy functional \eqref{con:18} to be computed numerically in terms of $\gamma$ and $\beta$. By plotting $E_z$[$\gamma$, $\beta$] for specific values of $N$, $g$, $x_0$ and $U_0$, the results show that $E_z$[$\gamma$, $\beta$] has a local minimum at $\gamma=\gamma^*$ and $\beta=\beta ^*$. The minimum energy, $E_z$ [$\gamma^*$, $\beta^*$], obtained in this way, is supposed to be the energy of the zero-speed state at $x_0$. Substituting $E_z = E_z$ [$\gamma^*$, $\beta^*$] into Eq.\;\eqref{con:20}, we get the initial speed that produces a zero-speed state at the given position $x_0$ in the quantum reflection process. The initial speed for generating the zero-speed state with energy ($E_z$[$\gamma^*$, $\beta^*$]) maximum in terms of $x_0$ gives the critical speed above which transmission occurs. A similar but position-independent variational solution has been used to accurately predict the critical speed between reflection and transmission of bright solitons in the reflectionless potential \cite{Khawaja2021}.

\begin{figure}[htbp]
\includegraphics[width=4.2cm]{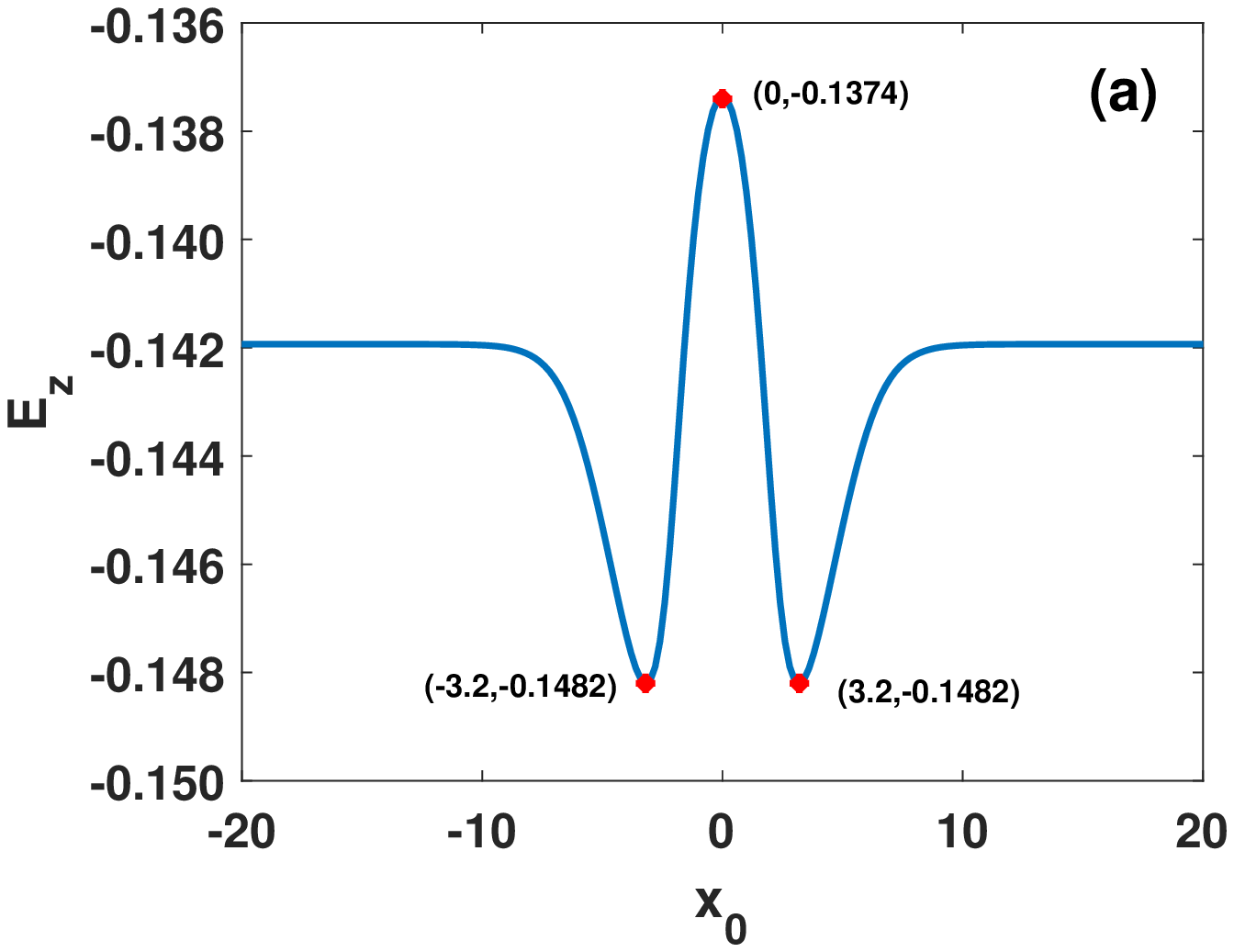}
\includegraphics[width=4.2cm]{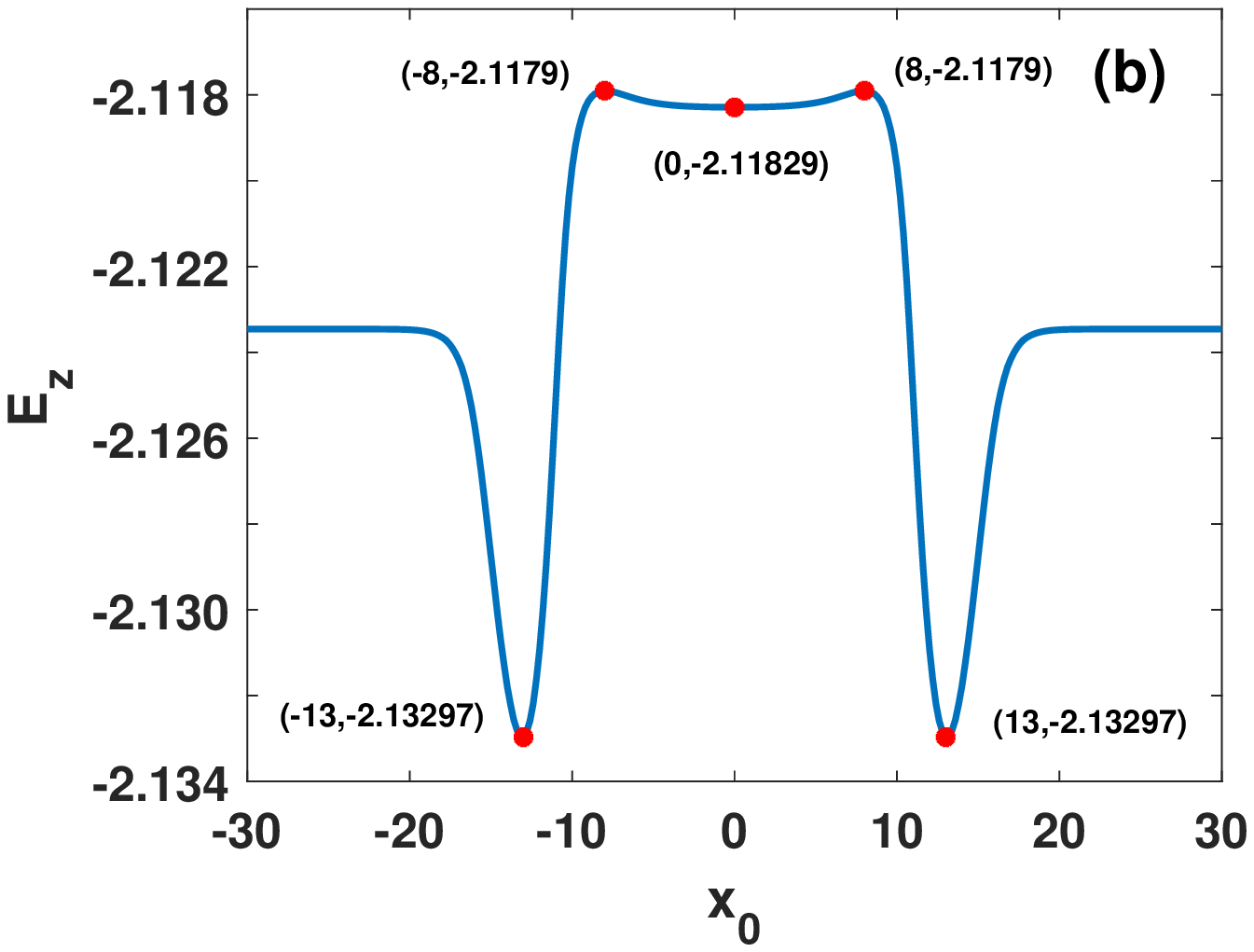}
    \caption{Dependence of the energy of zero-speed states on position $x_0$ derived from variational method using the position-dependent trial function \eqref{con:22}. (a) is for a small quantum droplet ($N=1$) and (b) is for a large quantum droplet ($N=10$). Other parameters: $U_0=4, g=1$. The energy maximum in terms of $x_0$ corresponds to the trapped mode (representing the critical state in the scattering): The marked point $(0,-0.1374)$ denotes the metastable trapped mode for small quantum droplet ($N=1$) with a balanced profile, while the point $(-8,-2.1179)$ denotes the metastable trapped mode for large quantum droplet ($N=10$) with an imbalanced profile. The energy of zero-speed states exhibits left-right symmetry, because the quantum droplet can be launched from either the left or the right.}
    \label{fig3}
\end{figure}
\begin{figure*}[htbp]
\includegraphics[width=4.3cm]{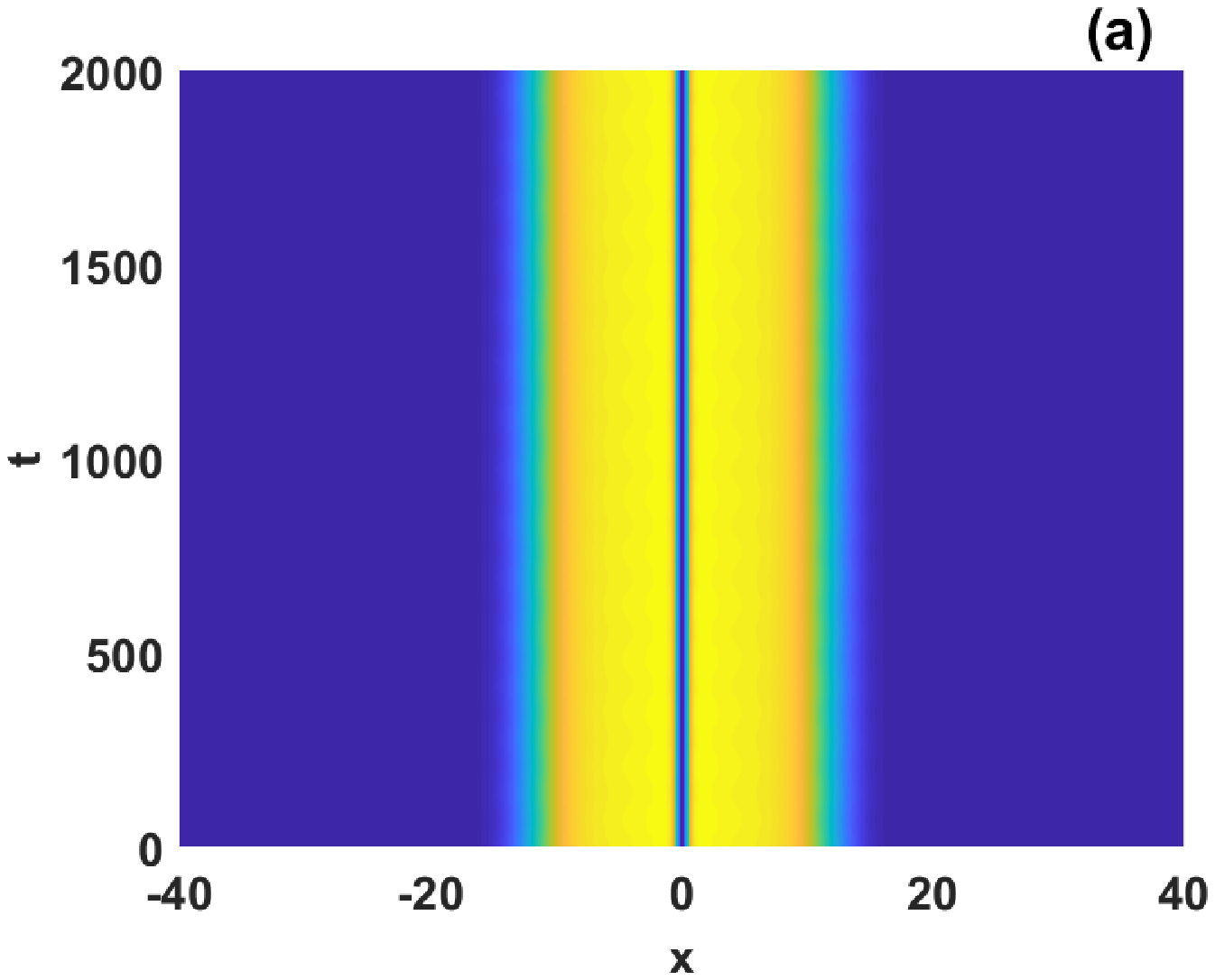}
\includegraphics[width=4.3cm]{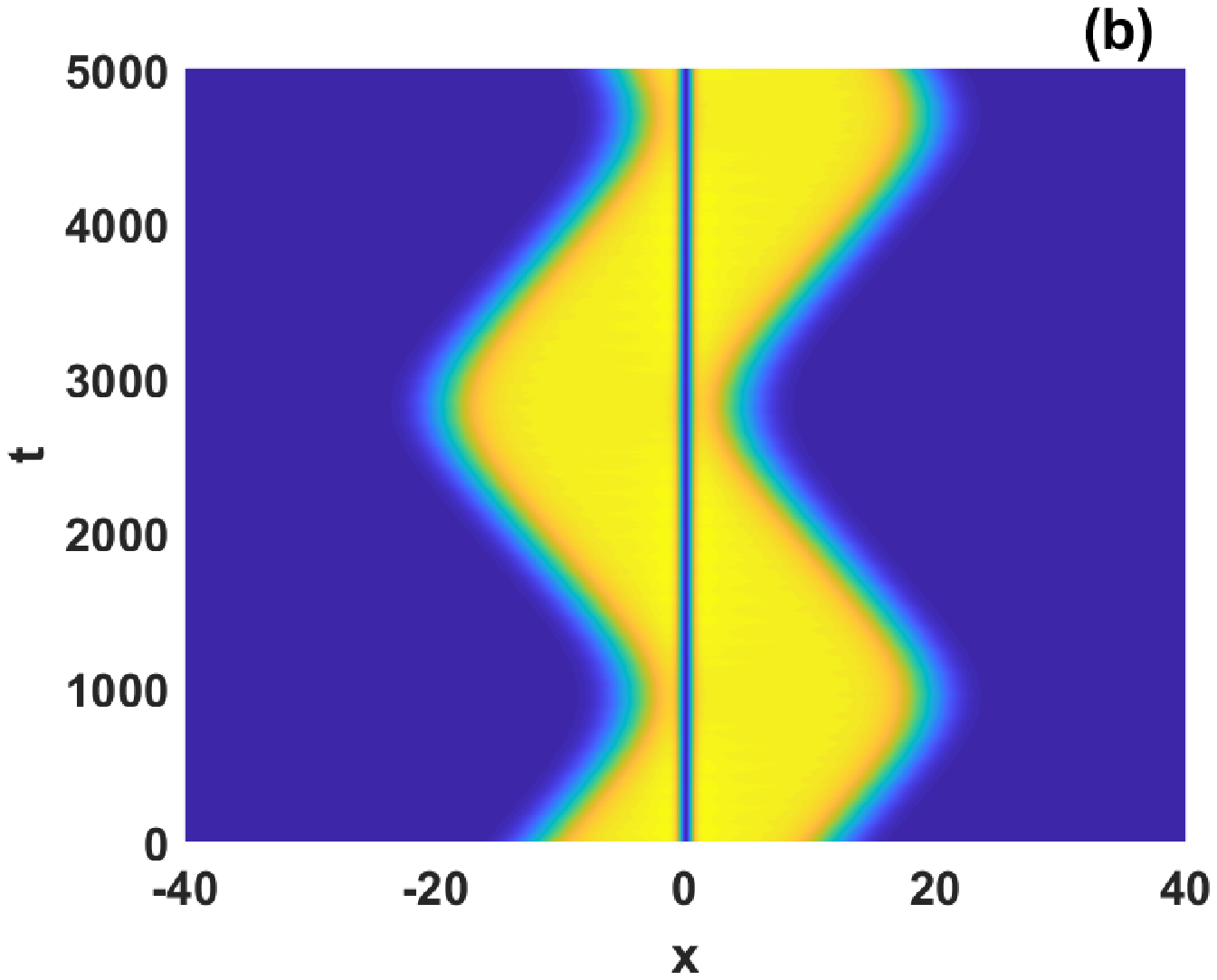}
\includegraphics[width=4.3cm]{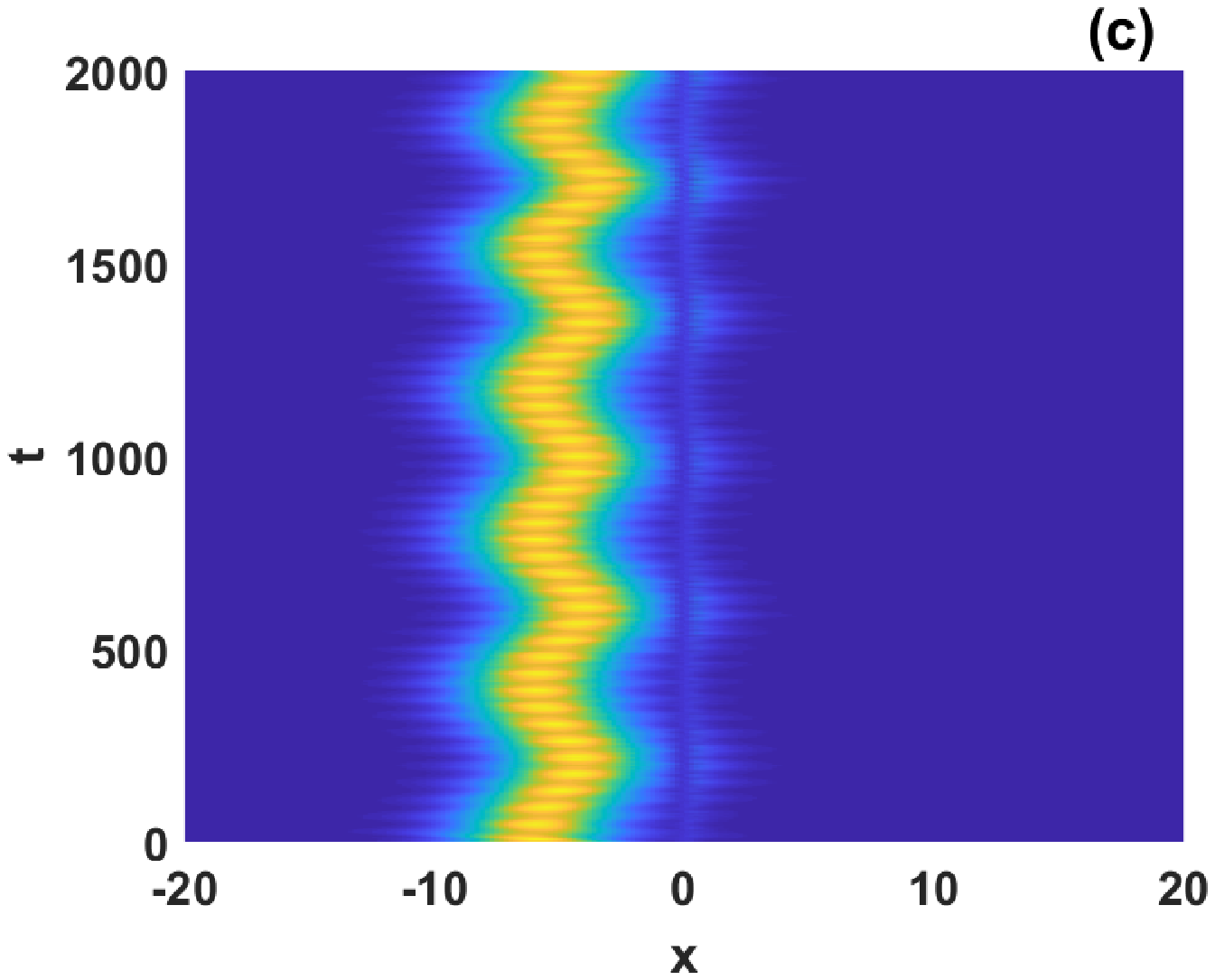}
\includegraphics[width=4.3cm]{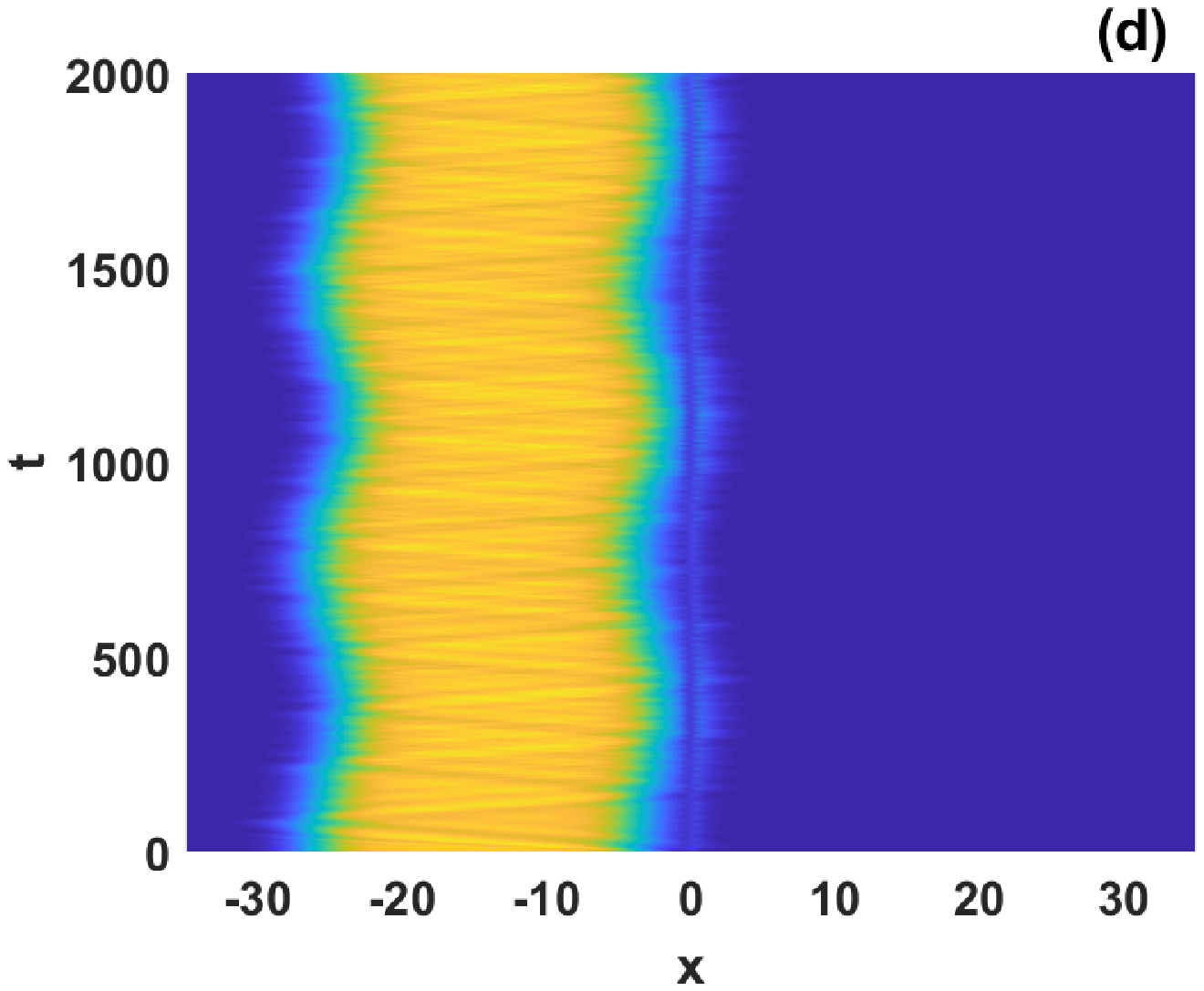}
    \caption{ The top view of the spatiotemporal evolution of the densities of the  quantum droplets with different initial profiles. (a) represents the evolution of the symmetric nonlinear stationary state of Eq.\;\eqref{con:14} with the chemical potential $\mu = -0.222222185$, obtained  by the square operator method, corresponding to the marked point $(0,-2.11829)$  for the large quantum droplet as shown in Fig.\;\ref{fig3}(b), obtained by the variational method. (b) represents the evolution of the symmetric nonlinear stationary state as shown in (a) with an incident speed $v = 0.01$, where we observe the oscillations around the reflectionless potential, which represents the trapping of the quantum droplet. (c) and (d) show the evolution of the low-energy zero-speed states derived by the variational method for a small quantum droplet ($N = 1$) initially located at $\tilde{x}_0 = -6$, and for a large quantum droplet ($N = 10$) initially located at $\tilde{x}_0 = -15$, respectively. Other parameters: $U_0 = 4, g = 1$.}
    \label{fig4}
\end{figure*}
In Fig.\;\ref{fig3}, we plot the energy of zero-speed state (corresponding to a minimum of the energy functional with  respect to $\gamma$ and $\beta$) versus $x_0$ using the variational method with the trial function \eqref{con:22}. Fig.\;\ref{fig3} shows that the energy of the zero-speed state does indeed have a maximum at certain values of $x_0$, indicating that the zero-speed state at these equilibrium positions is unstable against variations in its position. During the scattering, when the initial speed of the quantum droplet is $v \approx v_c$, the zero-speed state formed at the potential well corresponds to the metastable state (i.e., the state with energy maximum in terms of $x_0$) perturbed by small position shift. Thus, the duration of the zero-speed state is longer when the initial speed $v$ is closer to the critical speed $v_c$, which implies that the droplet is trapped by the potential for a considerably longer time. The metastable zero-speed states can exist for long periods of time in the absence of perturbations, thus representing nonlinear eigenstates of the system, also referred to as trapped modes in soliton scattering studies \cite{Khawaja2021}. As can be seen from Fig. \ref{fig3}, for small quantum droplets, the energy maximum is at the center of the potential, $x_0 = 0$, while  for large quantum droplets energy maximum point is displaced from the center of the potential. This means that when large quantum droplets are scattered, a metastable state that is spatially asymmetric (due to $x_0 \neq 0$) can form for the initial droplet speed just at the critical speed, in stark contrast to the soliton and small quantum droplet that both support the spatially symmetric trapped mode.

\begin{figure}[htbp]
\includegraphics[width=4.2cm]{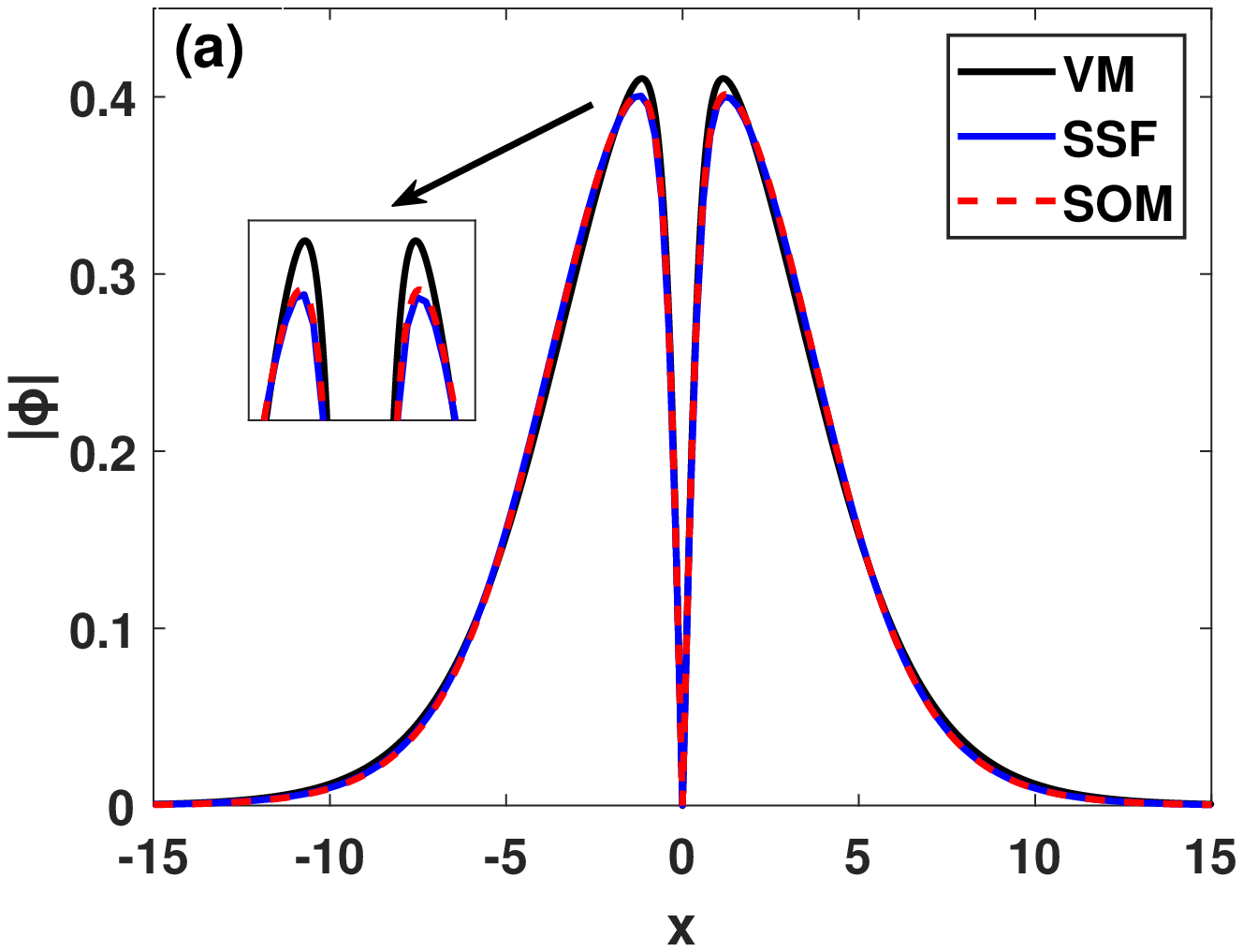}
\includegraphics[width=4.2cm]{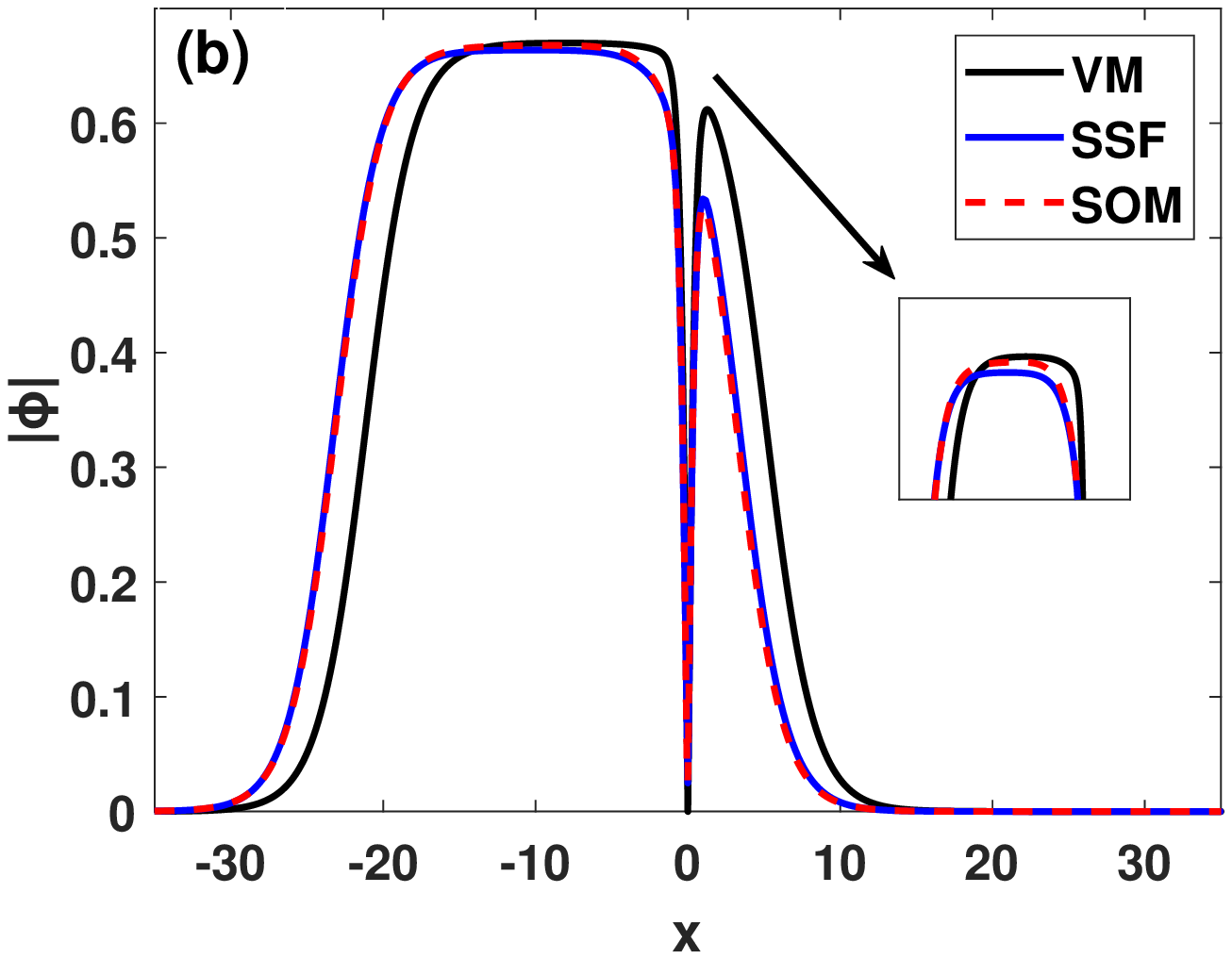}
\includegraphics[width=4.2cm]{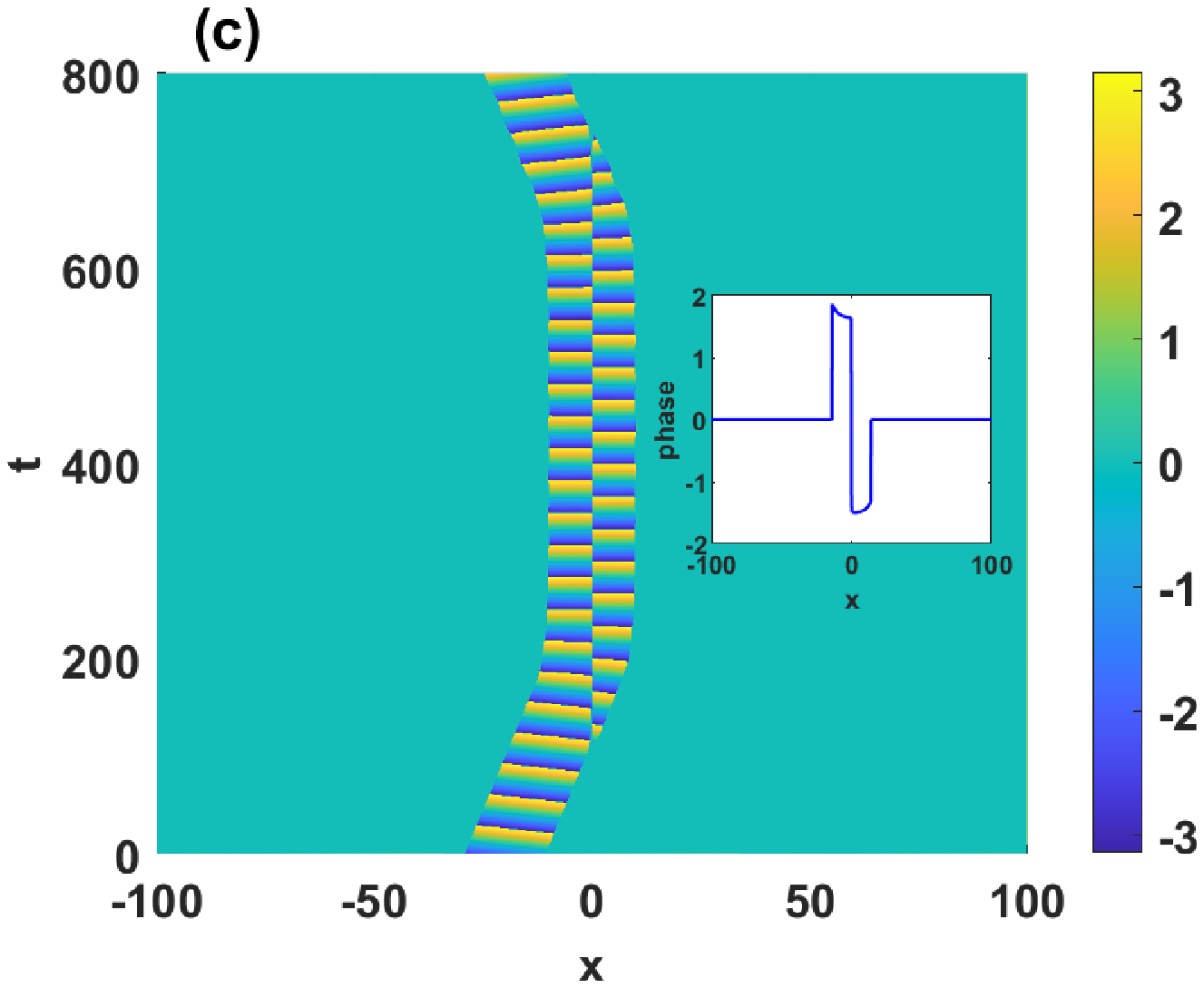}
\includegraphics[width=4.2cm]{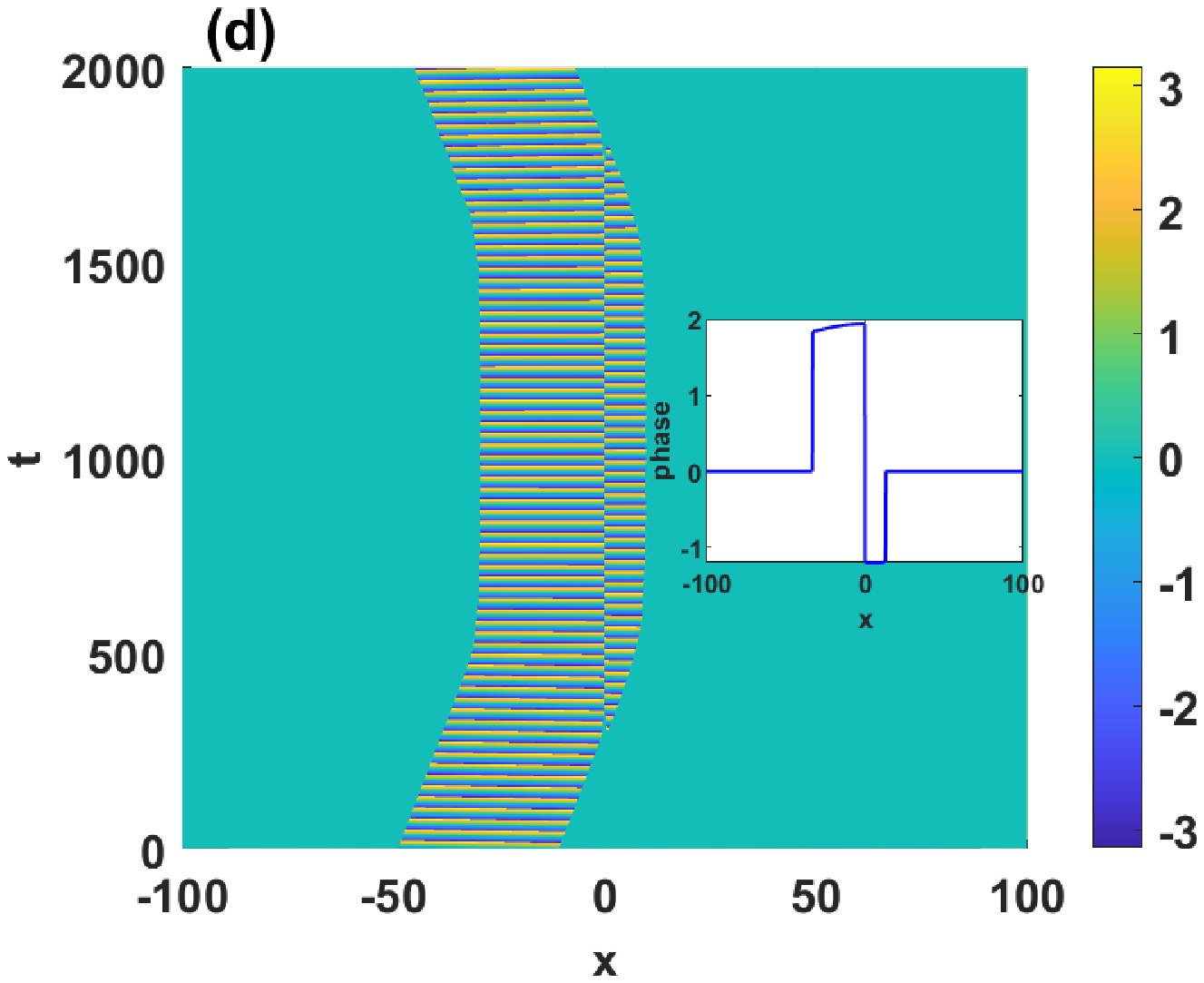}
    \caption{Profile of the trapped mode (upper panels) and phase evolution (lower panels) for scattering of the quantum droplets by a reflectionless potential centered at $x=0$. (a) and (c) are for small quantum droplet ($N=1$); (b) and (d) are for large droplet ($N=10$). In Figs.\;\ref{fig5}(a) and (b), the solid (blue) lines correspond to the trapped modes resulting from the scattering of a quantum droplet at the critical speed via direct simulation of the time-dependent Eq.\;\eqref{con:14} using the split-step Fourier (SSF) method. The dotted (red) lines are the numerical stationary solutions of Eq.\;\eqref{con:14} given by the square operator method (SOM), with the chemical potential $\mu$ fine-tuned to require the norm of the resulting state to be equal to that of the initial droplet in the quantum scattering. The solid (black) lines correspond to the results obtained by the variational method (VM). The lower panels show the phase evolution of the small quantum droplet (c) with initial speed $v = 0.0928757$ and the large quantum droplet (d) with $v = 0.03532$. Values of other parameters used are $g = 1$ and $U_0=4$.}
	\label{fig5}
\end{figure}
The results of the variational approach show the existence of zero-speed states with energies lower than the energy of the stationary quantum droplets ($E_{sd}$), as shown by the dips in Fig.\;\ref{fig3}. These low-energy states are not excited during the scattering process because the energy of the incident quantum droplets is higher than that of these zero-speed states obtained by the variational method. In Fig.\;\ref{fig3}(b), we are surprised to find that there is a local energy minimum at $x_0=0$ for large quantum droplet, corresponding to a stable spatially symmetric trapped mode. However, its energy is lower than the energy of the metastable state. The remarkable feature suggests that for quantum reflection of large droplets at the critical speed, a metastable state with an unbalanced intensity profile is excited instead of the balanced trapped mode, whereas for initial speed above the critical speed, the energy of the incoming droplet is large enough to pass through the potential and thus is not arrested in the balanced trapped mode.

To confirm the stable balanced trapped mode of the large droplet predicted by the variational method, we first numerically obtain the balanced stationary mode of Eq.\;(\ref{fig4}) by the square operator iteration method using the variational solution as the initial seed, and then give the propagation of the balanced stationary mode, which is invariant with time as expected, as illustrated in Fig.\;\ref{fig4}(a). By imposing the balanced stationary mode with an initial speed, as can be seen in Fig.\;\ref{fig4}(b), the propagation simulation shows that this state oscillates back and forth around the equilibrium point $x = 0$, showing that this state is trapped by the potential and neither reflection nor transmission occurs. The numerical results shown in Figs.\;\ref{fig4}(c) and (d) verify the existence of low energy states. When the quantum droplet is initialized in the low-energy stationary states, the interaction of the stationary quantum droplet with the potential well leads to the excitation and emission of the quantum droplet. The excited quantum droplet switches between the different low-energy states, resulting in local oscillations in the vicinity of the center of the potential well.

In Figs.\;\ref{fig5}(a) and (b), we plot the profile of the stationary state corresponding to the trapped mode for the small and large quantum droplets, using three numerical methods. Our first numerical procedure to calculate the trapped mode starts by solving Eq.\;\eqref{con:14} using the split-step Fourier (SSF) method, where the incident droplet speed is adjusted until the droplet is trapped in the potential for a substantial amount of time. The variational method (VM) is the second method of calculating the trapped mode based on the trial wave function \eqref{con:22} as described in detail at the beginning of this section, and the third method involves calculating the nonlinear stationary state of Eq.\;\eqref{con:14} using a squared operator method (SOM) \cite{Yang2007}. As can be seen in Fig.\;\ref{fig5}(a), for small quantum droplets, the trapped mode profiles obtained by the three methods are basically identical, and this trapped mode corresponds to a metastable balanced trapped mode. While for large quantum droplets, an unbalanced trapped mode profile is obtained as shown in Fig.\;\ref{fig5}(b), and the results obtained by SSF method and SOM are in good agreement. However, the result obtained by the VM shows slight discrepancies compared to those obtained by SSF method and SOM, which means that the VM is somewhat inaccurate in determining the position of the unbalanced trapped mode. Figs.\;\ref{fig5}(c) and (b) show the phase evolution of the quantum droplet for the initial speed at the critical speed, and the phase plots show a phase jump of $\pi$ at $x = 0$. This is consistent with our chosen trial wave function \eqref{con:22} with a node given by $\tanh(x)$ which satisfies this phase jump. In addition, in Table \ref{table:I}, we list the numerical values of the critical droplet speed calculated by  these three numerical methods.  A good match can be seen clearly in the table.
\begin{table}[htbp]
	\centering
	\caption{Three methods to solve the critical speed}
	\label{tab:1}
	\begin{tabular}{cccc}
		\hline\hline\noalign{\smallskip}
\toprule
\multirow{2}{*}\quad  & \multicolumn{3}{l}{\qquad \qquad Critical speed $v_c$}   \\
\cline{2-4}  
		 $N$ & SSF & VM & SOM  \\
		\noalign{\smallskip}\hline\noalign{\smallskip}
   0.1 & 0.0568 & 0.059008912 & 0.0567048776  \\
   0.3 & 0.0811 & 0.075073377 & 0.0811317970  \\
   0.5 & 0.0900 & 0.089720945 & 0.0899700200  \\
   0.6 & 0.0921 & 0.095240699 & 0.0920241325  \\
   0.8 & 0.0935 & 0.096590948 & 0.0935224949  \\
   1 & 0.0928 & 0.095695377 & 0.0928252791 \\
   2 & 0.078 & 0.080108024 & 0.0784391394  \\
   4 & 0.055 & 0.053647658 & 0.0555275841  \\
   6 & 0.045 & 0.043133578 & 0.0453212798  \\
   8 & 0.039 & 0.037331779 & 0.0390179123 \\
   10 & 0.035 & 0.033341549 & 0.0351731171  \\
   12 & 0.031 & 0.030435107 & 0.0312519825 \\
   14 & 0.029 & 0.028031223 & 0.0290528706  \\
   16 & 0.027 & 0.026326940 &  0.0273519847  \\
		\noalign{\smallskip}\hline
	\end{tabular}
\label{table:I}
\end{table}
\begin{figure}[htbp]
\includegraphics[width=4.2cm]{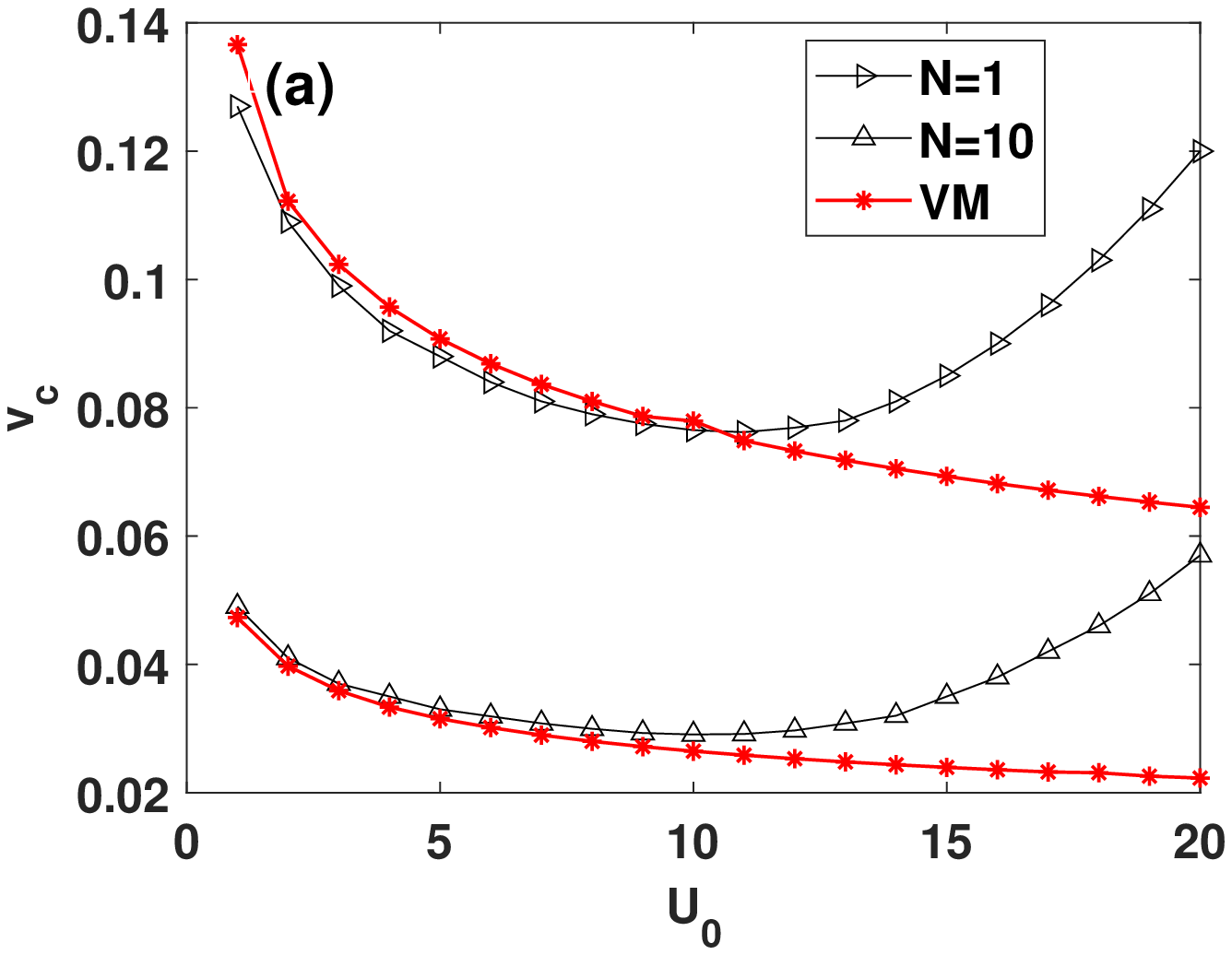}
\includegraphics[width=4.2cm]{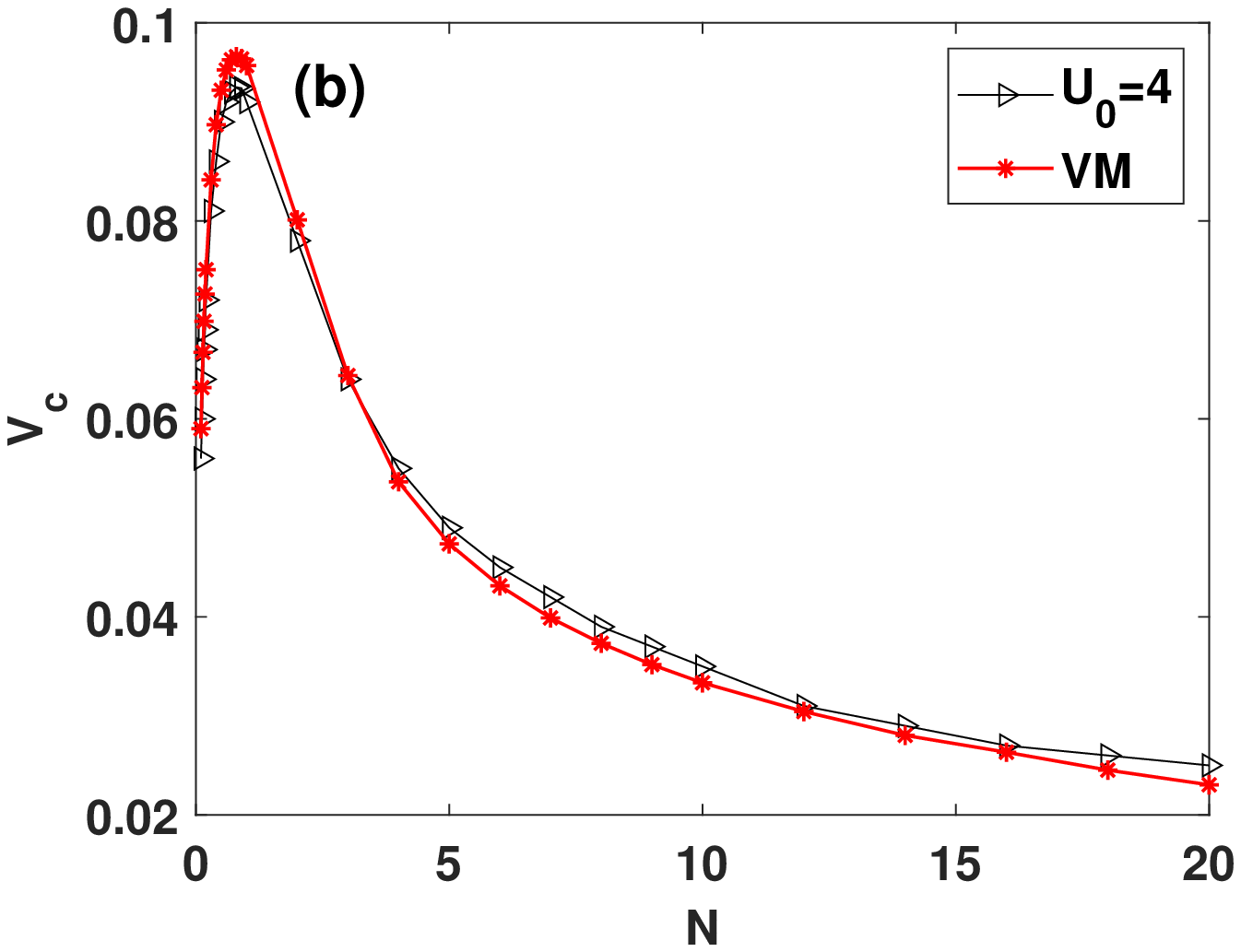}
\caption{(a) The critical speed $v_c$ versus potential well depth $U_0$; (b) The critical speed $v_c$ versus norm $N$. The black lines represent the results obtained from the numerical simulation of equation \eqref{con:14} using the SSF method, and the red lines represent the results obtained by the variational method.}
\label{fig6}
\end{figure}
\begin{figure*}[htbp]
\includegraphics[width=5.3cm]{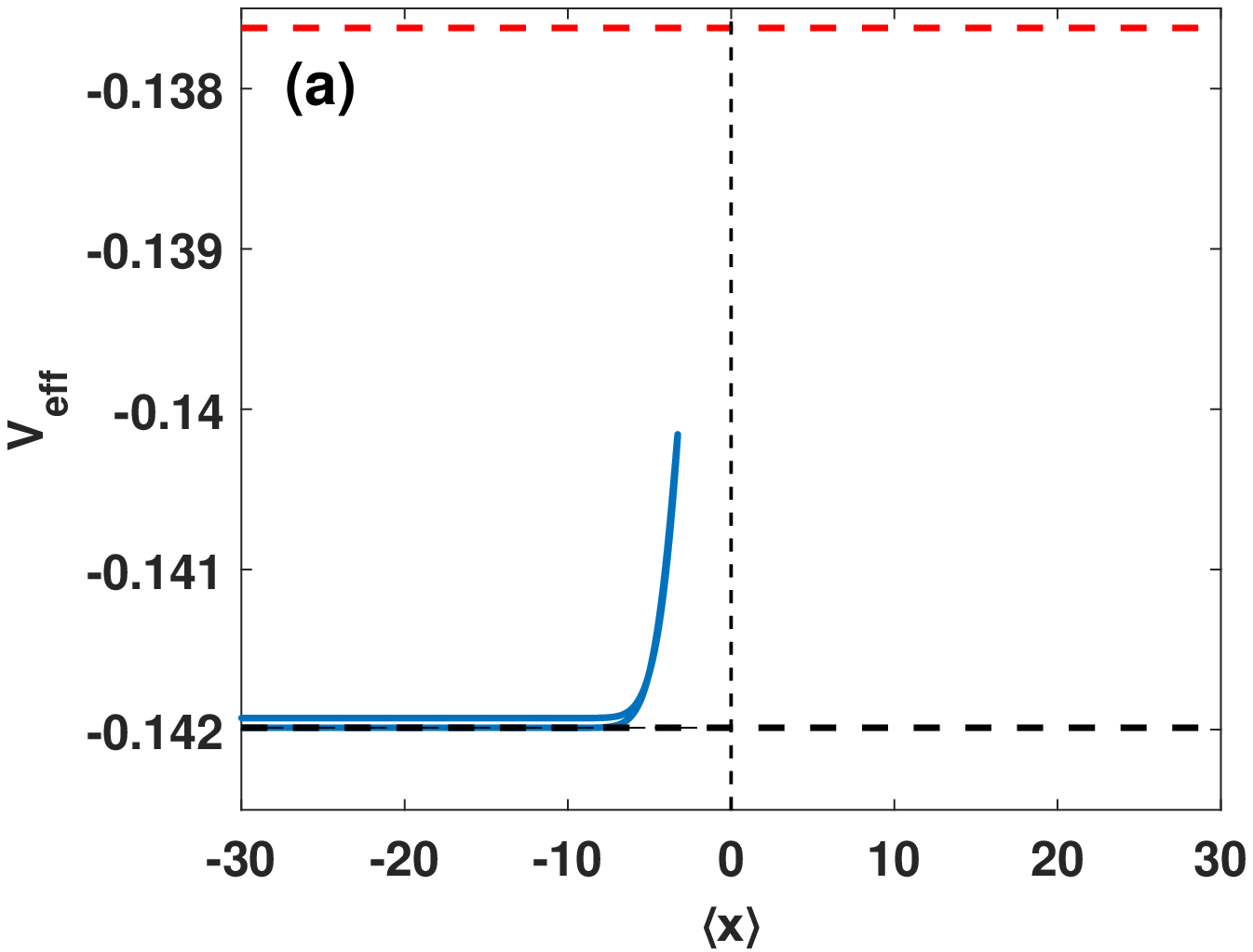}
\includegraphics[width=5.3cm]{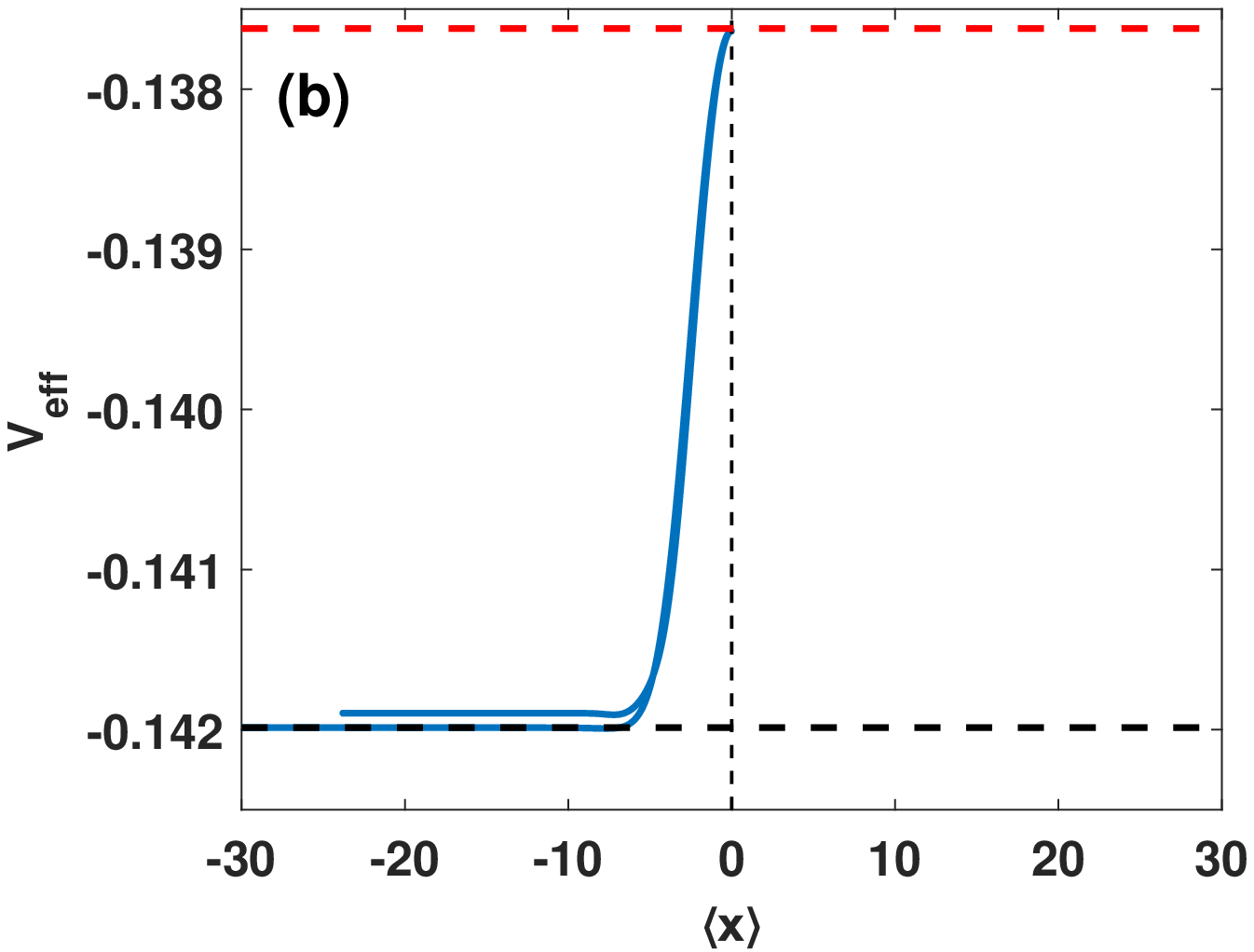}
\includegraphics[width=5.3cm]{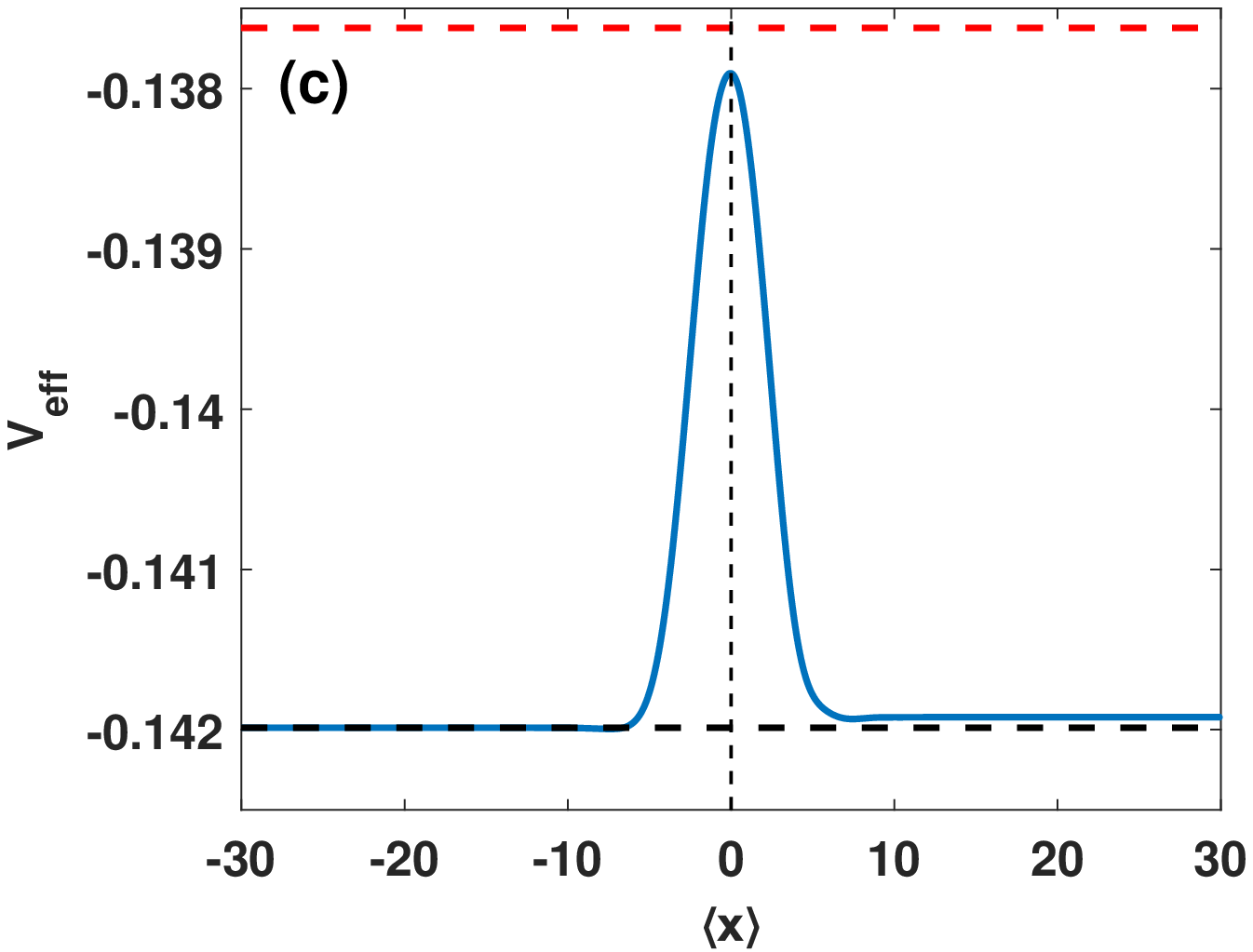}
\includegraphics[width=5.3cm]{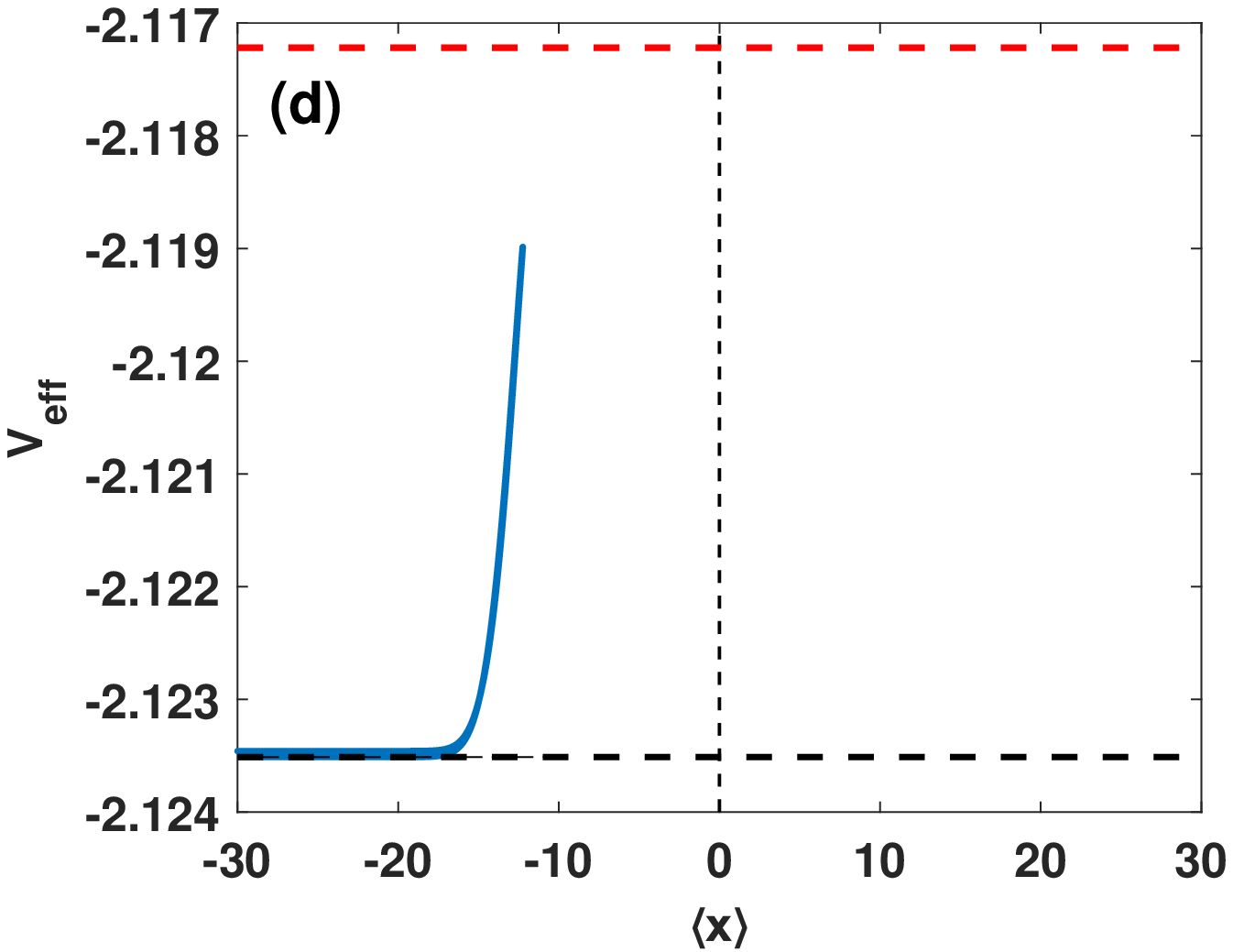}
\includegraphics[width=5.3cm]{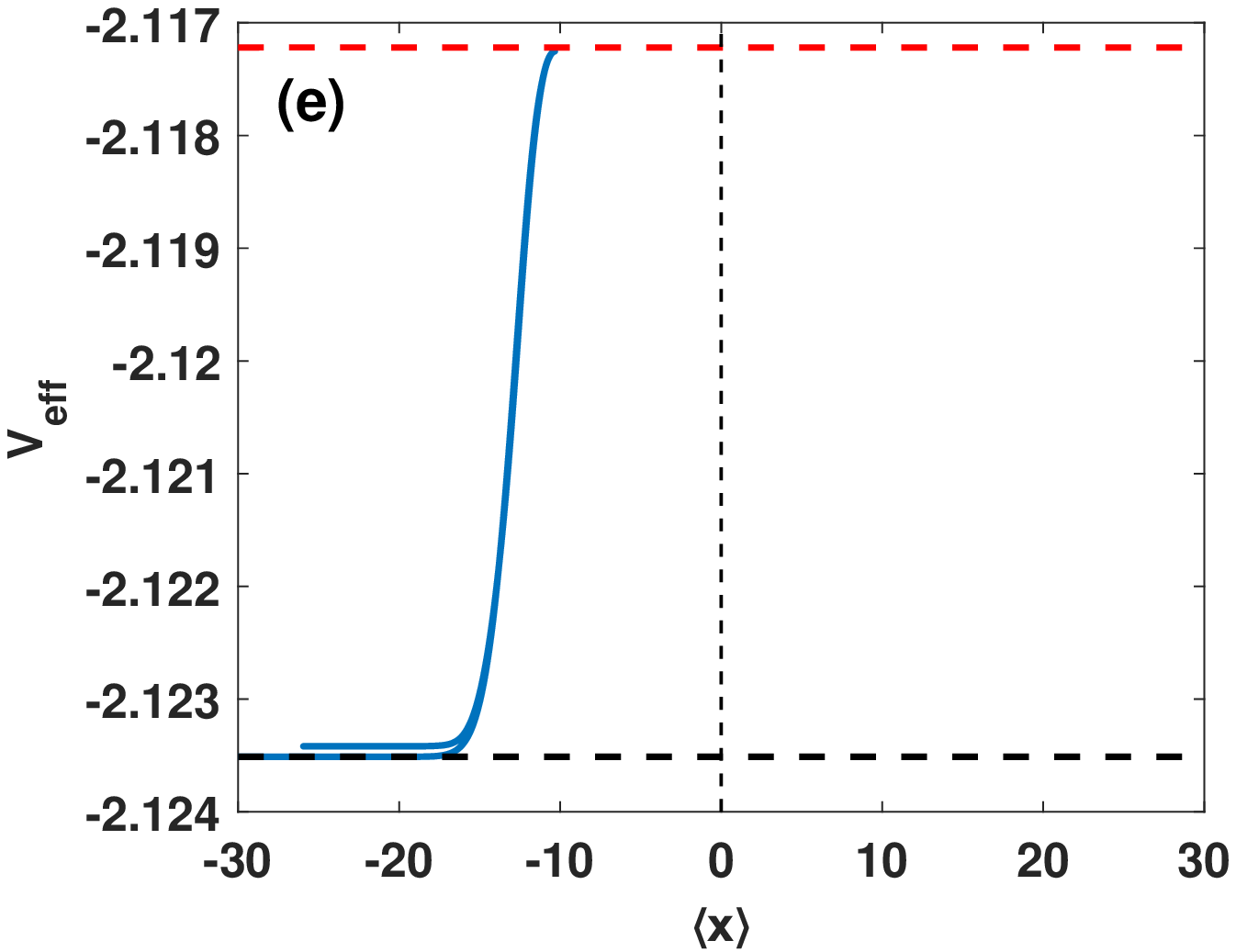}
\includegraphics[width=5.3cm]{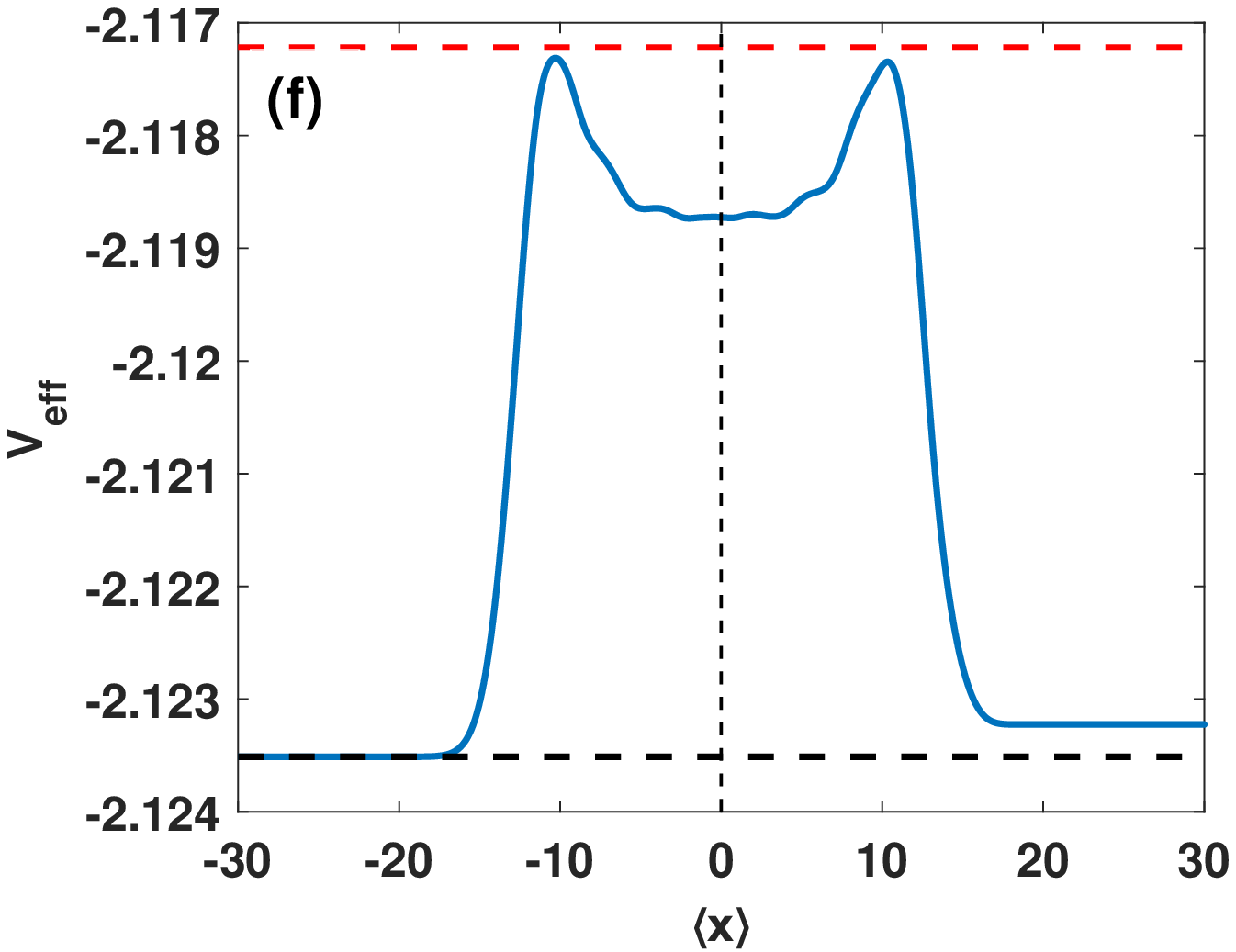}
     \caption{Effective potential plotted versus the position of the center of mass of the droplet, $\bar{x}(t)$, for the small ($N=1$, top row) and large ($N=10$, bottom row) quantum droplets scattered by the reflectionless potential with speeds less than [(a) and (d)], equal to [(b) and (e)], and greater than  [(c) and (f)] the critical speed. The horizontal dashed (red) and dotted (black) lines correspond to the energy of the trapped mode and the stationary initial quantum droplet energy, $E_{sd}$, respectively. Other parameters used are the same as those in Fig.\;\ref{fig2}.}
    \label{fig7}
\end{figure*}

In Fig.\;\ref{fig6}(a), we report the value of the critical velocity as a function of the potential well depth $U_0$, comparing the results for small and large $N$ derived from both the VM solution and the exact numerical solution using the SSF method. As expected, the critical speed is well captured by the VM for a wide range of relatively small $U_0$ (roughly $U_0<10$). As the potential well depth increases progressively, the VM becomes inapplicable. Fig.\;\ref{fig6}(b) shows the dependence of the critical speed on the size of the droplet. There is a nonmonotonous dependence, with the largest critical speed reached at $N \approx 1$. For small $N$, the critical speed increases with the increase of $N$, which is the same as in the case of bright solitons. However, the dependence of the critical speed of the droplet on the atomic number is reversed when the droplet is in the flat-top regime. It can be seen that for larger $N$, the critical speed of the droplet decreases with increasing $N$. This is a unique feature in comparison to the 1D soliton, where the critical speed is only found to increase with $N$. This peculiarity occurs because of a transition from small droplets with a $\sech^2$-shape to large droplets with a flat-top plateau. A small droplet has an approximately $\sech^2$-shape with decreasing width as $N$ increases, while in the opposite limit, unlike its 1D soliton counterpart, the size of large droplets grows with $N$. As can be seen in Fig.\;\ref{fig6}(b), the VM is able to accurately predict the critical speed even for large droplets with flat-top density profiles.

\section{the scattering process of quantum droplets}
\label{section:V}
Understanding the scattering of quantum droplets is facilitated by modeling their scattering process in terms of the motion of a classical object. The complexity of calculating the effective potential experienced by the droplet during scattering using the time-dependent variational method is due to the absence of an analytic form of the integral in the variational method. For this reason, we numerically determined the position of the center of mass of the droplet as $\bar{x}(t) = \langle x \rangle = \int_{-\infty}^{+\infty}x |\psi(x,t)|^2 dx$, where $\psi(x,t)$ is the time-evolved state given by the direct numerical simulation of Eq.\;\eqref{con:14}. From the equation of motion $N \ddot{{\bar{x}}}(t)=-\frac{d}{d \bar{x}} V_{\mathrm{eff}}\left[\bar{x}(t)\right]$, we obtain the effective potential

\begin{equation}\label{con:24}
V_{\mathrm{eff}}[\bar{x}(t)] = -N \int_{-\infty}^{+\infty}\ddot{\bar{x}}(t) d\bar{x}+V_{\mathrm{eff}}[\bar{x}(0)],\\
\end{equation}
where $V_{\mathrm{eff}}[\bar{x}(0)]=E_{sd}$. In this way, we can describe the quantum droplet as a classical particle with an effective mass $N$ moving in the effective potential \eqref{con:24}.

In Fig.\;\ref{fig7},  we plot the effective potential versus droplet position for small $N$ (top row) and large $N$ (bottom row), with initial speeds below, equal to, and above the critical speed. The scattering dynamics of a quantum droplet is similar to that of a classical object that is launched from the left and rises along a potential slope, the initial speed of which determines a different trajectory. When $v < v_c$, the object climbs along the potential ramp to the maximum height, the speed disappears, and then the object falls back down. This corresponds to the quantum reflection of the quantum droplet. At the critical speed, $v = v_c$, the maximum height of the potential ramp corresponds exactly to the energy of the trapped mode.  In this case, the object will reach exactly the highest point at which it can remain for a long time, which corresponds to the critical state of the droplet being completely trapped by the potential. When $v > v_c$, the initial kinetic energy is sufficient for the object to cross the maximum barrier height, corresponding to full transmission. The effective potential of the large droplet has two peaks that are shifted from the origin, corresponding to the spatially asymmetric trapped state. In this situation, the object crosses the maximum height of the potential barrier twice, as shown in Fig.\;\ref{fig7}(f).
\begin{figure}[htbp]
\includegraphics[width=4.2cm]{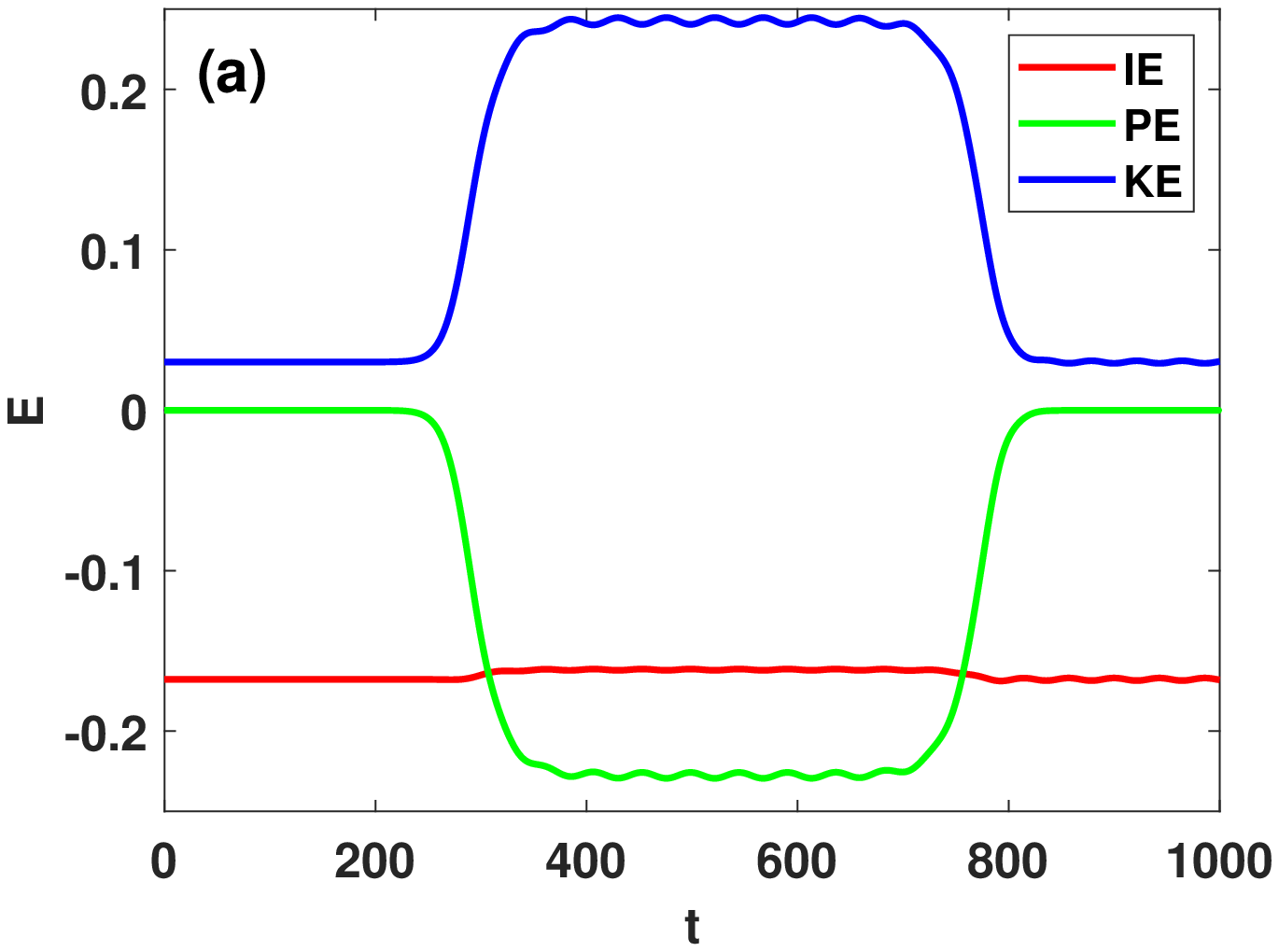}
\includegraphics[width=4.2cm]{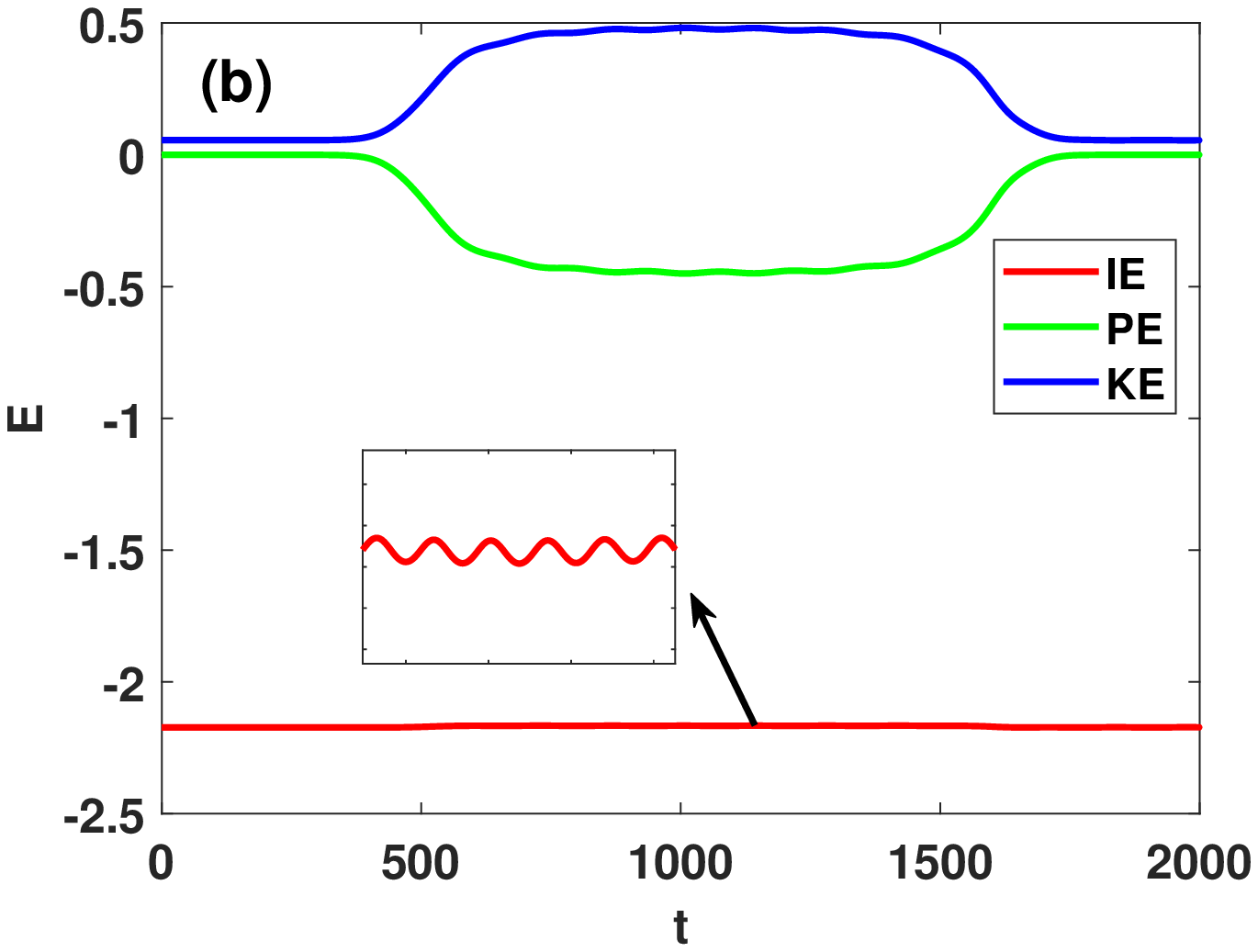}
 \caption{Time evolution of kinetic energy (KE), interaction energy (IE) and potential energy (PE) during the scattering of quantum droplets. (a) is for small quantum droplet ($N=1$) with the initial speed $v_c=0.0928757$ [corresponding to Fig.\;\ref{fig2}(b)], and (b) is for large quantum droplet ($N=10$) with initial speed $v_c=0.03532$ [corresponding to Fig.\;\ref{fig2}(e)]. Other parameters used are the same as those in Fig.\;\ref{fig2}.}
    \label{fig8}
\end{figure}

As we know, the quantum droplet's energy functional can be divided into three parts: kinetic energy, potential energy, and interaction energy. To understand quantum droplet scattering from a quantum mechanical wave perspective, we plotted the evolution of different quantum droplet energies over time. In particular, we consider the energies for the case of small [Fig.\;\ref{fig8}(a)] and large [Fig.\;\ref{fig8}(b)] quantum droplets with an initial speed close to the critical speed, whose spatiotemporal evolution of densities corresponds to the cases of Fig.\;\ref{fig2}(b) and Fig.\;\ref{fig2}(e), respectively. As can be seen in Fig.\;\ref{fig8}, when the incident quantum droplet is far from the potential, the potential energy is negligibly small, and thus the moving quantum droplet keeps its shape unchanged due to the balance between the different energy contributions. As the quantum droplet approaches the potential well, the moving quantum droplet interacts with the reflectionless potential, causing the potential energy to decrease and the kinetic energy to increase, while leaving the interaction energy essentially unchanged. The increase in kinetic energy here is due to the increase in quantum pressure caused by the dispersion of the quantum droplet, which represents a change from the initial quantum droplet state to the trapped mode. This transition is a consequence of the wave nature of quantum mechanics. At the critical speed, the conversion of potential energy into kinetic energy causes the quantum droplet to diffuse into the trapped mode, where the kinetic energy effect becomes the pure quantum pressure, without any excess energy acquiring the motion of the center of mass. In this case, the critical state, where the droplet is fully trapped by the potential,  occurs with a plateau of kinetic energy that includes only the contribution from the gradient of the wave function in space (i.e., the quantum pressure) due to the zero speed of the center of mass. Since the trapped mode is unstable with respect to its position, the droplet will remain in the trapped mode for a period of time and will be ejected from the potential, undergoing the opposite energy conversion and state transition as described above. Fig.\;\ref{fig8} also shows that the interaction energy oscillates after scattering, which is a manifestation of the collective excitation generated by the quantum droplets. In the next section, we will consider the collective excitation of the quantum droplets in the scattering process.

\section{Collective excitation of quantum droplets in the scattering process}
\label{section:VI}
\begin{figure}[htbp]
\includegraphics[width=4.2cm]{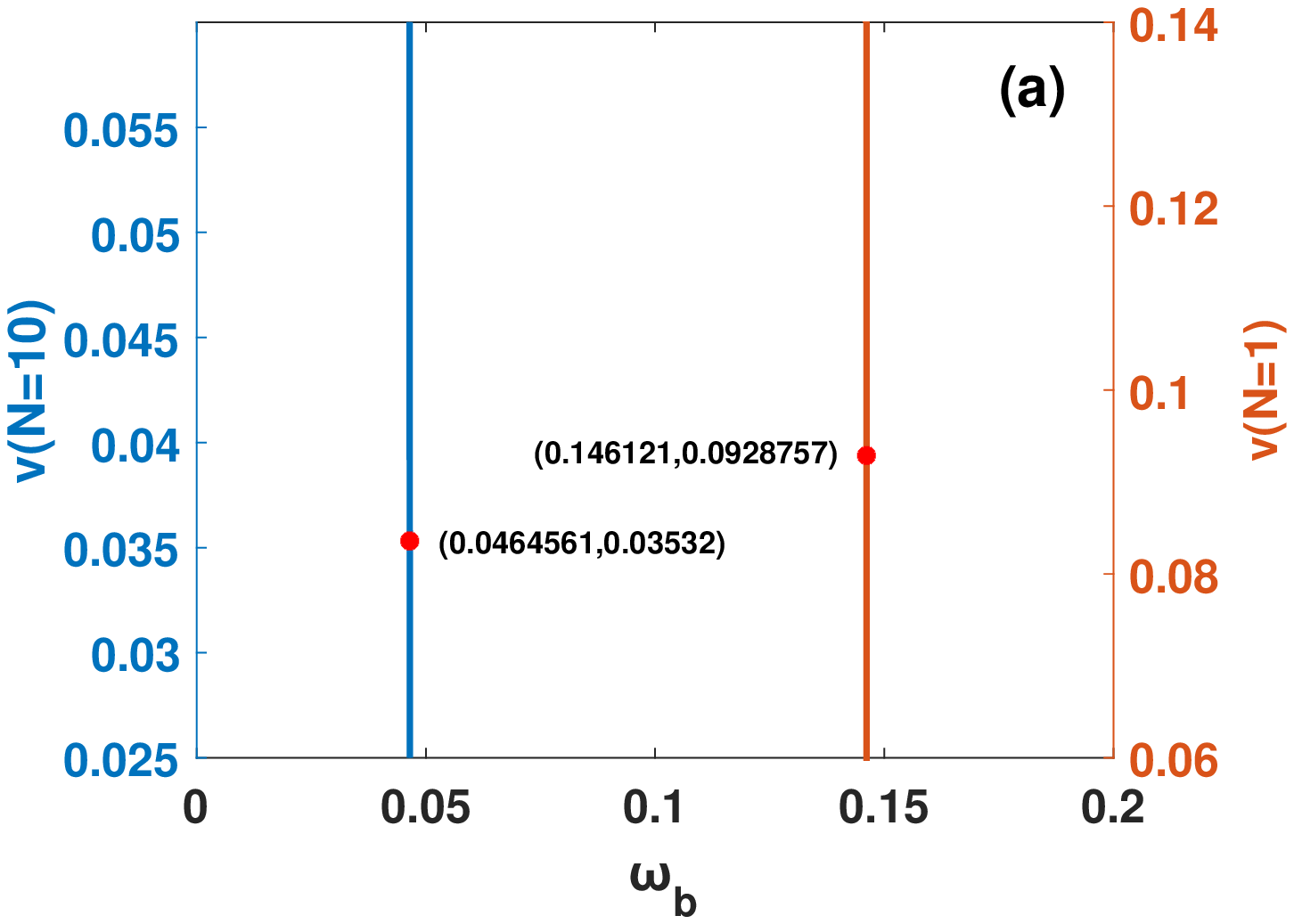}
\includegraphics[width=4.2cm]{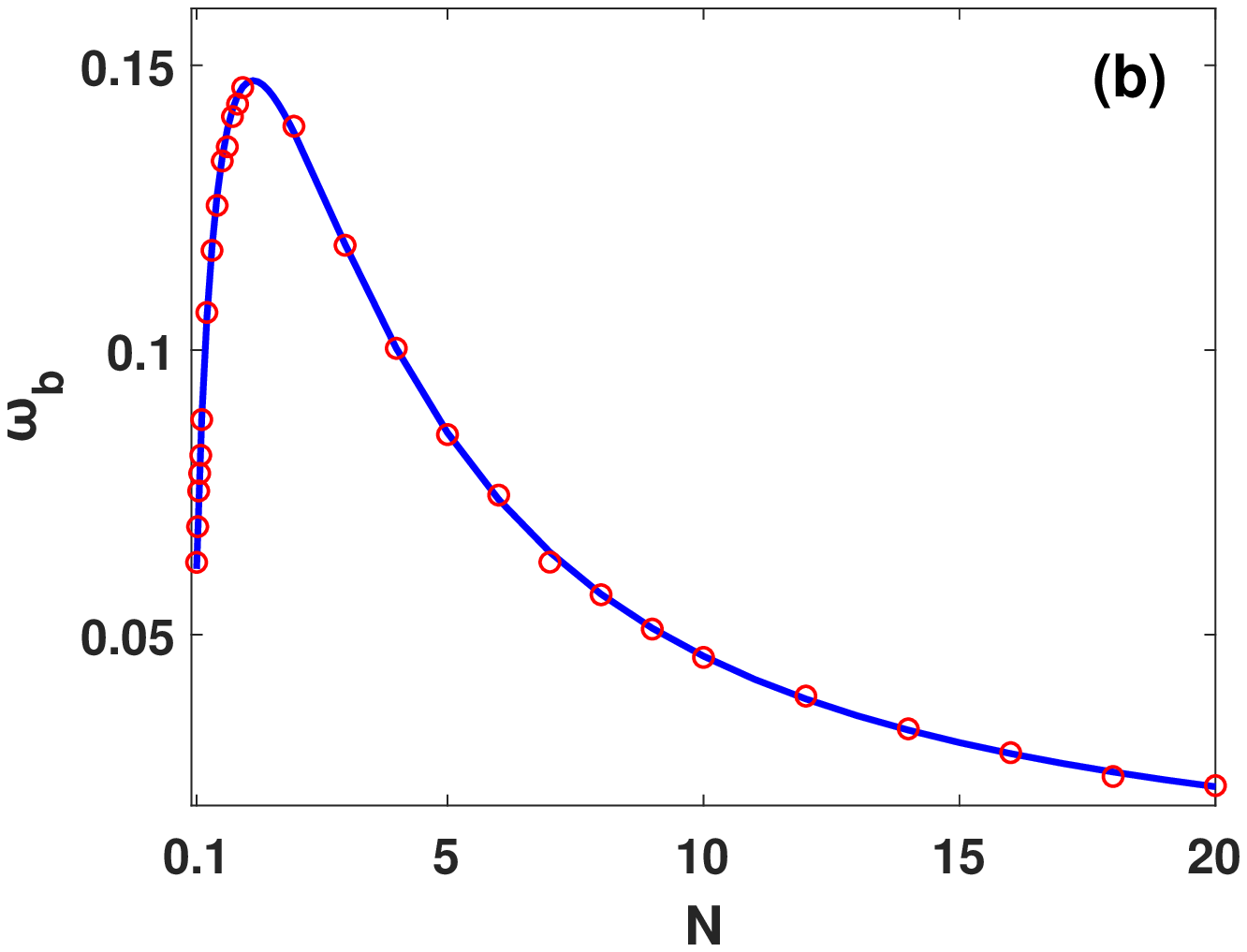}
\caption{(a) The initial speed versus breathing mode frequency $\omega_b$ with different norms $N=1$ (red line) and $N=10$ (blue line). The red solid dots mark the breathing mode frequencies at the critical speeds for the quantum reflection of the quantum droplets for the cases $N=1$ and $N=10$. (b) $\omega_b$ versus norm $N$. The red circle is the numerically calculated excitation frequency  [i.e., the oscillation frequency of $X(t)$] after scattering, and the blue solid line is the breathing mode frequency obtained by solving the BdG equation. Other parameters: $U_0=4, g=1$.}
    \label{fig9}
\end{figure}

As is well known, internal modes are linear eigenmodes inherent to stable localized nonlinear states in nonintegrable systems, which  are responsible for the internal oscillations of the nonlinear states. As observed in our numerical simulations, the quantum droplet size (and hence the interaction energy) exhibits internal oscillations after scattering, which should be explained by internal mode excitation. The breathing mode is the lowest internal mode which manifests itself as a periodic oscillation of the quantum droplet size. A deeper insight into the differences between the scattering states of the small and large quantum droplets can be gained by examining the collective excitation frequencies. To determine the frequency $\omega_b$ of the breathing mode, we numerically compute the time evolution of the standard deviation $X=\sqrt{\left \langle (x-\overline{x})^2 \right \rangle}$, which measures the size of the quantum droplet, and then use frequency spectrum analysis to obtain the oscillation frequency of $X(t)$. In Fig.\;\ref{fig9}(a), $\omega_b$ is plotted as a function of the initial speed for fixed values of $N = 1$ and $N = 10$. We observe that the frequency of the breathing mode excited during scattering is independent of the initial speed, and that the same breathing mode is excited whether the droplet is fully reflected or transmitted. Fig.\;\ref{fig9}(b) presents the dependence of the resultant breathing mode on the size of droplet. It can be seen that on the small droplet side the breathing mode frequency increases with $N$, whereas on the flat-top droplet side the breathing mode frequency decreases with increasing $N$.

To further demonstrate that the oscillatory behavior of a quantum droplet scattered from a localized potential is associated with the existence of internal mode, which can be treated as a localized linear excitation connected to the self-bound ground state of Eq.\;\eqref{con:2}, we linearize GPE \eqref{con:2} around the ground state given by Eq.\;\eqref{con:3}.Writing $\psi(x, t)=e^{-i \mu t}\left\{\psi_0(x)+\sum_{\eta}\left[u_{\eta}(x) e^{-i \omega_{\eta} t}+v_{\eta}^{*}(x) e^{i \omega_{\eta} t}\right]\right\}$ and substituting it into Eq.\;\eqref{con:2}, the Bogoliubov-de Gennes (BdG) equation is obtained by linearization,

\begin{equation}\label{con:25}
\left[\begin{array}{cc}
\mathcal{T} & \mathcal{M} \\
-\mathcal{M} & -\mathcal{T}
\end{array}\right]\left[\begin{array}{c}
u_{\eta}(x) \\
v_{\eta}(x)
\end{array}\right]=\omega_{\eta}\left[\begin{array}{c}
u_{\eta}(x) \\
v_{\eta}(x)
\end{array}\right],
\end{equation}
where
\begin{equation}\label{con:26}
\begin{aligned}
\mathcal{T}& = -\partial_{x}^{2} / 2-\mu+2 g \psi_0^{2}-\frac{3}{2}\psi_0,\\
\mathcal{M}& = g \psi_0^{2}-\frac{1}{2} \psi_0.
\end{aligned}
\end{equation}

We solve Eq.\;\eqref{con:24} numerically and find that there can be discrete eigenfrequencies $\omega_{\eta}$, denoted by the integer $\eta$, which are associated with the internal dynamics of the matter wave. The localized eigenfunctions $[u_{\eta}(x)$  $v_{\eta}(x)]^T$, corresponding to real values of $\omega_{\eta}$, are known as internal modes of the nonlinear system. The breathing mode with $\eta = 2$ is the lowest nontrivial collective mode in our setup.

\begin{figure}[htbp]
\includegraphics[width=4.2cm]{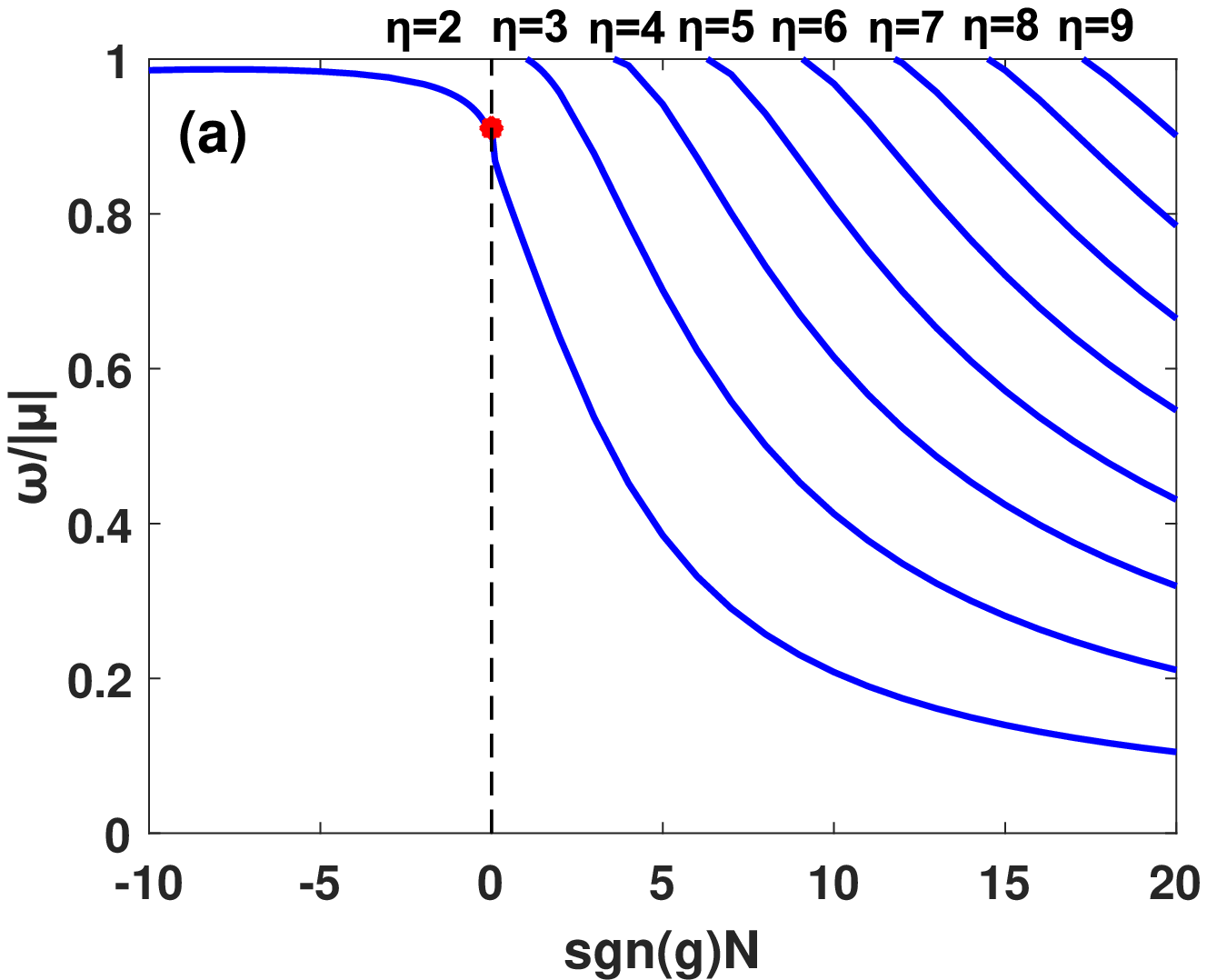}
\includegraphics[width=4.2cm]{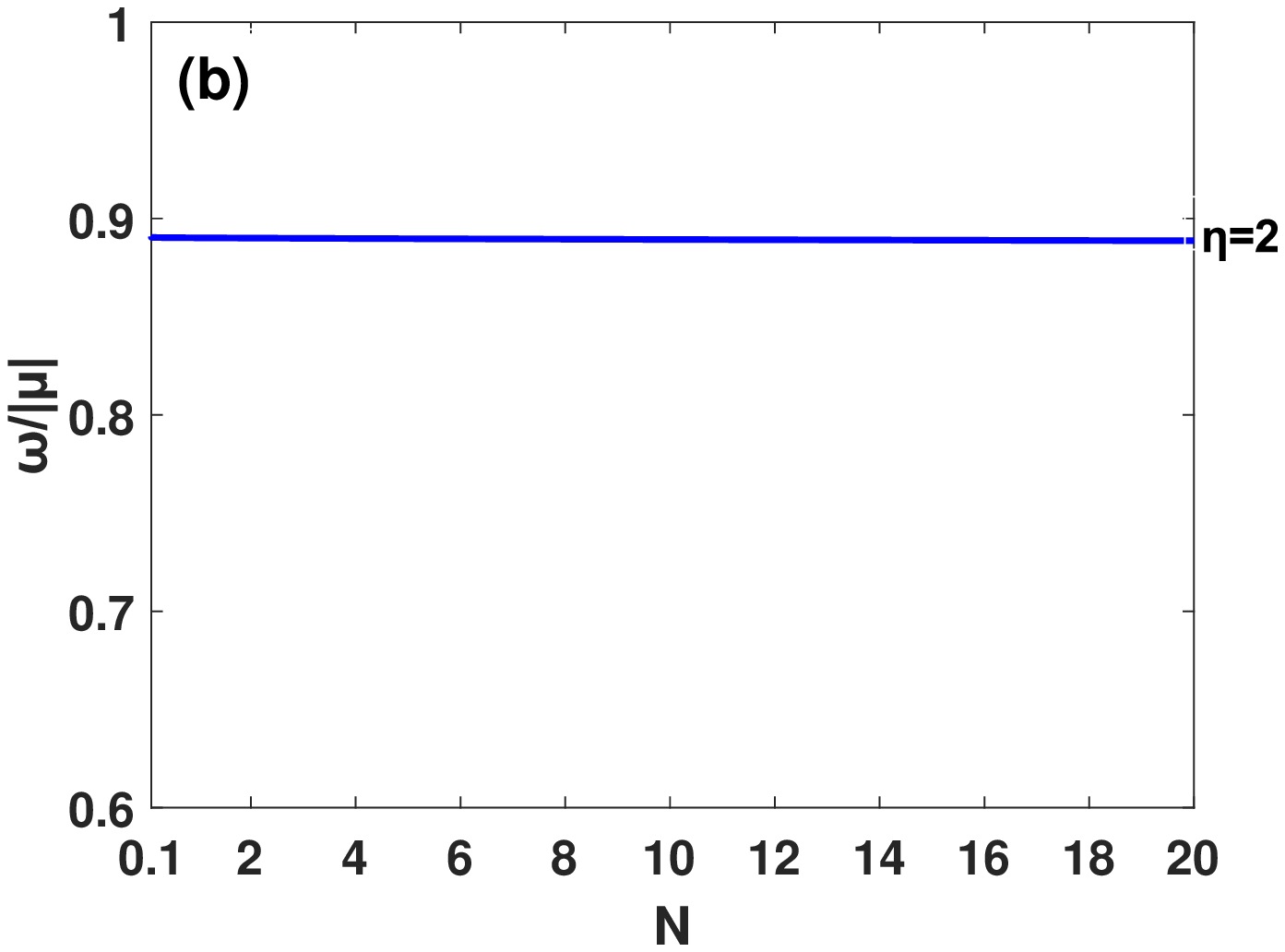}
\caption{(a) The ratio of the discrete Bogoliubov frequency $\omega_\eta$ to the particle emission threshold $-\mu$ as a function of sgn$(g)N$ [where sgn$(g)$ is the sign of $g$]. (b) The ratio of the discrete Bogoliubov frequency $\omega_2$ to the particle emission threshold $-\mu$ as a function of $N$ at $g=0$. }
    \label{fig10}
\end{figure}

\begin{figure*}[htbp]
\includegraphics[width=5.3cm]{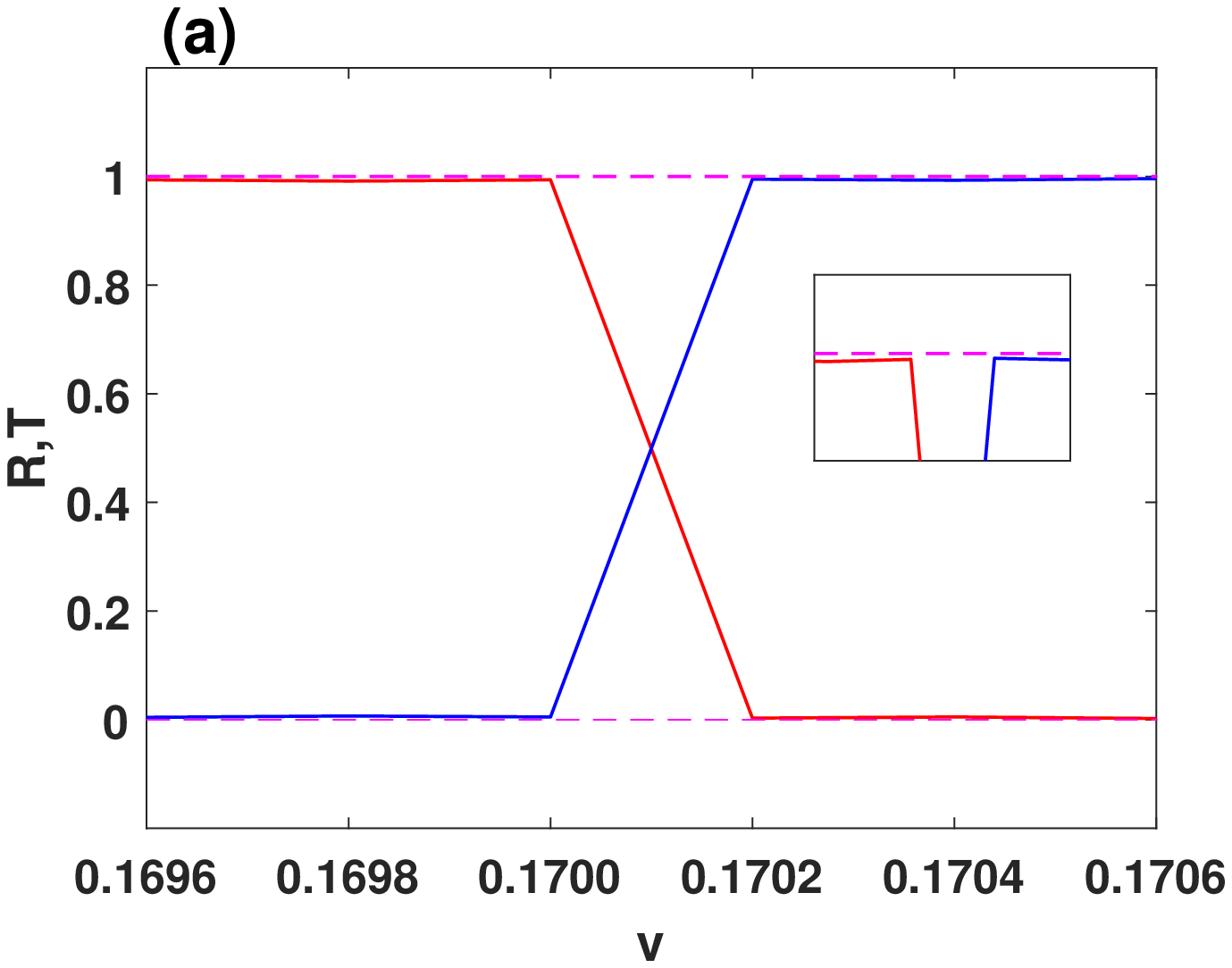}
\includegraphics[width=5.3cm]{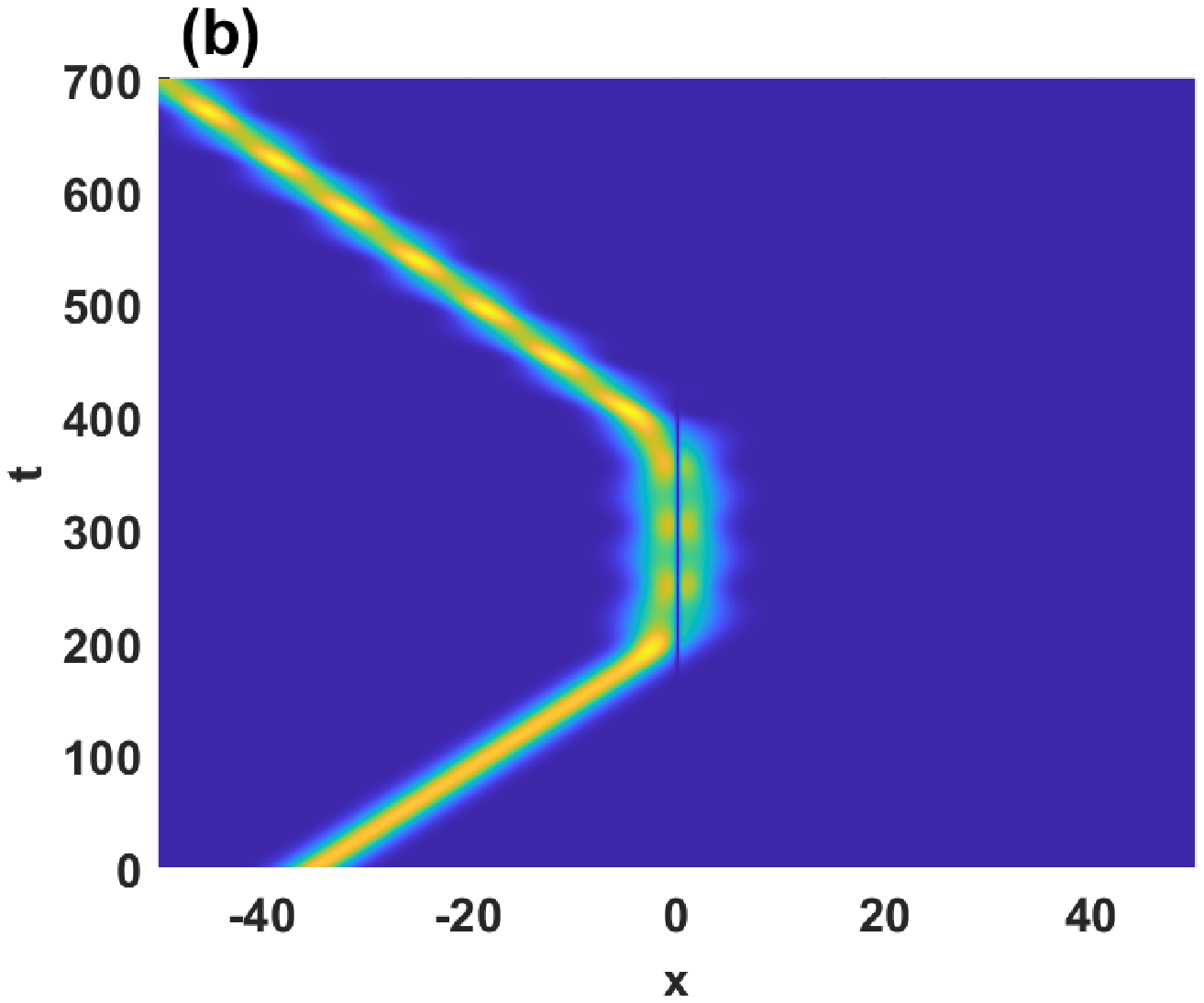}
\includegraphics[width=5.3cm]{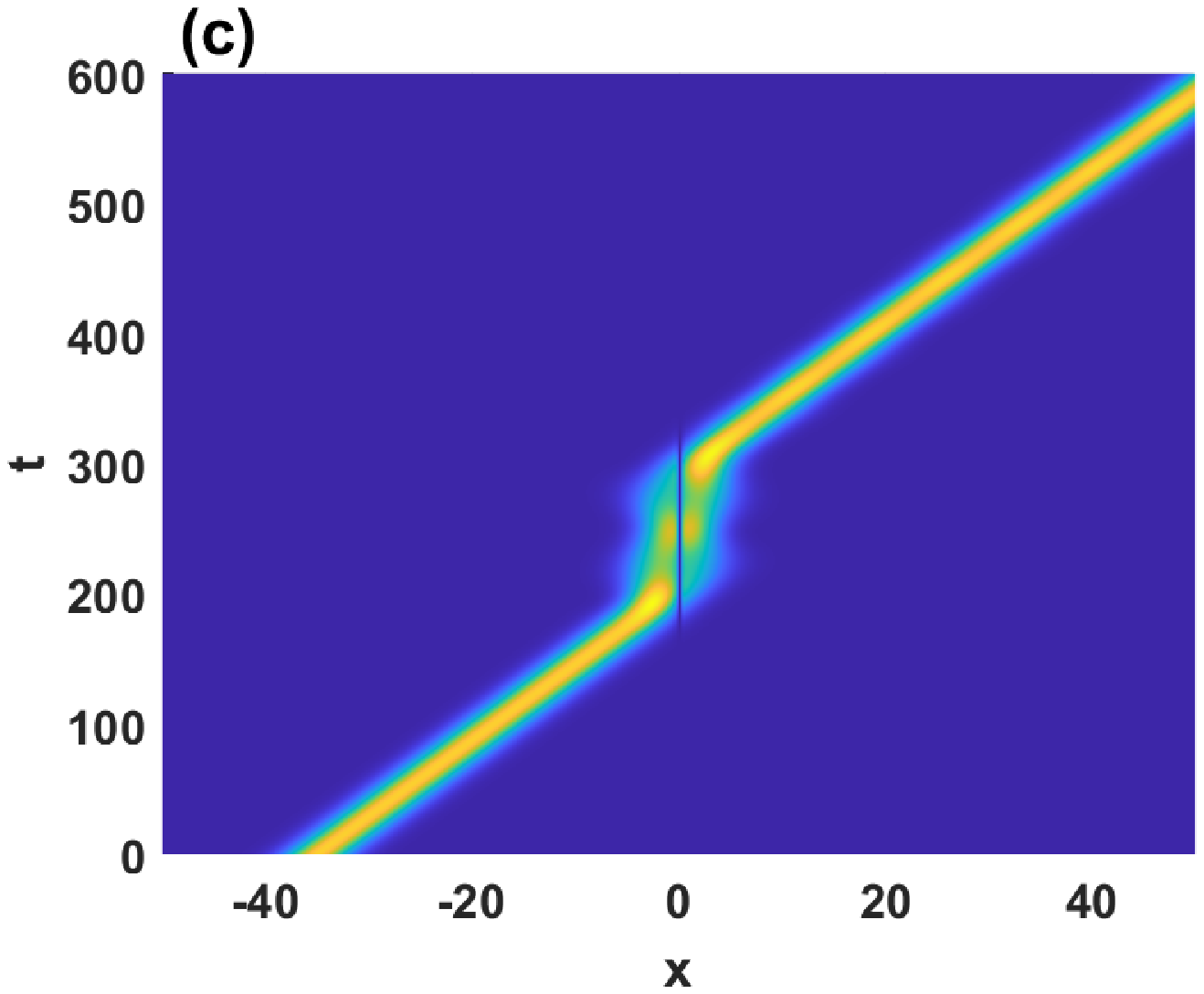}
\includegraphics[width=5.3cm]{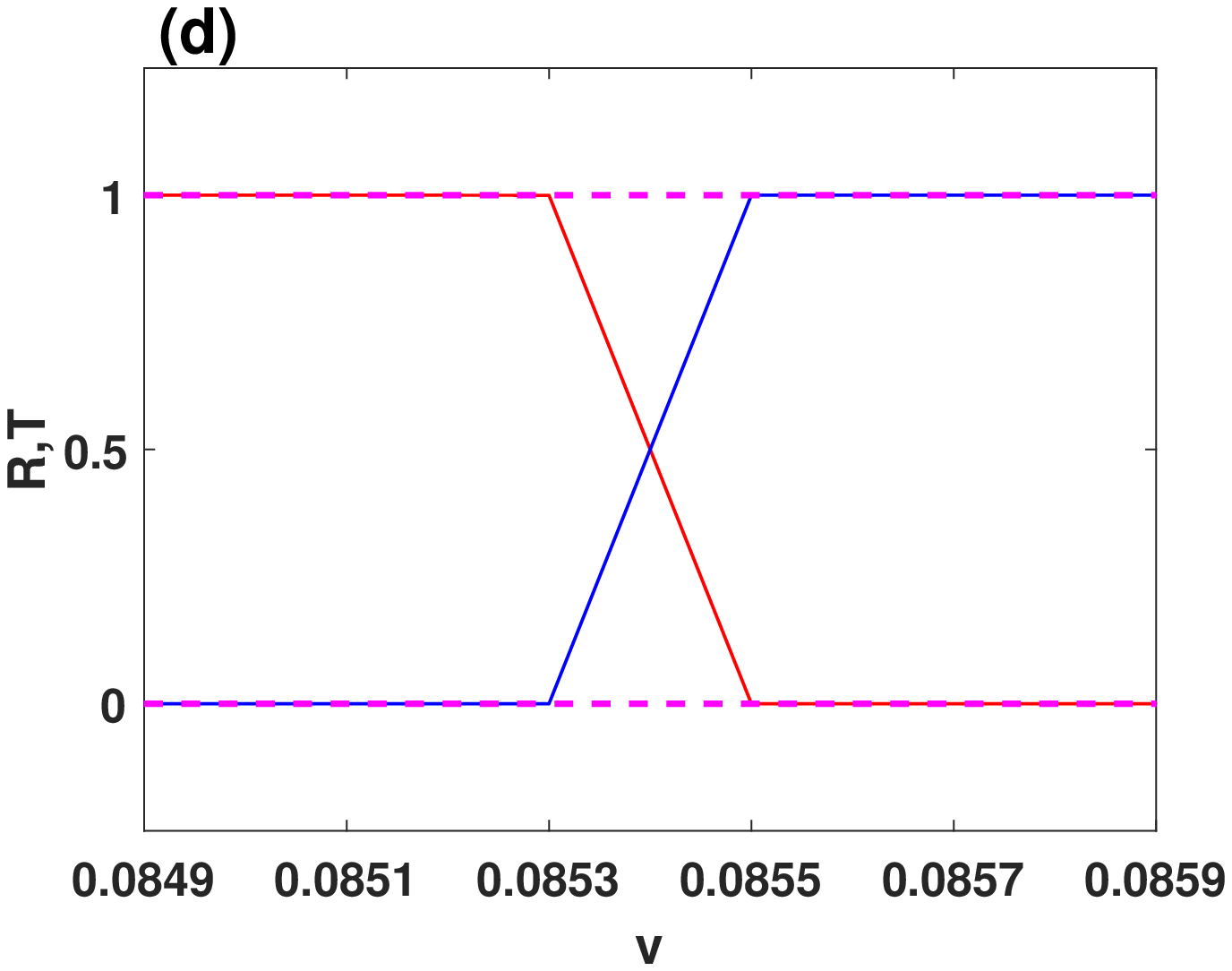}
\includegraphics[width=5.5cm]{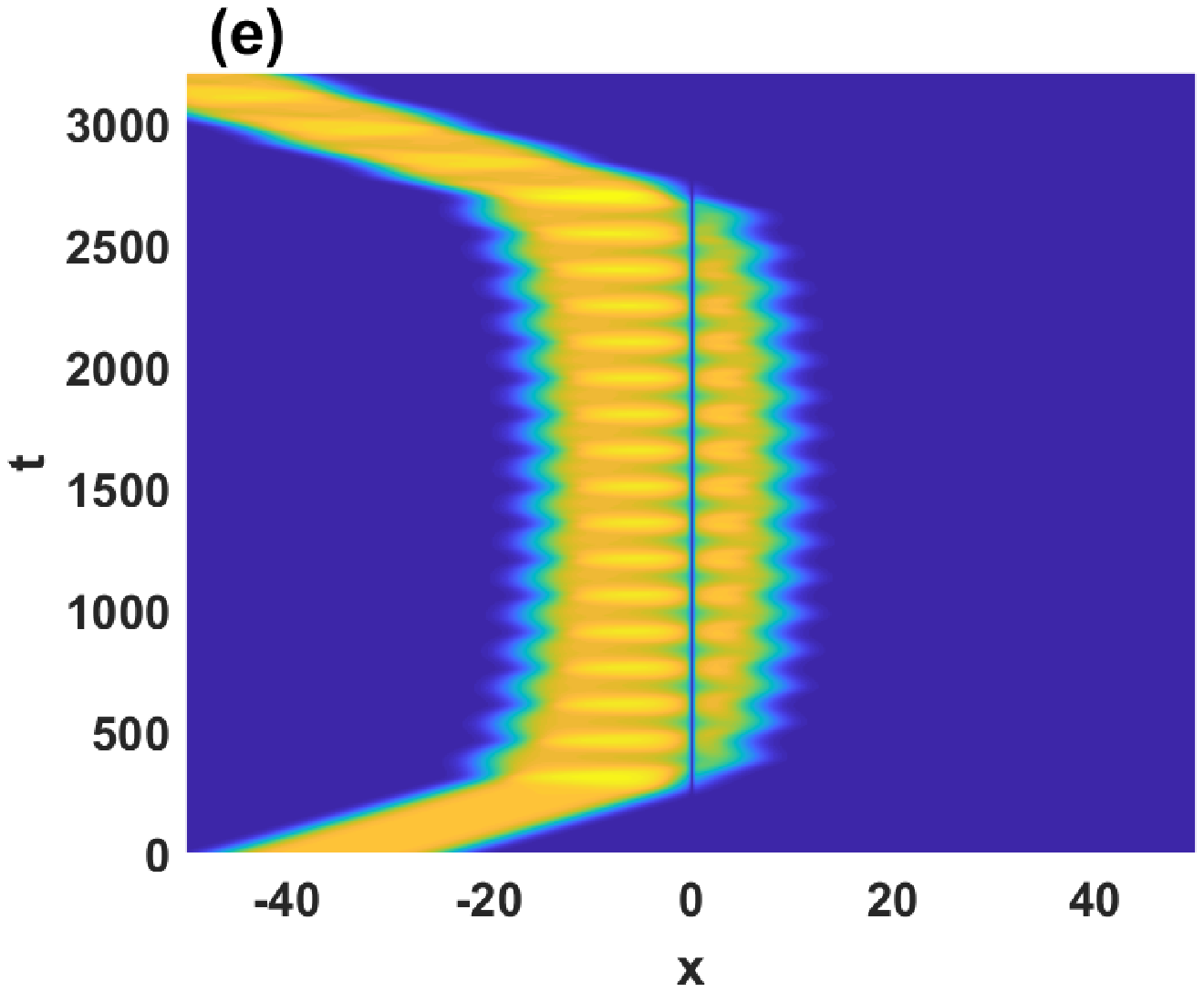}
\includegraphics[width=5.5cm]{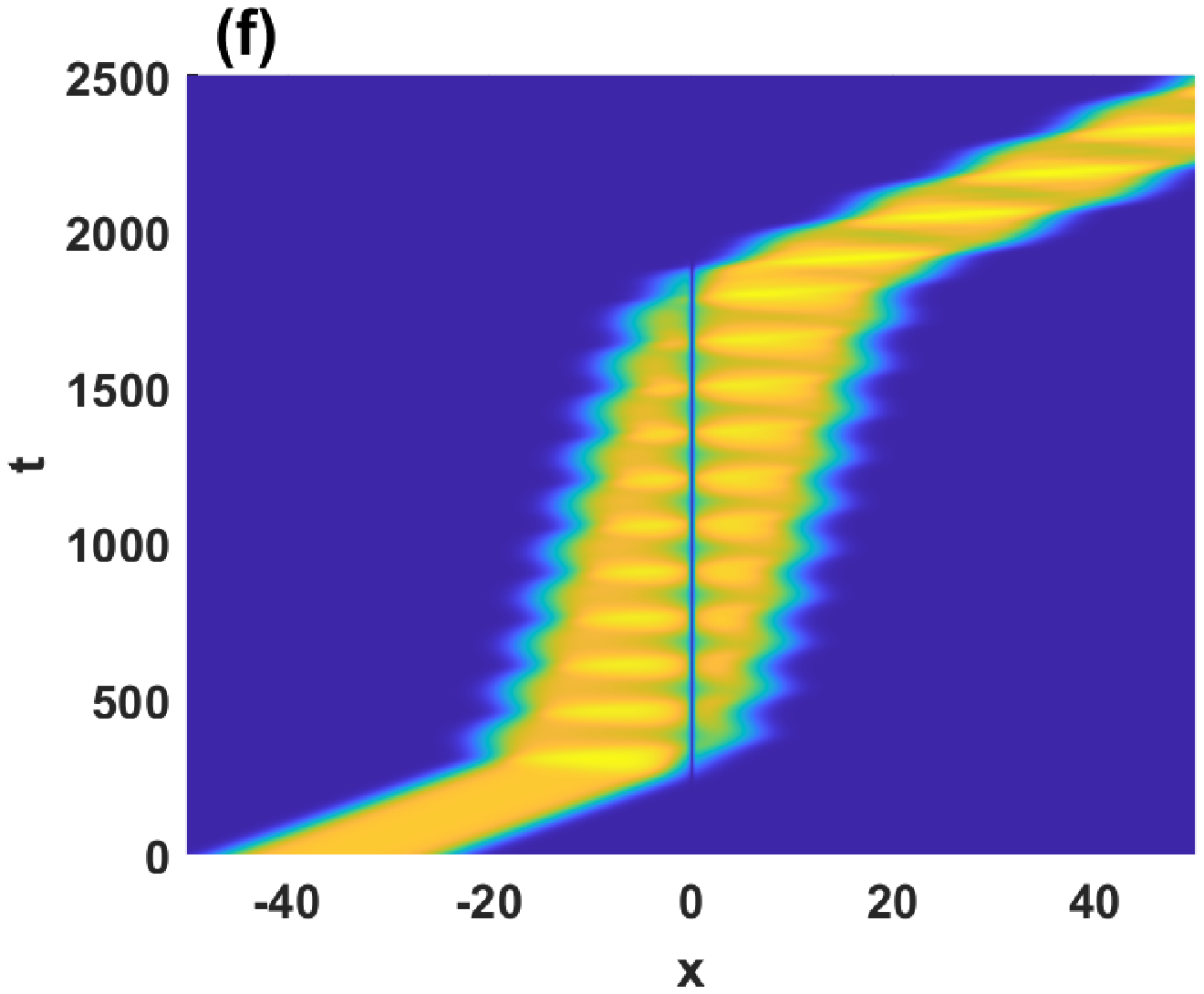}
       \caption{ Scattering of quantum droplets by a reflectionless potential well with a large potential depth for the small ($N=1$, upper row) and the large ($N=10$, lower row) quantum droplets. (a) and (d) Reflectance (red) and transmittance (blue) versus the initial speed of quantum droplet. Upper row: (b) quantum reflection of the small droplet with the initial speed $v=0.17$; (c) transmission with $v=0.171$. Lower row: (e) quantum reflection of the large droplet  with the initial speed $v=0.0853$; (f) transmission with $v=0.086$. Other parameters: $U_0=25$, $\alpha=\sqrt{U_{0}}$, $g=1$, $\tilde{x}_0=-35$. Here the potential depth $U_0$ is very large in comparison to Fig.\;\ref{fig2}.}
	 \label{fig11}
\end{figure*}

The ratio of the discrete Bogoliubov frequencies $\omega_\eta$ to the particle emission threshold -$\mu$ as a function of sgn$(g)N$ is shown in Fig.\;\ref{fig10}(a). Here the sgn$(\delta g)N$ is introduced, because the properties of the droplet are governed  by the combination of the rescaled atom number $N$ and the sign of $\delta g$. The main results of the excitation spectrum are: (i) the breathing mode of the quantum droplet ($g>0$) is always below the particle emission threshold; (ii) there are more internal modes below the particle emission threshold with increasing $N$ for positive $g$, and thus higher internal modes are easily excited for large quantum droplets; (iii) the ratio $\omega_\eta/|\mu|$ tends to +1 for large negative sgn$(g)N$ where the droplet crosses over to soliton, indicating that 1D solitons do not sustain small-amplitude collective (internal) modes, but only the continuum spectrum. When $g = 0$, we arrive at a GPE with a rather unusual quadratic-only nonlinearity, where the ratio of the breathing mode frequency to the particle emission threshold is equal to 0.8904, independent of $N$, as shown in Fig.\;\ref{fig10}(b). These results have been reported in Ref. \cite{Tylutki2020}. In Fig.\;\ref{fig9}(b), we compare the breathing mode frequency $w_2$ with the oscillation frequency extracted from the periodic oscillation of the droplet width quantified by $X(t)$, and find a good agreement between them. This agreement confirms that the scattering of the quantum droplets results in the excitation of the internal mode.

As shown in Fig.\;\ref{fig6}, when the quantum droplet scatters off a sufficiently deep reflectionless potential, the variational method no longer works. However, even for large potential depth $U_0=25$, there is still a sharp transition between full transmission and full (quantum) reflection for both small ($N=1$) and large ($N=10$) quantum droplets, as illustrated in Figs.\;\ref{fig11}(a) and (d) , respectively. We also examine the spatiotemporal evolution of the density of quantum droplets for the initial speed slightly below [Figs.\;\ref{fig11}(b) and (e)] and above [Figs.\;\ref{fig11}(c) and (f)] the respective critical speed for small droplets [Figs.\;\ref{fig11}(b) and (c)] and large droplets [Figs.\;\ref{fig11}(e) and (f)]. We observe that the droplets are more excited when the potential well depth is large. Nevertheless, the droplets almost maintain their integrity after scattering. In this case, the trapped mode cannot be correctly captured by the trial function used in the variational method. For small quantum droplets, the condition $\omega_\eta< -\mu$ is marginally satisfied, i.e. only the breathing mode exists below the particle emission threshold. In general, when the droplet scatters from the reflectionless potential, the radiation is completely absent. This is due to the fact that the internal mode always occurs for quantum droplets. Only when the potential depth is sufficiently large can the small quantum droplets be excited to the continuum spectrum, resulting in vanishingly small emission of particles out of the droplet, as shown in the inset of Fig.\;\ref{fig11}(a). For large quantum droplets, however, more internal modes will stay below the particle emission threshold. The excitation of higher internal modes inhibits the loss of particle number, allowing the droplet to traverse the scattering region with probability one even for sufficiently large potential depths. This property of complete transmission and complete reflection is quite different from the scattering of solitons with a large number of particles, where such a large object may become radiated, and partially be trapped by the potential due to the fact that the soliton width shrinks as $N$ increases. As a matter of fact, for soliton scattering, it is impossible to completely avoid the radiation, because the 1D soliton supports no small-amplitude collective (internal) modes.
\section{Collisions between two droplets at reflectionless potential}
\label{section:VII}
\begin{figure*}[htbp]
	\centering
        \begin{minipage}[b]{.3\linewidth}
			\centering
\includegraphics[width=5.3cm]{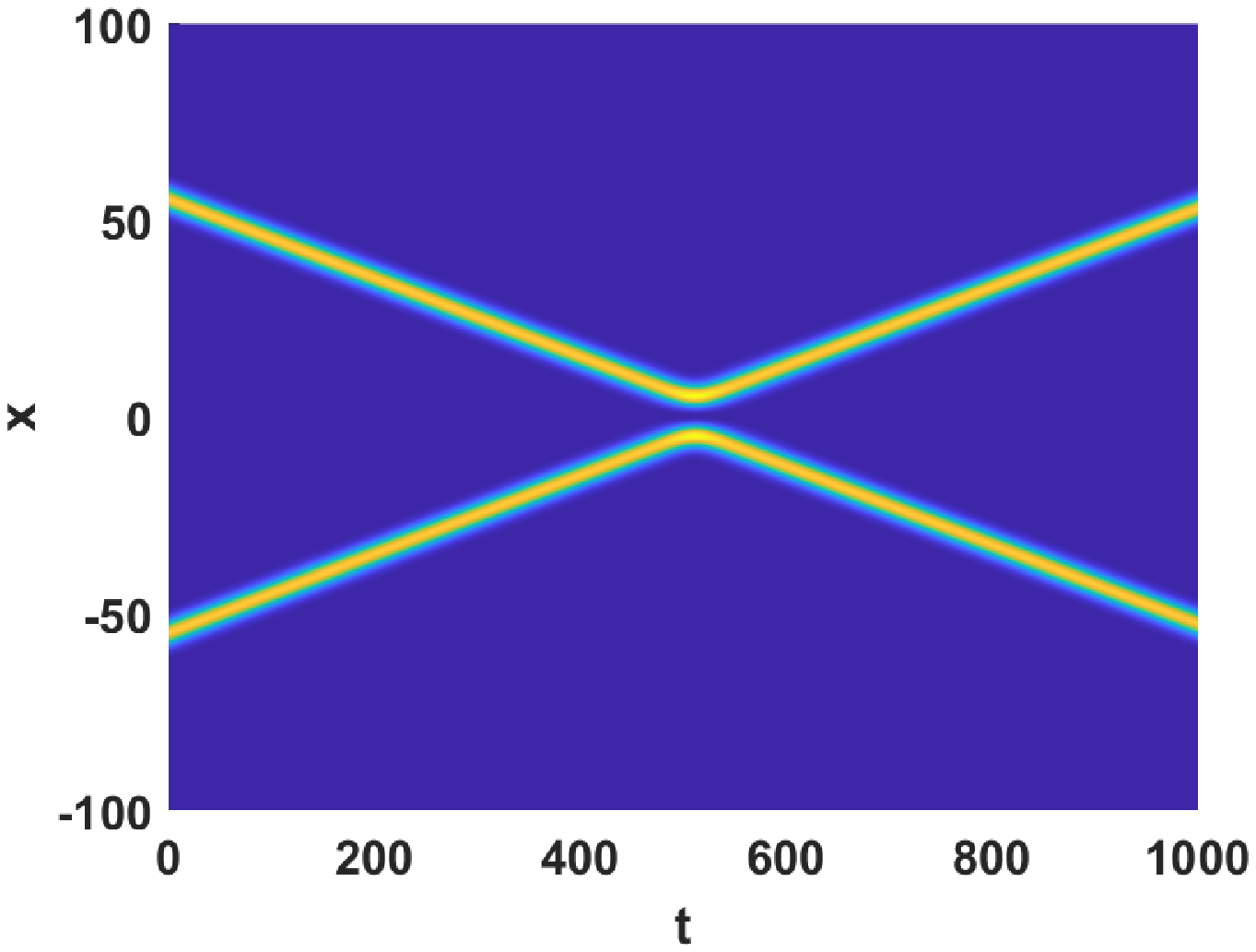}
\scriptsize { (a) $N_1=N_2=1$, $\varphi=\pi$, $U_0=0$}
		\end{minipage}
\begin{minipage}[b]{.3\linewidth}
			\centering
\includegraphics[width=5.3cm]{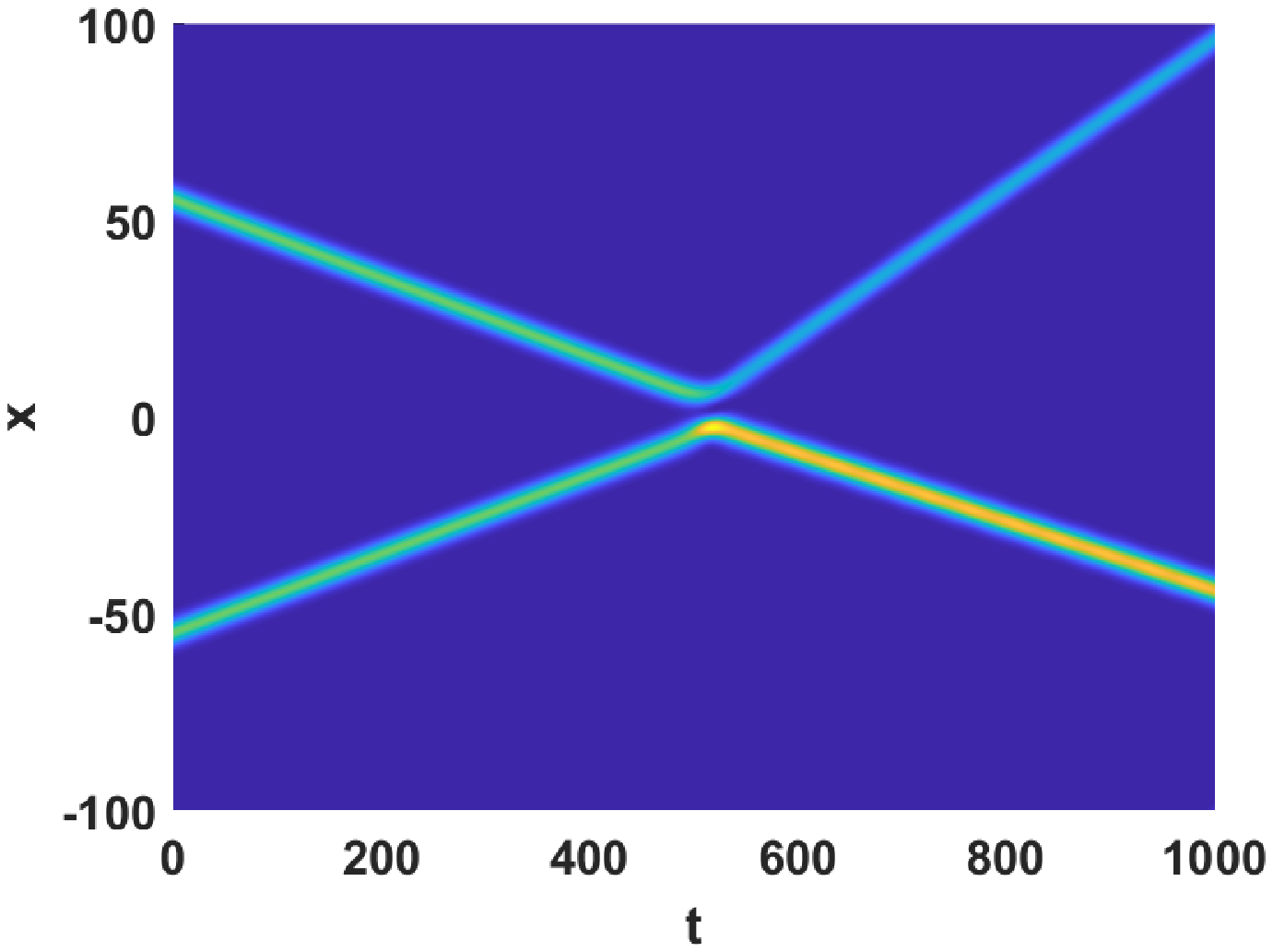}
\scriptsize { (b) $N_1=N_2=1$, $\varphi=\frac{3\pi}{2}$, $U_0=0$}
		\end{minipage}
\begin{minipage}[b]{.3\linewidth}
			\centering
\includegraphics[width=5.3cm]{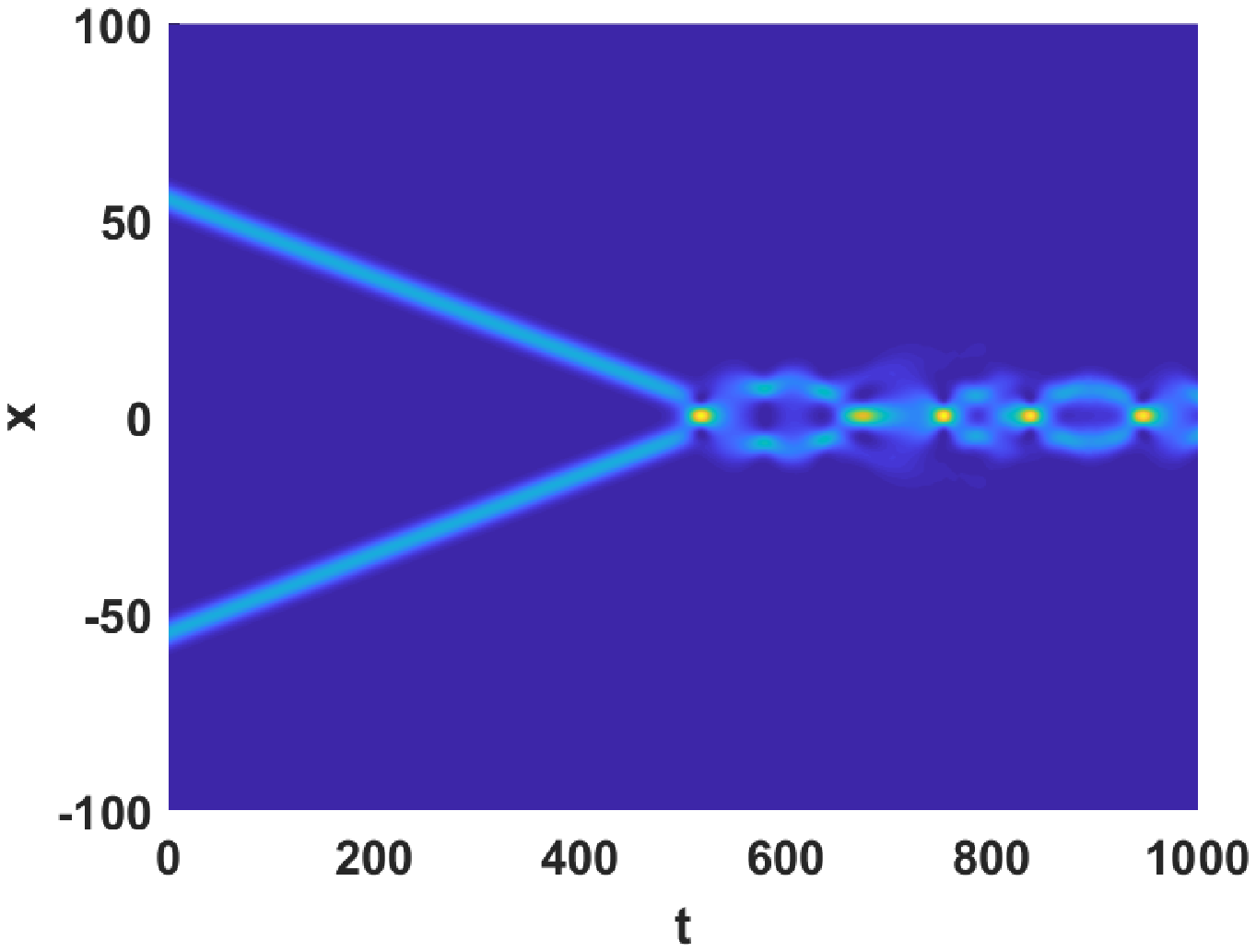}
\scriptsize { (c) $N_1=N_2=1$, $\varphi=0$, $U_0=0$}
		\end{minipage}
\\
\begin{minipage}[b]{.3\linewidth}
			\centering
\includegraphics[width=5.3cm]{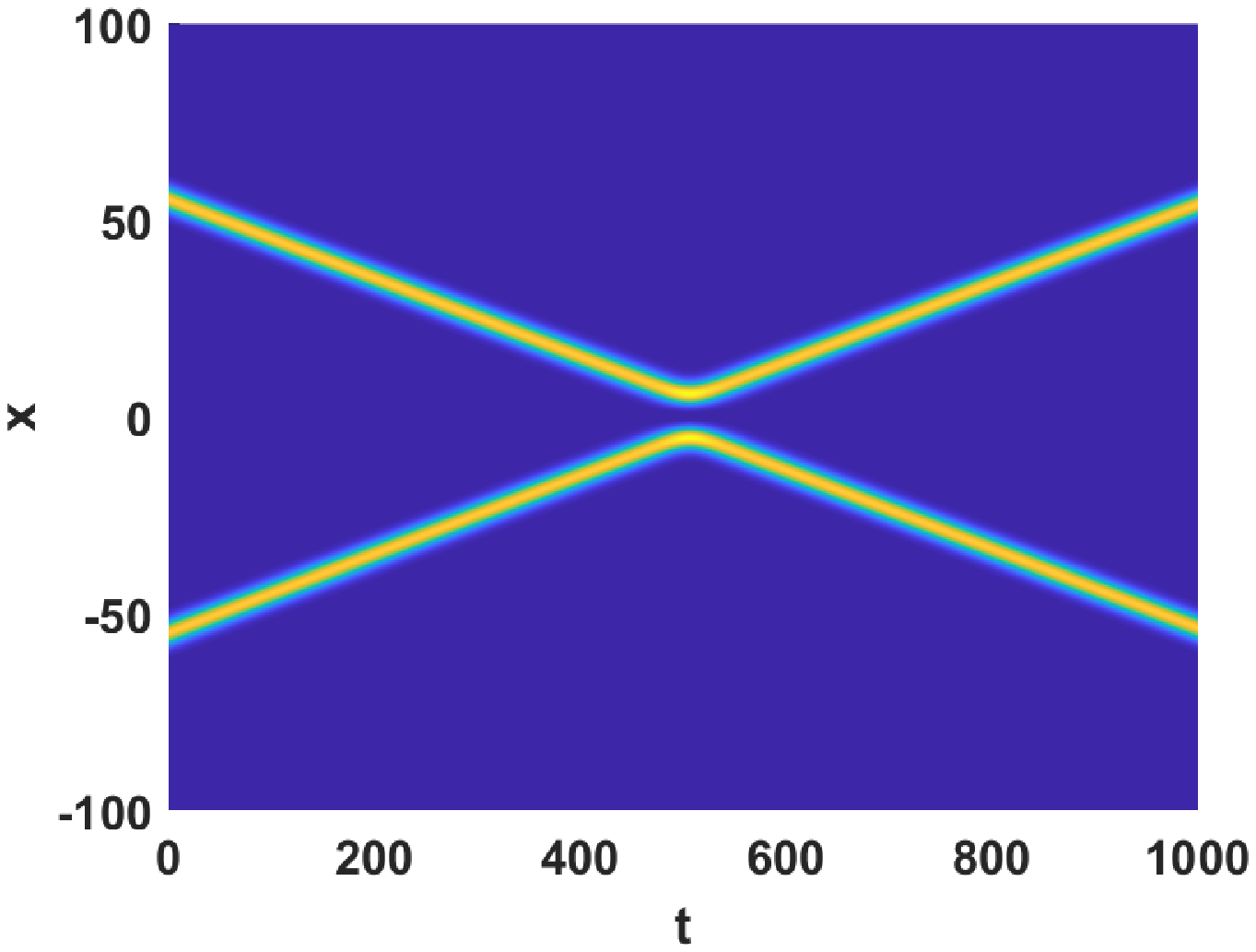}
\scriptsize { (d) $N_1=N_2=1$, $\varphi=0$, $U_0=4$}
		\end{minipage}
\begin{minipage}[b]{.3\linewidth}
			\centering
\includegraphics[width=5.3cm]{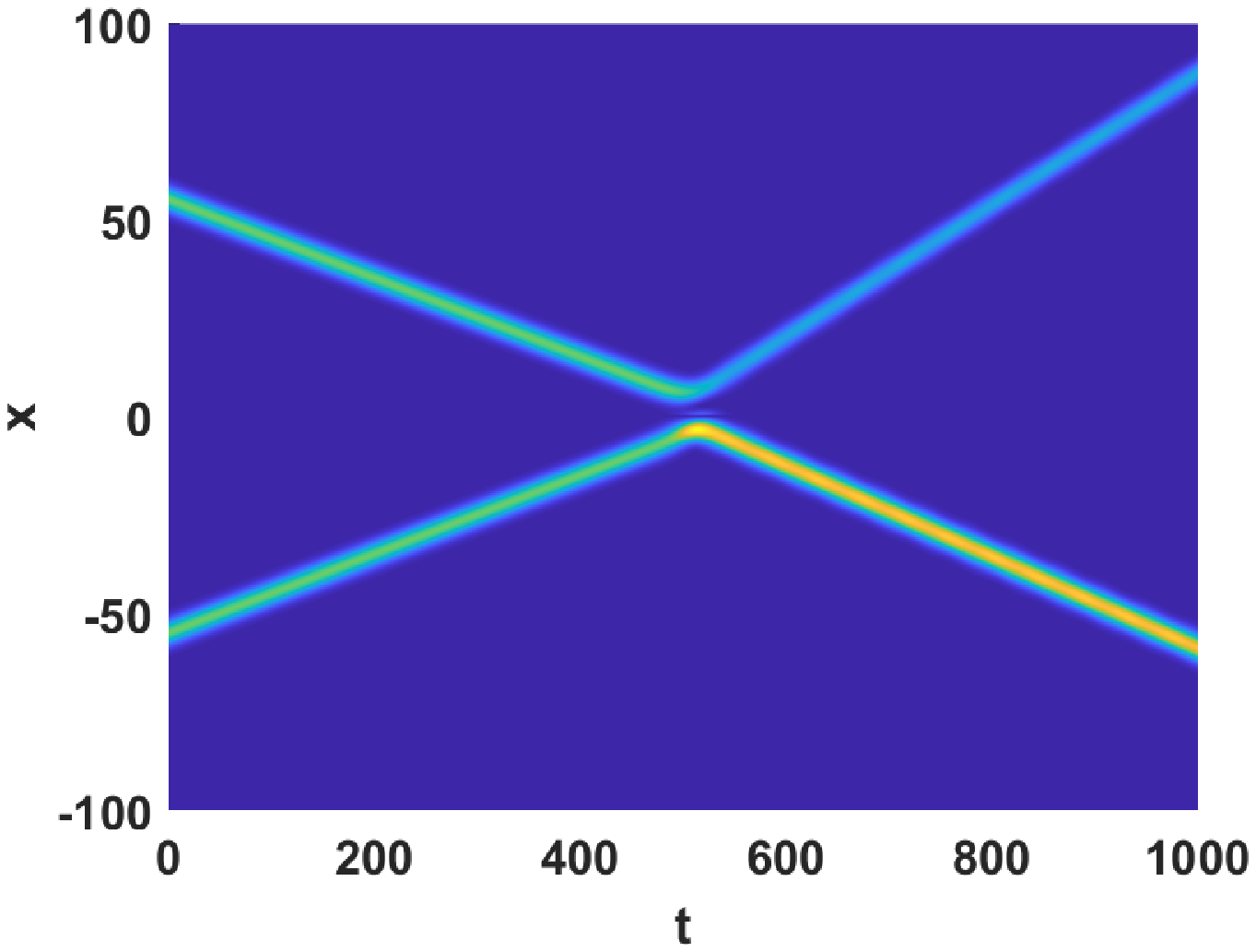}
\scriptsize { (e) $N_1=N_2=1$, $\varphi=\frac{\pi}{2}$, $U_0=4$}
		\end{minipage}
\begin{minipage}[b]{.3\linewidth}
			\centering
\includegraphics[width=5.3cm]{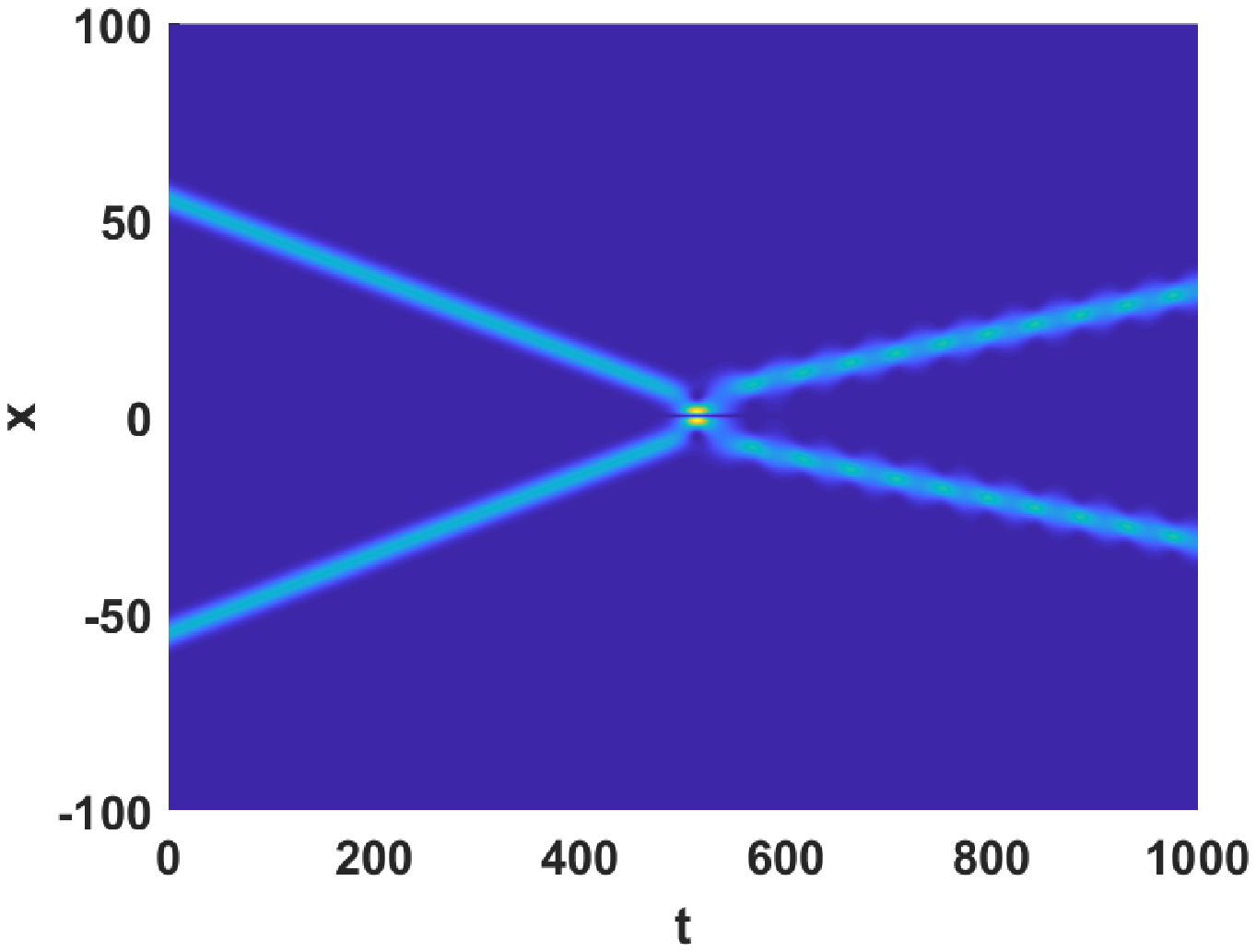}
\scriptsize { (f) $N_1=N_2=1$, $\varphi=\pi$, $U_0=4$}
		\end{minipage}
 \caption{Upper row: collisions between two quantum droplets ($N_1=N_2=1$) in free space, launched as Eq.\;\eqref{con:27}. (a) $\phi=\pi$; (b) $\phi=3\pi/2$; (c) $\phi=0$. Lower row: collisions between two quantum droplets ($N_1=N_2=1$) in the presence of reflectionless potential well centered at $x=0$. (d) $\phi=0$; (e) $\phi=\pi/2$; (f) $\phi=\pi$. In these figures, $v=0.1$ is fixed.}
    \label{fig12}
\end{figure*}

Low-energy collisions of two interacting quantum droplets can cause them to merge, repel, or evaporate by manipulating the quantum phases \cite{Pylak2022}. The dynamics of interacting quantum droplets is closely related to the relative phases of the two droplets. The scattering of quantum droplets at the reflectionless potential induces a phase change, which is expected to play an important role in the collision dynamics when the quantum droplets collide at the reflectionless potential. To probe the effect of the reflectionless potential on quantum droplet collisions, we simulated Eq.\;\eqref{con:14} using SSF method, selecting two oppositely moving droplets as the initial condition,

\begin{equation}\label{con:27}
\psi(x, 0)=e^{i v x +\phi} \psi_{1}\left(x+\tilde{x}_{0}\right)+e^{-i v x } \psi_{2}\left(x-\tilde{x}_{0}\right),
\end{equation}
where $\psi_1(x)$ and $\psi_2(x)$ are the stationary profiles of the quantum droplets normalized to $N_1$ and $N_2$, respectively, taken from Eq.\;\eqref{con:3}. This ansatz Eq.\;\eqref{con:26} sets two initial droplets, separated by distance $2\tilde{x}_0$, with speeds of $\pm v$ and an initial phase difference $\phi$ between them.

Fig.\;\ref{fig12} shows the density profiles of the collisions between a pair of slowly moving small droplets with equal norm ($N_1=N_2=1$) and equal and opposite speed ($v=0.1$) in free space (top row) and at the reflectionless potential centered at $x=0$ (bottom row). As illustrated in Figs.\;\ref{fig12} (a)-(c), two small droplets colliding in the free space repel each other at $\phi=\pi$ [Fig.\;\ref{fig12}(a)], experience mass transfer between the two droplets at $\phi=3\pi/2$ [Fig.\;\ref{fig12}(b)], and merge (small-amplitude repeated coalescence) at a relatively small value of $v$ [Fig.\;\ref{fig12}(c)] or pass through each other at large $v$ (not shown) when the initial relative phase is $\phi=0$. For comparison, we also explore the two small droplets colliding in the presence of a reflectionless potential, as shown in Figs.\;\ref{fig12}(d)-(f). The comparison shows that when the initial relative phase is changed by $\pi$, the picture of the collision between two droplets with and without potential is basically the same, indicating that the phase jump acquired by the scattering plays a key role in the collisions. The only exception is that, as can be seen by comparing Fig.\;\ref{fig12}(f) and Fig.\;\ref{fig12}(c), when the speed is $v=0.1$, the collision outcome of two small droplets of $\phi=\pi$ with reflectionless potential is different from that of two small droplets of $\phi=0$ without reflectionless potential: The former pass through each other and the latter merge. However, if we vary the speed, two small out-of-phase droplets colliding in the presence of the reflectionless potential will undergo a transition from merging to passing through each other, which is the same as the collision of two small in-phase droplets in free space. These results imply that the reflectionless potential in quantum droplet collisions corresponds to a $\pi$ phase difference ``generator''.
\begin{figure*}[htbp]
\centering
        \begin{minipage}[b]{.3\linewidth}
			\centering
\includegraphics[width=5.3cm]{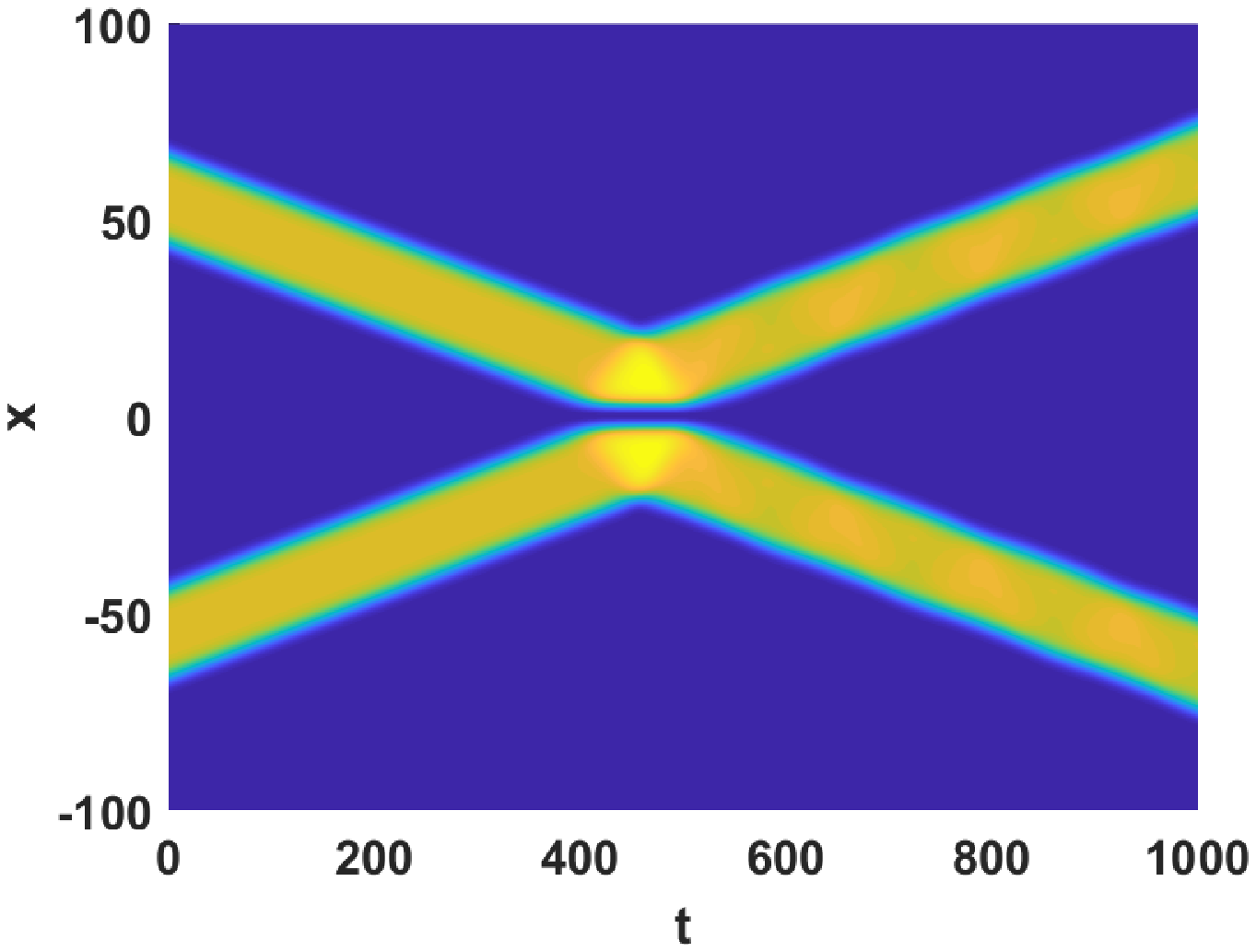}
\scriptsize { (a) $N_1=N_2=10$, $\varphi=\pi$, $U_0=0$}
		\end{minipage}
\begin{minipage}[b]{.3\linewidth}
			\centering
\includegraphics[width=5.3cm]{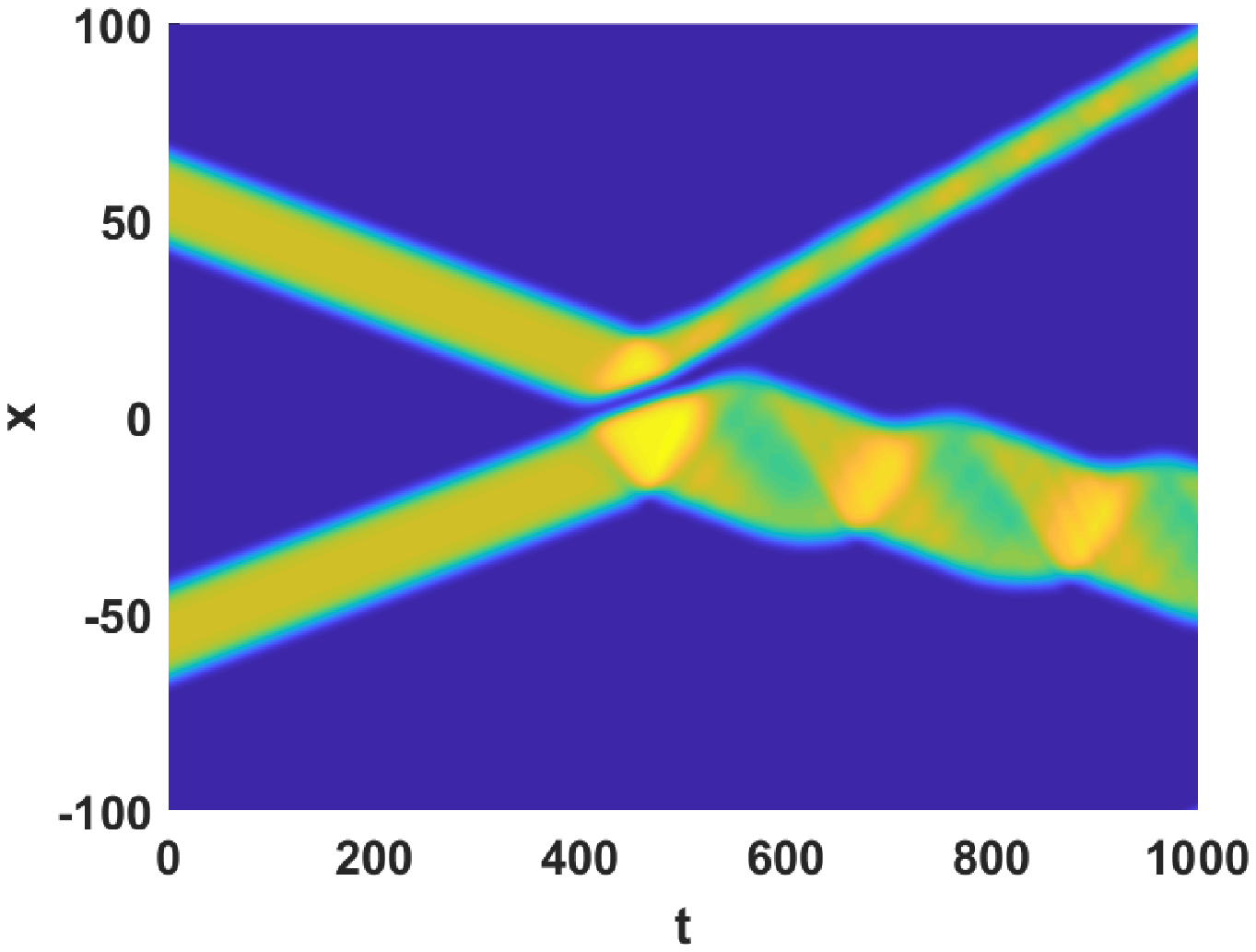}
\scriptsize { (b) $N_1=N_2=10$, $\varphi=\frac{3\pi}{2}$, $U_0=0$}
		\end{minipage}
\begin{minipage}[b]{.3\linewidth}
			\centering
\includegraphics[width=5.3cm]{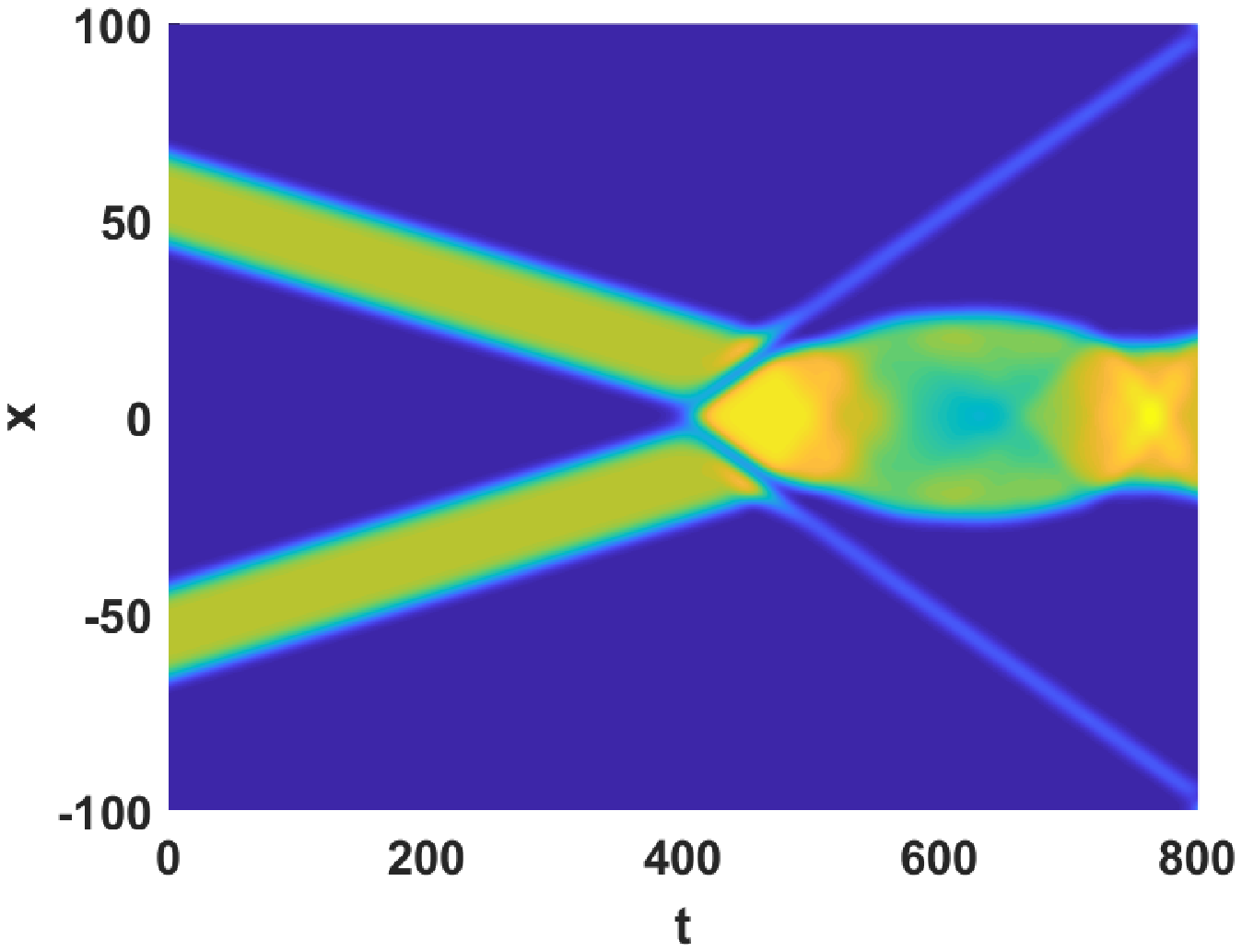}
\scriptsize { (c) $N_1=N_2=10$, $\varphi=0$, $U_0=0$}
		\end{minipage}
\\
\begin{minipage}[b]{.3\linewidth}
			\centering
\includegraphics[width=5.3cm]{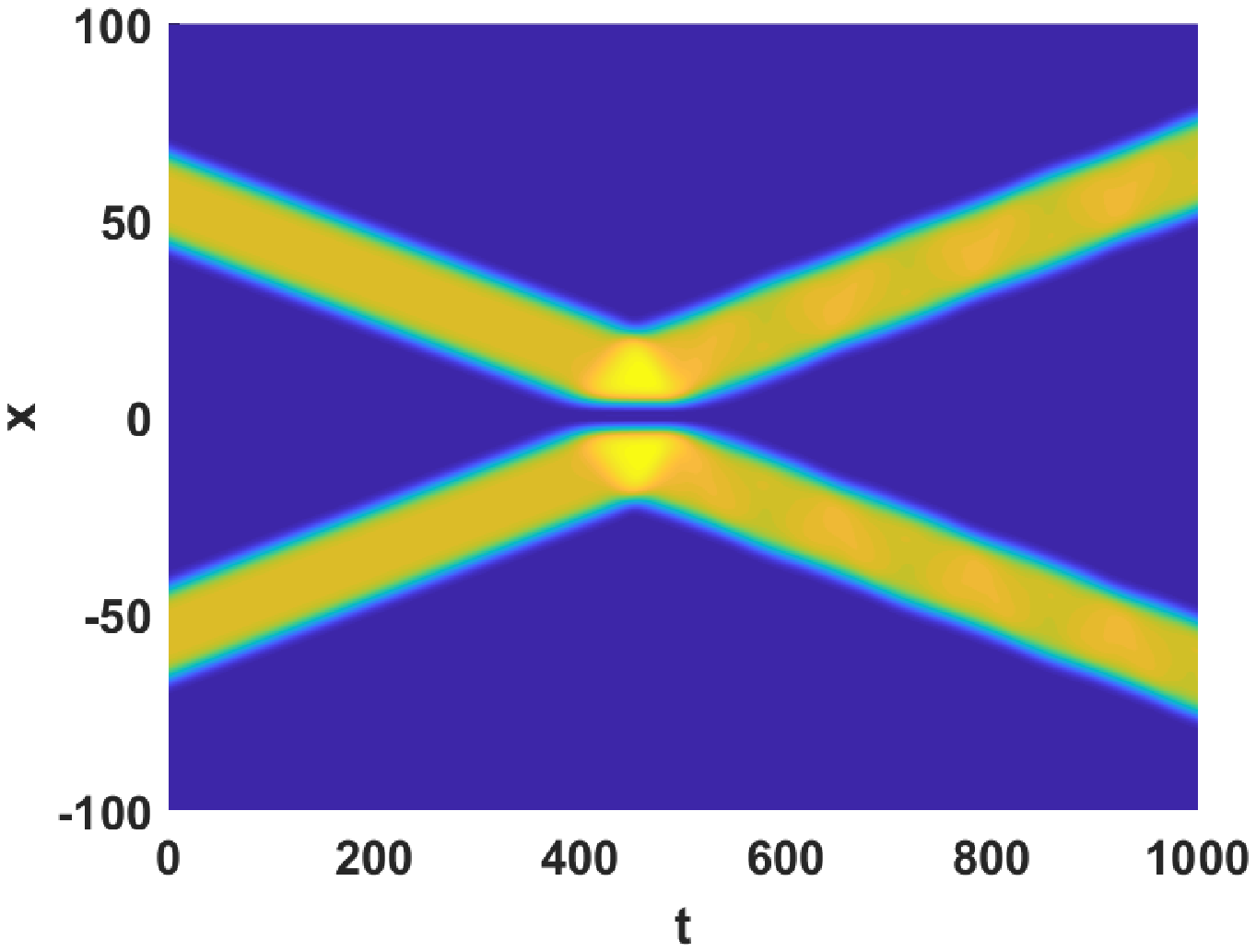}
\scriptsize { (d) $N_1=N_2=10$, $\varphi=0$, $U_0=4$}
		\end{minipage}
\begin{minipage}[b]{.3\linewidth}
			\centering
\includegraphics[width=5.3cm]{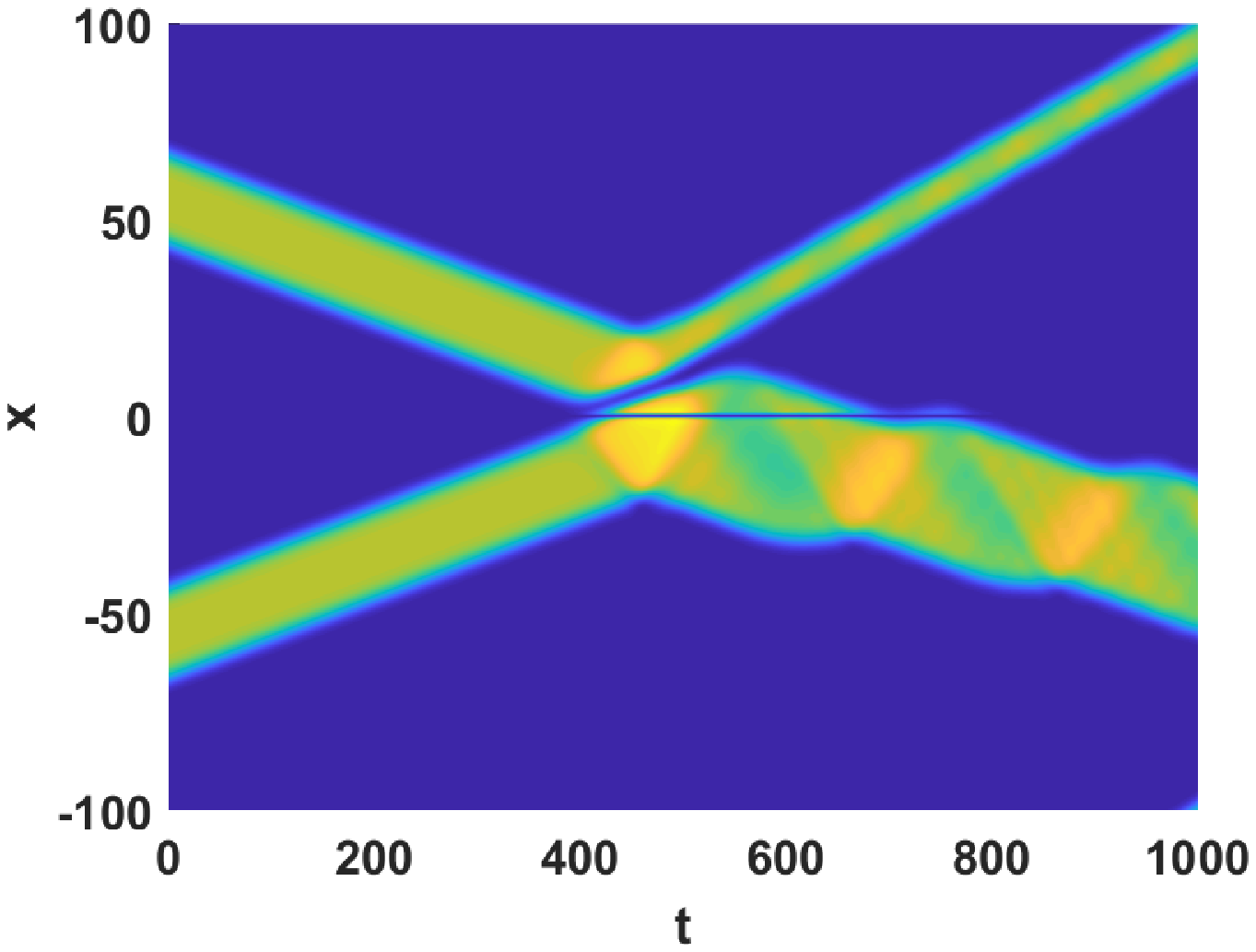}
\scriptsize { (e) $N_1=N_2=10$, $\varphi=\frac{\pi}{2}$, $U_0=4$}
		\end{minipage}
\begin{minipage}[b]{.3\linewidth}
			\centering
\includegraphics[width=5.3cm]{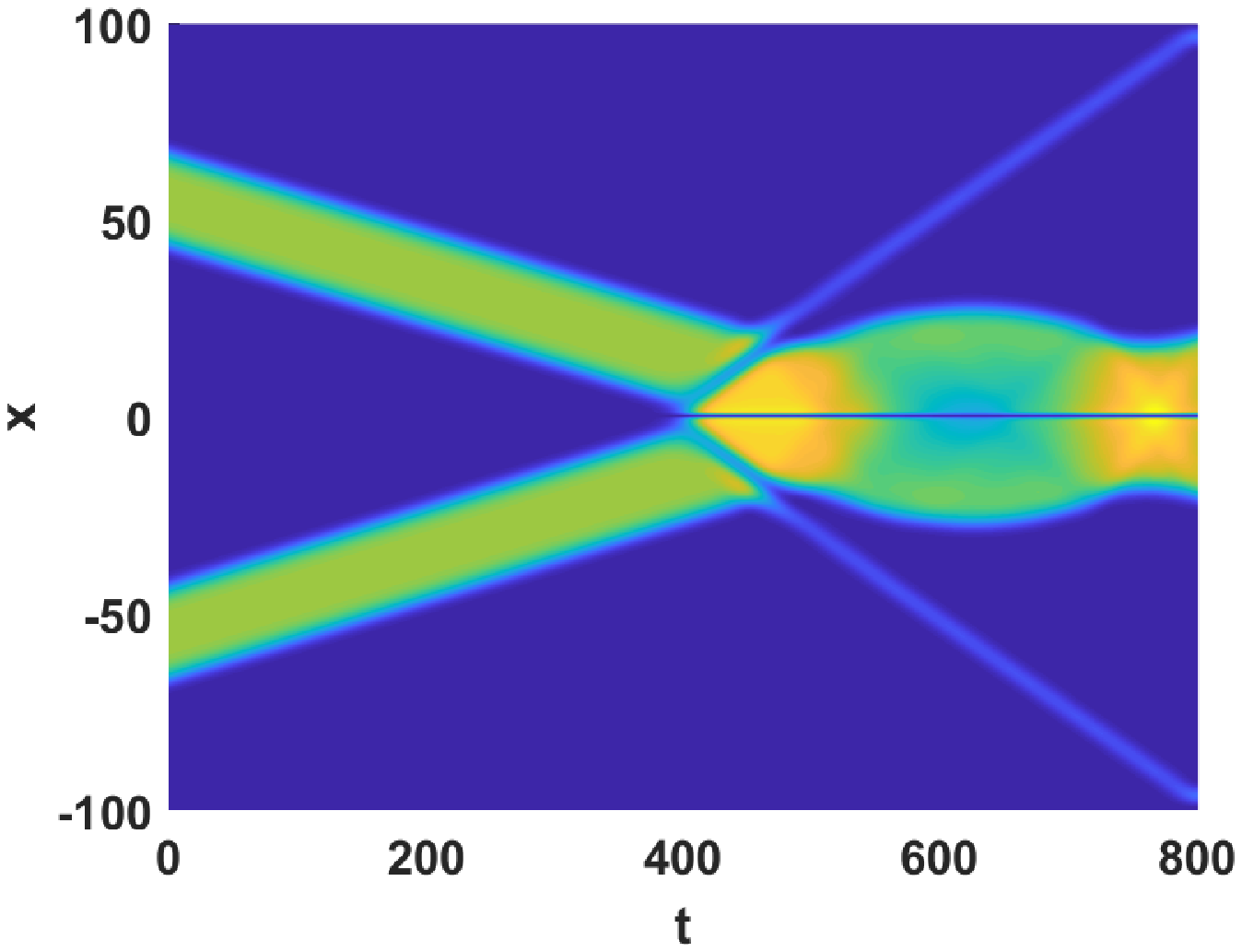}
\scriptsize { (f) $N_1=N_2=10$, $\varphi=\pi$, $U_0=4$}
		\end{minipage}
   \caption{Upper row: collisions between two quantum droplets ($N_1=N_2=10$) in free space, launched as Eq.\;\eqref{con:27}. (a) $\phi=\pi$; (b) $\phi=3\pi/2$; (c) $\phi=0$. Lower row: collisions between two quantum droplets ($N_1=N_2=10$) in the presence of reflectionless potential well centered at $x=0$. (d) $\phi=0$; (e) $\phi=\pi/2$; (f) $\phi=\pi$. In these figures, $v=0.1$ is fixed.}
\label{fig13}
\end{figure*}

Now we turn our attention to the case of collisions between two large quantum droplets with equal norm  $N_1= N_2=10$. The outcomes of the collision between two large droplets have been elaborated in Ref. \cite{Astrakharchik2018}, the distinctive features of which are this collision of such extended droplets leads to fragmentation and merger, accompanied by visible excitation of a single droplet as long as the initial phase difference is not equal to $\pi$. Comparing the results without external potential [Figs.\;\ref{fig13}(a)-(c)] and with external potential [Figs.\;\ref{fig13}(e)-(f)], we also observe that if we assign a $\pi$-shift to the initial relative phase, the outcomes of the collisions between the two slowly moving large droplets with external potential are basically the same as those without external potential. The only difference between the cases of without potential ($\phi = 0$) and  with potential ($\phi=\pi$) is that the presence of the reflectionless potential results in a central node in the newly formed central quiescent droplet. So far, we have dealt only with the collision of two droplets moving at a relatively low speed. Our numerical simulations (not shown) reveal that the conclusion about the role of the phase difference generated by the reflectioless potential in the quantum droplet collisions is unaffected when we consider two fast-moving droplets. Finally, we emphasize that the effect of the initial relative phase on the collision smears out when the reflectionless potential is deep, naturally because the droplet-potential interaction rather than the droplet-droplet interaction dominates in this case.

\section{Conclusion}
\label{section:VIII}
In this paper, we study the scattering of one-dimensional binary Bose gases forming self-bound quantum droplets from a reflectionless P\"{o}schl-Teller potential well, by solving the modified GPE with cubic (MF) and attractive quadratic (BMF) nonlinearities. Using direct numerical calculations and variational methods with a position-dependent wave function, the critical speed between quantum reflection and transmission and the corresponding trapped modes are accurately accounted for, and two types of physically distinct scattering of quantum droplets are identified. The scattering of small droplet exhibits soliton-like behaviour, where a spatially symmetric trapped mode is formed at the critical speed. However, the situation is quite different for large flat-top droplets$:$ the trapped mode formed at the critical speed turns out to be the nonlinear stationary state of the system with a density profile asymmetric with respect to the center of the potential, although the system with large atom numbers also admits spatially symmetric stationary bound states. Furthermore, we observe a nonmonotonic dependence of the critical speed on the droplet size: For small droplets, the critical speed increases with the atom number, as in the case of bright solitons, whereas for large droplets the critical speed decreases with increasing atom number. We also discuss in detail the energy exchange mechanism of quantum droplets during scattering, both from the classical particle-like motion and quantum matter-wave viewpoints, and clarify the sharp transition between full reflection and full transmission at the critical initial speed. Another exotic property associated with the scattering of quantum droplets is that the scattering excites the internal modes below the particle-emission threshold, making it possible to completely prevent the quantum droplets from emitting particles, which is impossible for soliton scattering since the bright solitons have no internal modes but only the continuum spectrum. Analysis of the small-amplitude excitation spectra shows that as the number of particles increases, the droplets support more internal modes below the particle emission threshold, and thus excitation of higher internal modes prevents particle loss and makes it easier for the droplets to pass through the scattering region without radiation. This property is quite different from the scattering of solitons with a large number of particles, where such a large object may be radiated, and partially trapped by the potential because the width of the solitons, unlike the large flat-top droplets, decreases as the number of particles increases. Finally, we study the collisions of quantum droplets at the reflectionless potential, and find that if we impose a $\pi$-shift on the relative phase between the initial two droplets, the outcomes of the collisions between the quantum droplets with the reflectionless potential are basically the same as those in free space. The above results show that quantum droplets have good scattering properties in the P\"{o}schl-Teller reflectionless potential well. Some relevant questions along this line need to be further investigated, for example, what is the scattering behaviour of quantum droplets in other potential wells, such as Gaussian and square potential wells, etc.? are these good scattering properties preserved? The study of these questions will contribute to a better understanding of the interaction between the quantum droplets and the external potentials.

\acknowledgments
The work was supported by  the National
Natural Science Foundation of China (Grant No. 11975110), the Natural Science Foundation of Zhejiang Province (Grant No. LY21A050002), and Zhejiang Sci-Tech University Scientific Research
Start-up Fund (Grant No. 20062318-Y). Yu Guo was also supported by the National Natural Science Foundations of China (Grant No. 12275033), and the Natural Science Foundation of Hunan Province (Grant No. 2022JJ30582).



\end{document}